\documentclass[prd,superscriptaddress,amsfonts,amssymb,amsmath,showpacs,twocolumn]{revtex4-2}
\usepackage{bm}
\usepackage{amsfonts}
\usepackage{latexsym}
\usepackage[latin1]{inputenc}
\usepackage{graphicx}
\usepackage{amsmath}
\usepackage{palatino}
\usepackage{mathpazo}
\usepackage{textcomp}
\usepackage{float}
\usepackage{booktabs}
\usepackage{dcolumn}
\usepackage{ragged2e}
\usepackage{hyperref}
\hypersetup{colorlinks=true,linkcolor=red,filecolor=magenta,    urlcolor=blue}
\usepackage{amsmath}
\usepackage{xcolor}
\usepackage{orcidlink}
\usepackage[caption=false]{subfig}
\usepackage{commath}
\def\jnl@style{\it}
\def\aaref@jnl#1{{\jnl@style#1}}

\def\aaref@jnl#1{{\jnl@style#1}}

\def\aj{\aaref@jnl{AJ}}                   
\def\apj{\aaref@jnl{ApJ}}                 
\def\apjl{\aaref@jnl{ApJ}}                
\def\apjs{\aaref@jnl{ApJS}}               
\def\apss{\aaref@jnl{Ap\&SS}}             
\def\aap{\aaref@jnl{A\&A}}                
\def\aapr{\aaref@jnl{A\&A~Rev.}}          
\def\aaps{\aaref@jnl{A\&AS}}              
\def\mnras{\aaref@jnl{Mon.~Not.~Roy.~Astron.~Soc.}}             
\def\prd{\aaref@jnl{Phys.~Rev.~D}}        
\def\prc{\aaref@jnl{Phys.~Rev.~C}}  
\def\prl{\aaref@jnl{Phys.~Rev.~Lett.}}    
\def\qjras{\aaref@jnl{QJRAS}}             
\def\skytel{\aaref@jnl{S\&T}}             
\def\ssr{\aaref@jnl{Space~Sci.~Rev.}}     
\def\zap{\aaref@jnl{ZAp}}                 
\def\nat{\aaref@jnl{Nature}}              
\def\aplett{\aaref@jnl{Astrophys.~Lett.}} 
\def\apspr{\aaref@jnl{Astrophys.~Space~Phys.~Res.}} 
\def\physrep{\aaref@jnl{Phys.~Rep.}}      
\def\physscr{\aaref@jnl{Phys.~Scr}}       
\def\commat{\aaref@jnl{Comm.~Math.~Phys.}}              
\def\science{\aaref@jnl{Science}}               
\def\cqg{\aaref@jnl{Classical Quant.~Grav.}}            
\def\jpcs{\aaref@jnl{JPCS}}                                     
\def\ijmpd{\aaref@jnl{Int.~J.~Mod.~Phys.~D}}                    
\def\grg{\aaref@jnl{Gen.~Relat.~Gravit.}}               
\def\rpp{\aaref@jnl{Rep.~Prog.~Phys.}}          
\def\npa{\aaref@jnl{Nucl.~Phys.~A}}        
\def\lrr{\aaref@jnl{Living Rev.~Rel.}}                   
\def\jcap{\aaref@jnl{J.~Cosmology Astropart.~Phys.}}    
\def\rmp{\aaref@jnl{Rev.~Mod.~Phys.}}   
\def\epjc{\aaref@jnl{Eur.~Phys.~J.~C}}  
\def\arxiv{\aaref@jnl{arxiv.org}}

\allowdisplaybreaks[1]

\begin{document}
\color{black}       
\title{Traversable wormhole geometries in $f(Q)$ gravity}
\author{Zinnat Hassan\orcidlink{0000-0002-6608-2075}}
\email{zinnathassan980@gmail.com}
\affiliation{Department of Mathematics, Birla Institute of Technology and
Science-Pilani,\\ Hyderabad Campus, Hyderabad-500078, India.}
\author{Sanjay Mandal\orcidlink{0000-0003-2570-2335}}
\email{sanjaymandal960@gmail.com}
\affiliation{Department of Mathematics, Birla Institute of Technology and
Science-Pilani,\\ Hyderabad Campus, Hyderabad-500078, India.}

\author{P.K. Sahoo\orcidlink{0000-0003-2130-8832}}
\email{pksahoo@hyderabad.bits-pilani.ac.in}
\affiliation{Department of Mathematics, Birla Institute of Technology and
Science-Pilani,\\ Hyderabad Campus, Hyderabad-500078, India.}
%
\date{\today}
\begin{abstract}
The current interests in the universe motivate us to go beyond Einstein's General theory of relativity. One of the interesting proposals comes from a new class of teleparallel gravity named symmetric teleparallel gravity, i.e., $f(Q)$ gravity, where the non-metricity term $Q$ is accountable for fundamental interaction. These alternative modified theories of gravity's vital role are to deal with the recent interests and to present a realistic cosmological model. This manuscript's main objective is to study the traversable wormhole geometries in $f(Q)$ gravity. We construct the wormhole geometries for three cases: (i) by assuming a relation between the radial and lateral pressure, (ii) considering phantom energy equation of state (EoS), and (iii) for a specific shape function in the fundamental interaction of gravity (i.e. for linear form of $f(Q)$). Besides, we discuss two wormhole geometries for a general case of $f(Q)$ with two specific shape functions. Then, we discuss the viability of shape functions and the stability analysis of the wormhole solutions for each case. We have found that the null energy condition (NEC) violates each wormhole model which concluded that our outcomes are realistic and stable. Finally, we discuss the embedding diagrams and volume integral quantifier to have a complete view of wormhole geometries.
\end{abstract}

\maketitle
\section{Introduction}
In the last few decades, the evolution of the universe is described by a large amount of astrophysical observational data such as Laser Interferometer Gravitational-Wave Observatory (LIGO) \cite{Gourgoulhon/2019}, Virgo \cite{Abuter/2020}, Event Horizon Telescope (EHT) \cite{Chael/2016, Akiyama/2019}, Advanced Telescope for High-Energy Astrophysics (ATHENA) \cite{Barcons/2017}, International Gamma-Ray Astrophysics Laboratory (INTEGRAL) \cite{Winkler/2003}, Imaging x-ray Polarimetry mission (IXPE) \cite{Soffitta/2013}, XMM-Newton \cite{Beckwith/2004, Tomsick/2014}, and Swift \cite{Burrows/2005}. These observations motivate the research community to explore and develop more insights and advanced strategies to test the gravity in strong gravitational fields. The concept of wormholes (WHs) reserved special attention, which are the hypothetical tunnels connecting different asymptotically flat regions of the spacetime or universes. These are unusual astrophysical objects provided with no singularities and horizons \cite{Visser/1995}. To explore these objects challenges us in the fundamental physics and draws interest to understand the unique form of matter called `exotic matter' and how gravity shapes its' geometry.

In general relativity and modified theories of gravity, WHs are the solutions of the field equations \cite{Capozziello/2011}. These solutions would act as a short-cut way between two distant universes and be used to make a time machine for which a stable traversable WH is required \cite{Morris/1988, Morris/1988a}. The WH-like solution was first coined by Einstein and Rosen in their collaborative work \cite{Einstein/1935}. Later on, some studies showed the WHs features like the Einstein-Rosen bridge coming from the connection of two Schwarzschild-solutions \cite{Ovgun/2019}. Moreover, these studies are done in the presence of event horizon, which results, anyone trying to escape the WH throat (the wormhole short-cut path is traversable through a minimal surface area called WH throat), always falls into the singularity \cite{Einstein/1935}. To resolve this issue, one can make a prior assumption on the metric. Besides, one can adopt the Birkhoff theorem approach to put bounds at the WH throat \cite{Morris/1988}. In this case, the radial tension might be large enough to exceed the total mass-energy density, i.e., $\tau_0> \rho_0 c^2$ must hold. In consequence, the energy-momentum tensor violates the null-energy condition (NEC) at the throat i.e. $T_{\mu\nu}k^{\mu}k^{\nu}$ \cite{Capozziello/2015}. Besides, the existence of this type of exotic matter is possible as the dark energy or phantom energy shows the same kind of features to explain the accelerated expansion of the universe \cite{Eiroa/2010}. Several wormhole geometries have been studied using phantom energy, see, e.g. \cite{Sushkov/2005}.

In this view, there are several approaches have been proposed to relieve the problem. But, we cannot impose the above condition directly on the matter. Therefore, we can either consider the exotic form of matter or modified theories of gravity where the higher-order curvature terms provide the WHs properties to deal with this problem. In this context, the traversable WHs and thin-shell WHs with their features have been studied in $f(R)$ gravity \cite{Lobo/2009}. The wormhole geometries in $f(R,T)$ gravity have been studied by presuming different radial and lateral pressure relations in \cite{Moraes/2017}. Also, they obtained the solutions for shape functions and discussed their properties with energy conditions. Moraes and his collaborators studied the WH solutions in $R^2$-gravity and exponential $f(R,T)$ formalism \cite{Moraes/2019}. Moreover, WHs are discussed widely in teleparallel gravity and other extended theories of gravity \cite{Capozziello/2012}. Besides, one of the interesting work done by M. Zubair et al. \cite{Saira/2016}, where they have studied the wormhole solution for three separated cases such as isotropic, anisotropic and barotropic fluids in $f(R,T)$ gravity. Also they have concluded that anisotropic matter presents a realistic and stable wormhole model.

This manuscript focused on exploring the WH geometries in symmetric teleparallel gravity ($f(Q)$ gravity). As the WHs are supported by the exotic matter, and that is an entirely unsolved problem. This issue motivates us to study the WH geometries through modified theories where curvature explains the WHs and retains standard matter. Among several motivations to explore the WHs in modified theories, we highlight $f(Q)$ gravity, introduced by Jimenez et al. \cite{Jimenez/2018}, where the gravitational interaction is described by the non-metricity term $Q$. Recently, the study on $f(Q)$ gravity has been developed rapidly in theoretical and observational fields (see details in \cite{Harko/2018}). Sahoo and his group studied the cosmic acceleration of the universe in the presence bulk viscous fluid with the latest Pantheon dataset \cite{solanki/2021}. Therefore, the study on WHs in $f(Q)$ gravity may bring new insights into cosmology as it is a novel approach.

Here, we have studied three types of WH geometries by considering (1) a relation between the radial and lateral pressure, (2) phantom energy equation of state (EoS), and (3) a specific shape-function for $b(r)$. We have calculated and discussed the properties of $b(r)$ for three cases. To do the stability analysis of the wormhole solutions, we have tested the energy conditions. Besides this, we have measured the exotic matter for all WH geometries.

The outlines of this manuscript layered as follows. In Sec.  \ref{sec2}, we have presented the basic formulation for $f(Q)$ gravity. And, basic conditions and remarks required for a traversable wormhole have been discussed in Sec. \ref{sec3}. In Sec \ref{sec4}, we discussed the framework of the traversable wormhole geometries in $f(Q)$ gravity. We also discussed the energy conditions and three types of wormhole solutions. In Sec. \ref{sec5}, we construct the motion equations for two different form of $f(Q)$. Then, we discuss three wormhole solutions for the linear form of $f(Q)$ and two wormhole solutions for a non-linear form of $f(Q)$ in Sec. \ref{sec6} and \ref{sec7}, respectively. Embedding diagrams for wormhole solutions have been discussed in Sec \ref{sec8}. In Sec. \ref{sec9}, we discussed the volume integral quantifier to measure the exotic matter. Finally we conclude our outcomes in Sec. \ref{sec10}.

\section{Basic Field Equations in $f(Q)$ gravity}\label{sec2}
Here, we have considered the action for symmetric teleparallel gravity is given by \cite{Jimenez/2018}
\begin{equation}
\label{1}
\mathcal{S}=\int\frac{1}{2}\,f(Q)\sqrt{-g}\,d^4x+\int \mathcal{L}_m\,\sqrt{-g}\,d^4x\,
\end{equation}
where $f(Q)$ represents the function form of Q, $g$ is the determinant of the metric $g_{\mu\nu}$, and $\mathcal{L}_m$ is the matter Lagrangian density.\\
The non-metricity tensor and its traces can be written as\\
\begin{equation}
\label{2}
Q_{\lambda\mu\nu}=\bigtriangledown_{\lambda} g_{\mu\nu}
\end{equation}
\begin{equation}
\label{3}
Q_{\alpha}=Q_{\alpha}\;^{\mu}\;_{\mu},\; \tilde{Q}_\alpha=Q^\mu\;_{\alpha\mu}
\end{equation}
Also, the non-metricity tensor helps us to write the  superpotential as 
\begin{equation}
\label{4}
P^\alpha\;_{\mu\nu}=\frac{1}{4}\left[-Q^\alpha\;_{\mu\nu}+2Q_{(\mu}\;^\alpha\;_{\nu)}+Q^\alpha g_{\mu\nu}-\tilde{Q}^\alpha g_{\mu\nu}-\delta^\alpha_{(\mu}Q_{\nu)}\right]
\end{equation}
where the trace of non-metricity tensor \cite{Jimenez/2018} has the form
\begin{equation}
\label{5}
Q=-Q_{\alpha\mu\nu}\,P^{\alpha\mu\nu}
\end{equation}
Again, by definition, the energy-momentum tensor for the fluid description of the spacetime cab be written as
\begin{equation}
\label{6}
T_{\mu\nu}=-\frac{2}{\sqrt{-g}}\frac{\delta\left(\sqrt{-g}\,\mathcal{L}_m\right)}{\delta g^{\mu\nu}}
\end{equation}
Now, one can write the motion equations by varying the action \eqref{1} with respect to metric tensor $g_{\mu\nu}$, which can be written as
\begin{multline}
\label{7}
\frac{2}{\sqrt{-g}}\bigtriangledown_\gamma\left(\sqrt{-g}\,f_Q\,P^\gamma\;_{\mu\nu}\right)+\frac{1}{2}g_{\mu\nu}f \\
+f_Q\left(P_{\mu\gamma i}\,Q_\nu\;^{\gamma i}-2\,Q_{\gamma i \mu}\,P^{\gamma i}\;_\nu\right)=-T_{\mu\nu},
\end{multline}

where $f_Q=\frac{df}{dQ}$. Also varying \eqref{1} with respect to the connection,one obtains
\begin{equation}
\label{8}
\bigtriangledown_\mu \bigtriangledown_\nu \left(\sqrt{-g}\,f_Q\,P^\gamma\;_{\mu\nu}\right)=0.
\end{equation}

\section{Basic Conditions for Traversable wormholes}\label{sec3}

Now, we consider a general spherically symmetric, static wormhole spacetime of the Morris Thorne class. This spacetime is generically written as\\
\begin{equation}
\label{9}
ds^2=-e^{2\Phi(r)}dt^2+\left(1-\frac{b(r)}{r}\right)^{-1}dr^2+r^2d\theta^2+r^2\text{sin}^2\theta d\phi^2
\end{equation}
where $\Phi(r)$ is the redshift function of radial co-ordinate $r$ $(0<r_0\leq r\leq \infty)$ and its value always finite everywhere to avoid the event horizons. $b(r)$ is the shape function that determines the shape of the wormhole.  To investigate the wormhole geometry, $b(r)$ has to satisfy the following conditions:

\begin{itemize}
\item Throat condition: $b(r_0)=r_0$ and $b(r)$ should be less than $r$ for $r>r_0$.
\item Flaring out condition: $b'(r_0)<1$ i.e. $\frac{b(r)-rb'(r)}{b^2(r)}>0$, where $'$ represents derivative w.r.t. $r$.
\item Asymptotically Flatness condition: $\frac{b(r)}{r}\rightarrow o$ as $r\rightarrow \infty$.
\end{itemize}
Another important criterion is the proper radial distance $l(r)$, defined as\\
\begin{equation}
\label{10}
\ell(r)=\pm\int_{r_0}^{r}\frac{dr}{\sqrt{1-\frac{b(r)}{r}}}
\end{equation}

To have a proper description of traversable WH, $\ell (r)$ must be finite over the radial coordinate. Thus, it is a decreasing function. First, it falls down from the upper universe to the wormhole's throat, i.e., $\ell=+\infty$ to $\ell=0$, and then from wormhole's throat to lower universe, i.e., $\ell=0$ to $\ell=-\infty$. Besides, $\ell$ should be greater than or equal to the radial coordinate distance i.e., $\lvert\ell(r)\rvert\geq{r-r_0}$. The signature of $\ell$ represents the lower and upper parts of the wormhole. The positive and negative sign of $\ell$ denotes the WH's upper and lower sections, and both sections are connected by the wormhole's throat.

For the present interest, let us consider the matter is described by an anisotropic stress-energy tensor of the form
\begin{equation}
\label{13}
T_{\mu}^{\nu}=\left(\rho+P_t\right)u_{\mu}\,u^{\nu}-P_t\,\delta_{\mu}^{\nu}+\left(P_r-P_t\right)v_{\mu}\,v^{\nu}
\end{equation}
where $u_{\mu}$ is the four-velocity, $v_{\mu}$ the unitary space-like vector in the radial direction, $\rho$ is the energy density, $P_r$ is the pressure in the direction of $u_{\mu}$ (radial pressure) and $P_t$ is the pressure orthogonal to $v_{\mu}$(tangential pressure). Here, $P_r$ and $P_t$ are functions of redial component $r$.

\section{Wormhole Geometries in $f(Q)$ Gravity}\label{sec4}

In this section, we discuss the different kinds of wormhole solutions with their self-stability. The trace of the non-metricity tensor $Q$ for the wormhole metric in \eqref{9} takes the form below,
\begin{equation}
\label{12}
Q=-\frac{2}{r}\left(1-\frac{b(r)}{r}\right)\left(2\phi^{'}(r)+\frac{1}{r}\right).
\end{equation}

Now, by substituting \eqref{9} and \eqref{13} in \eqref{7} one can find the following field equations
\begin{widetext}
\begin{equation}
\label{14}
\left[\frac{1}{r}\left(-\frac{1}{r}+\frac{rb^{'}(r)+b(r)}{r^2}-2\phi^{'}(r)\left(1-\frac{b(r)}{r}\right)\right)\right]f_Q-\frac{2}{r}\left(1-\frac{b(r)}{r}\right)\dot{f}_Q-\frac{f}{2}=-\rho,
\end{equation}
\begin{equation}
\label{15}
\left[\frac{2}{r}\left(1-\frac{b(r)}{r}\right)\left(2\phi^{'}(r)+\frac{1}{r}\right)-\frac{1}{r^2}\right]f_Q+\frac{f}{2}=-P_r,
\end{equation}
\begin{multline}
\label{16}
\left[\frac{1}{r}\left(\left(1-\frac{b(r)}{r}\right)\left(\frac{1}{r}+\phi^{'}(r)\left(3+r\phi^{'}(r)\right)+r\phi^{''}(r)\right)-\frac{rb^{'}(r)-b(r)}{2r^2}\left(1+r\phi^{'}(r)\right)\right)\right]f_Q+\\
\frac{1}{r}\left(1-\frac{b(r)}{r}\right)\left(1+r\phi^{'}(r)\right)\dot{f}_Q+\frac{f}{2}=-P_t.
\end{multline}
\end{widetext}
Using \eqref{14}-\eqref{16}, one can study different wormhole models and their properties.\\

\subsection{Energy Conditions}

Energy conditions are discussed about the physically realistic matter configuration that developed from the Raychaudhuri equations. The Raychaudhuri equations state the temporal evolution of expansion scalar ($\theta$) for the congruences of timelike ($u^\mu$) and null ($\eta_\mu$) geodesics as \cite{Raychaudhuri/1955}
\begin{equation}
\label{1a}
\frac{d\theta}{d\tau}-\omega_{\mu\nu}\,\omega^{\mu\nu}+\sigma_{\mu\nu}\sigma^{\mu\nu}+\frac{1}{3}\theta^2+R_{\mu\nu}u^\mu\,u^\nu=0
\end{equation}
\begin{equation}
\label{1b}
\frac{d\theta}{d\tau}-\omega_{\mu\nu}\,\omega^{\mu\nu}+\sigma_{\mu\nu}\sigma^{\mu\nu}+\frac{1}{2}\theta^2+R_{\mu\nu}\eta^\mu\eta^\nu=0
\end{equation}
where $\sigma^{\mu\nu}$ and $\omega_{\mu\nu}$ are the shear and the rotation associated with the vector field $u^\mu$ respectively. For attractive nature of gravity ($\theta<0$) and neglecting the quadratic terms, the Raychaudhuri equations \eqref{1a} and \eqref{1b} satisfy the following conditions
\begin{equation}
\label{3a}
R_{\mu\nu}u^\mu\,u^\nu\geq0
\end{equation}
\begin{equation}
\label{4a}
R_{\mu\nu}\eta^\mu\eta^\nu\geq0
\end{equation}
As we are working with anisotropic fluid matter distribution, the energy condition recovered from standard General Relativity (GR) are\\
$\bullet$ Strong energy conditions (SEC) if $\rho+P_j\geq0$, $\rho+\sum_jP_j\geq0$, $\forall j$.\\
$\bullet$ Dominant energy conditions (DEC) if $\rho\geq0$, $\rho \pm P_j\geq0$, $\forall j$.\\
$\bullet$ Weak energy conditions (WEC) if $\rho\geq0$, $\rho+P_j\geq0$, $\forall j$.\\
$\bullet$ Null energy condition (NEC) if $\rho+P_j\geq0$, $\forall j$.\\
where $\rho$ and $P$ describe the energy density and pressure, respectively.

\section{Wormhole solutions in different forms of $f(Q)$}\label{sec5}

In this study we will consider two different form of $f(Q)$, (i) Linear form of $f(Q)$, and (ii) Non-linear form of $f(Q)$. By using these forms of $f(Q)$'s, we will get different field equations, and subsequently, we shall discuss by assuming particular models.
\subsection{Field equations in $f(Q)=\alpha Q$}

To proceed further, we have presumed the linear functional form  of $Q$ as
\begin{equation}\label{b}
f(Q)=\alpha Q,
\end{equation}
 where `$\alpha$' is constant, which is the teleparallel gravitational term. The linear form of $f(Q)$ recovers the symmetric teleparallel equivalent of general relativity (STEGR), which helps us to compare our WH solutions to its' fundamental level. Further, the redshift function $\Phi(r)$ in \eqref{9} must be finite and non-vanishing at the throat $r_0$. So one can consider $\Phi(r)=constant$ to achieve the de Sitter and anti-de Sitter asymptotic behaviour. Therefore the field equations in \eqref{14}-\eqref{16} reads
\begin{equation}
\label{17}
-\frac{b^{'}(r)}{r^2}\alpha=\rho,
\end{equation}
\begin{equation}
\label{18}
\frac{b(r)}{r^3}\alpha=P_r,
\end{equation}
\begin{equation}
\label{19}
\left(\frac{b^{'}(r)}{2\,r^2}-\frac{b(r)}{2\,r^3}\right)\alpha=P_t.
\end{equation}
Now, in the next section, we are going to discuss three special cases of wormhole solutions for our study.

\subsection{Field equations in $f(Q)=a\,Q^2+B$}

In this study, we have assumed a particular power-law form of $f(Q)$ i.e. $f(Q)=a\,Q^2+B$, where $a$ and $B$ are constants. Researchers have already shown that the power-law model can oblige the regular thermal extending history including the cold dark matter-dominated stage and the radiation. Also the redshift function $\phi(r)$ is finite everywhere therefore we have considered $\phi(r)=constant$ in this study. Therefore using the model we developed the field equations \eqref{14}-\eqref{16} as follows

\begin{equation}
\label{b1}
\frac{2 a \left( b(r)-r\right) \left(11 b(r)-(6\,b^{'}(r)+7)\,r\right)}{r^6}+\frac{B}{2}=\rho
\end{equation}
\begin{equation}
\label{b2}
\frac{2 a \left(3 (b(r))^2-4\,r\, b(r)+r^2\right)}{r^6}-\frac{B}{2}=P_r
\end{equation}
\begin{equation}
\label{b3}
-\frac{6\,a \left(b(r)-r\right) \left(2 b(r)-(b^{'}(r)+1) r\right)}{r^6}-\frac{B}{2}=P_t
\end{equation}

\section{Wormhole Models With Linear $f(Q)$}\label{sec6}

In this section, we have considered simplest linear form of $f(Q)$ i.e. $f(Q)=\alpha Q$. With this form, we shall discuss three special cases of wormhole solutions.

\subsubsection{\textbf{Wormhole (WH1) solution with $P_t=mP_r$}}
\justifying
In first model, we assume the pressures $P_r$ and $P_t$ are related as (for more details see Ref. \cite{Moraes/2017})
\begin{equation}
\label{20}
P_t=mP_r
\end{equation}
where $m$ is an arbitrary constant.

\begin{figure}[H]
\centering
	\includegraphics[scale=0.4]{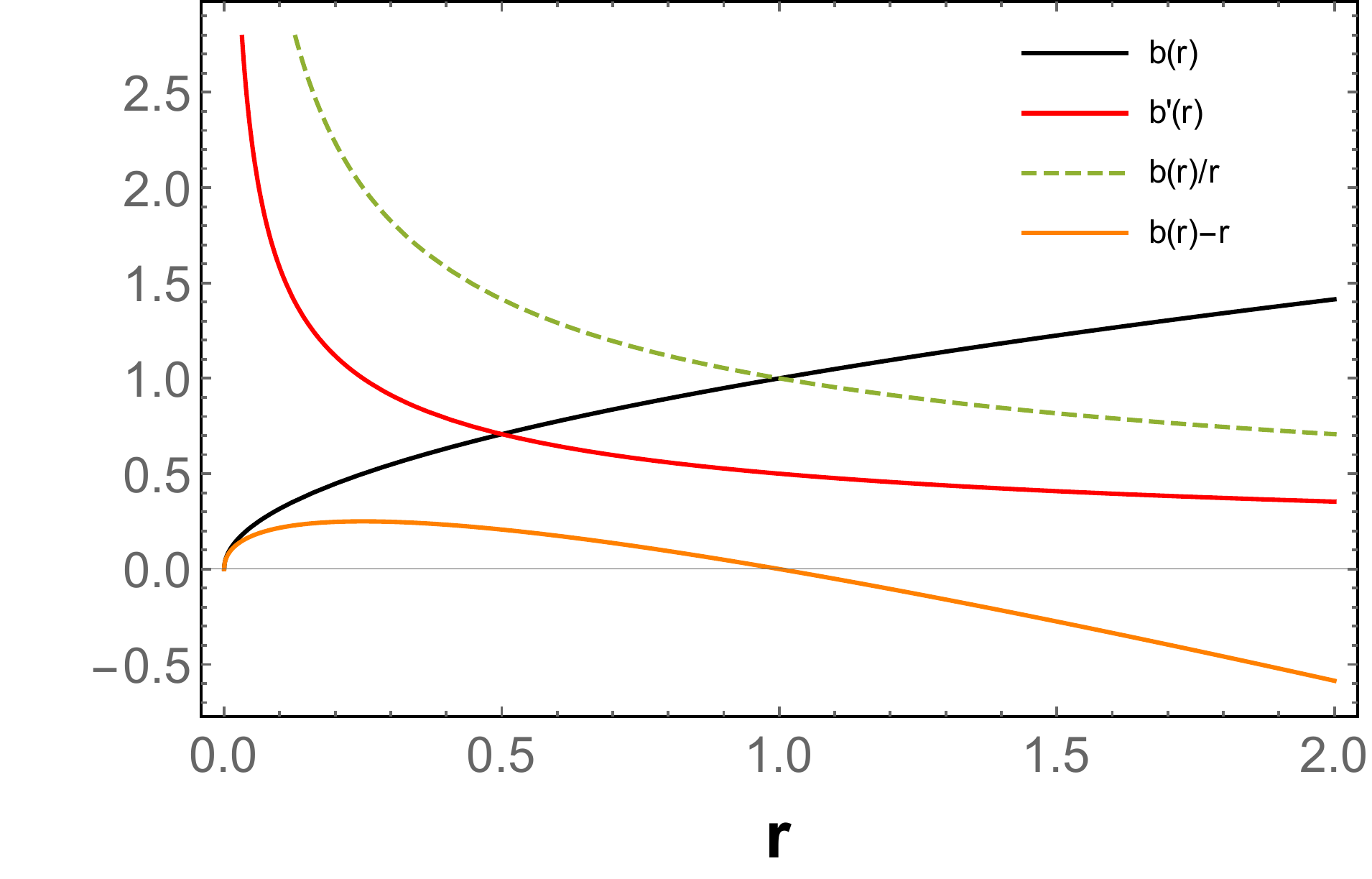}
	\caption{The profile of shape function $b(r)$, Flaring out condition $b'(r)<1$, throat condition $b(r)-r<0$, and asymptotically flatness condition $\frac{b(r)}{r}\rightarrow0$ as $r\rightarrow\infty$ with varying $r$, for $m=-0.25$ (for WH1).}
	\label{f1}
	\end{figure}

By using equations \eqref{18} and \eqref{19} in equation \eqref{20}, one can obtain
\begin{equation}
\label{21}
b(r)=kr^{(1+2m)}
\end{equation}
where $k$ is an integrating constant.
Without loss of generality we consider $k=1$. For $m<0$, Eqn. \eqref{21} retained the asymptotically flatness condition i.e. $b(r)/r\rightarrow 0$ for $r\rightarrow \infty$. In Fig. \ref{f1}, we depict the quantities $b(r), b(r)/r, b(r)-r$ and $b^{'}(r)$ with varying the radial component $r$, for $m=-0.25$. One can clearly see that $b(r)-r$ cuts the $r$-axis at $r_0=1$ in Fig. \ref{f1}. Note that, for a stable wormhole, the shape function $b(r)$ need to obey flaring, throat, and asymptotically conditions. Profiles in Fig. \ref{f1} shows that, shape function satisfied all the conditions which are required for a stable traversable wormhole.
Now using equation \eqref{20}, we can rewrite the energy density $\rho$, radial pressure $P_r$ and lateral pressure $P_t$ from \eqref{17}-\eqref{19} as follows
\begin{equation}
\label{22}
\rho=-\alpha (1+2m)\,r^{2(m-1)},
\end{equation}
\begin{equation}
\label{23}
P_r=\alpha\,r^{2(m-1)},
\end{equation}
\begin{equation}
\label{24}
P_t=m\,\alpha\,r^{2(m-1)}.
\end{equation}
Now from the eqns. \eqref{22}-\eqref{24}, we have
\begin{equation}
\label{25}
\rho+P_r=-2\,m\,\alpha\,r^{2(m-1)},
\end{equation}
\begin{equation}
\label{26}
\rho+P_t=-(m+1)\,\alpha\,r^{2(m-1)},
\end{equation}
\begin{equation}
\label{27}
\rho-P_r=-2\alpha(m+1)r^{2(m-1)},
\end{equation}
\begin{equation}
\label{28}
\rho-P_t=-\alpha(1+3m)r^{2(m-1)}.
\end{equation}
As we know, energy conditions are the best geometrical tool to test the cosmological models' self-stability. So we adopted this technique to test our models. Moreover, we have presumed the linear functional form of $f(Q)$. Hence, our models will retain the standard energy conditions.

\begin{figure}[H]
\centering
    \includegraphics[scale=0.3]{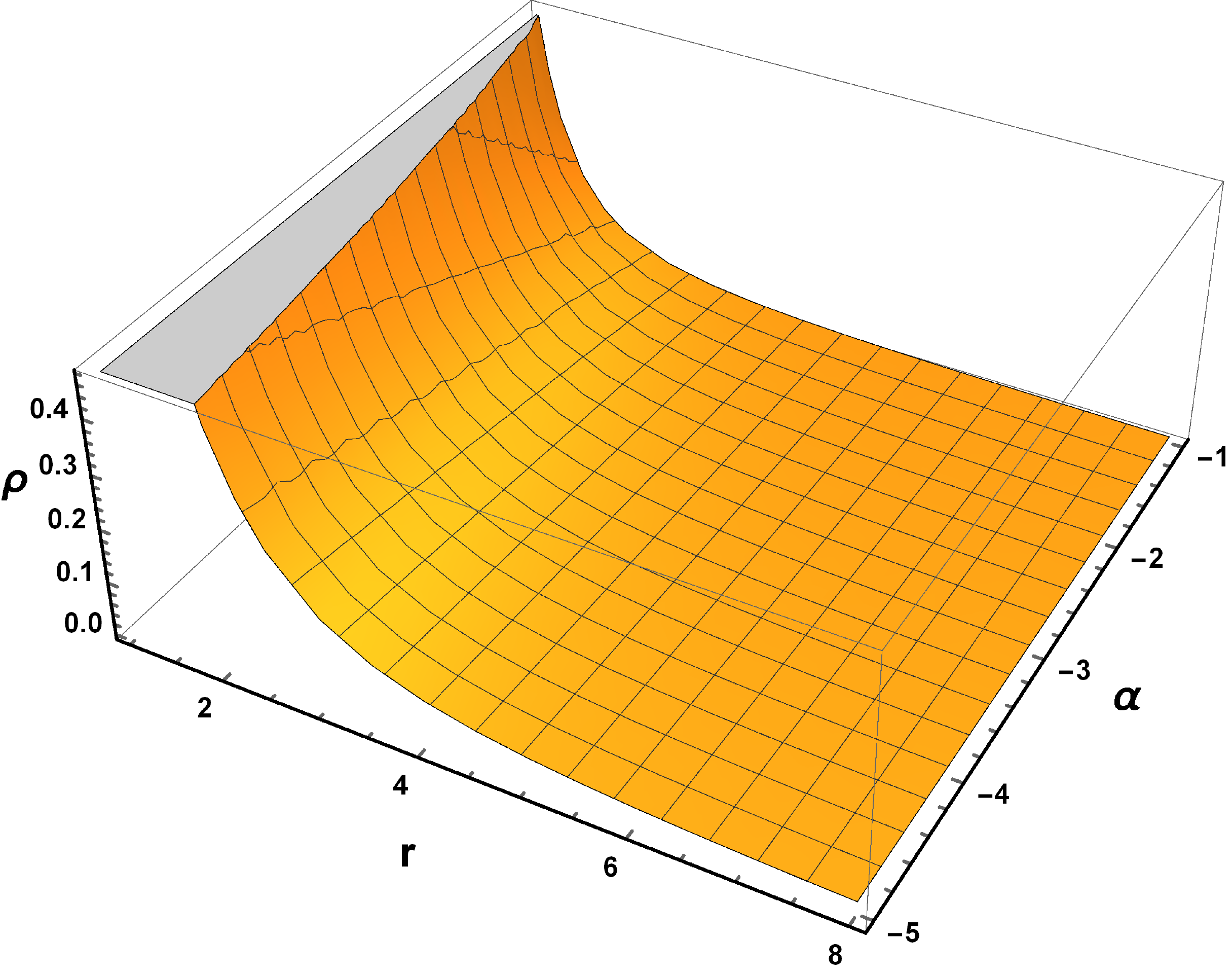}
    \caption{Profile of energy density, $\rho(r)$ w.r.t. $r$  and $\alpha$ with $m=-0.25$ for WH1.}
    \label{f2}
\end{figure}

\begin{figure}[]
\subfloat[$\rho+P_r$\label{sfig:testa}]{
  \includegraphics[scale =0.26]{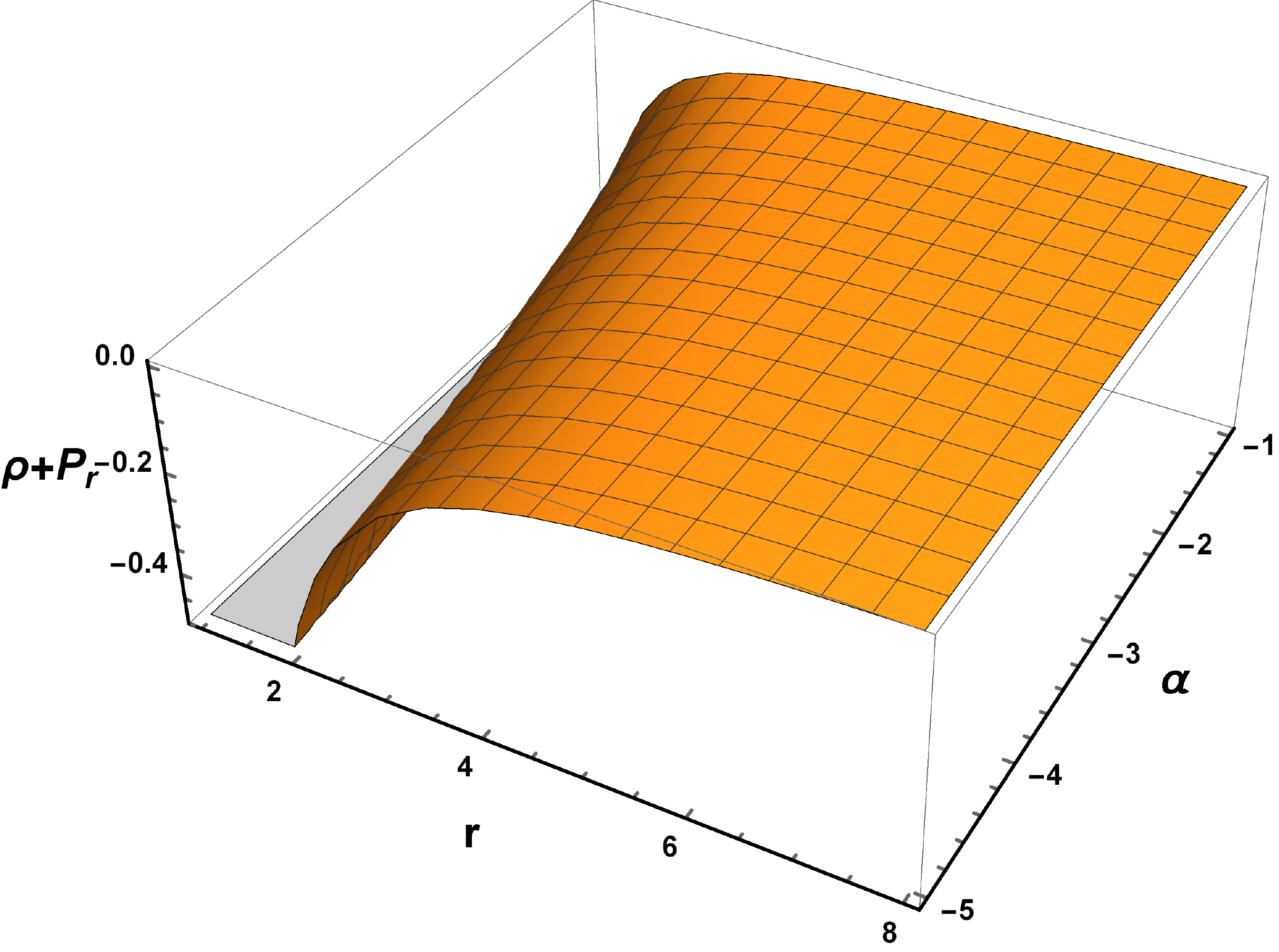}
}\hfill
\subfloat[$\rho+P_t$\label{sfig:testa}]{
  \includegraphics[scale =0.25]{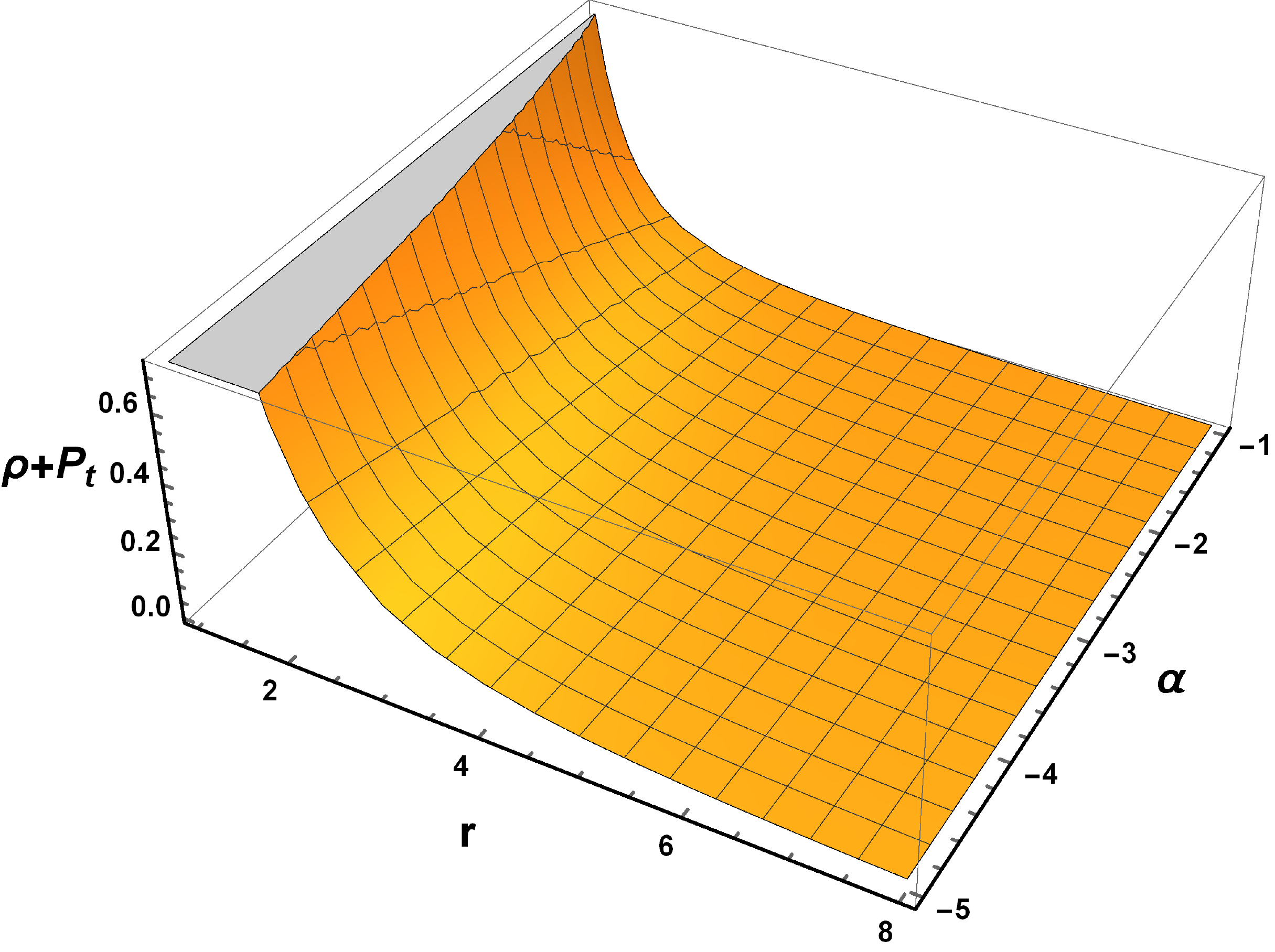}
}\hfill
\subfloat[$\rho-P_r$\label{sfig:testa}]{
  \includegraphics[scale =0.26]{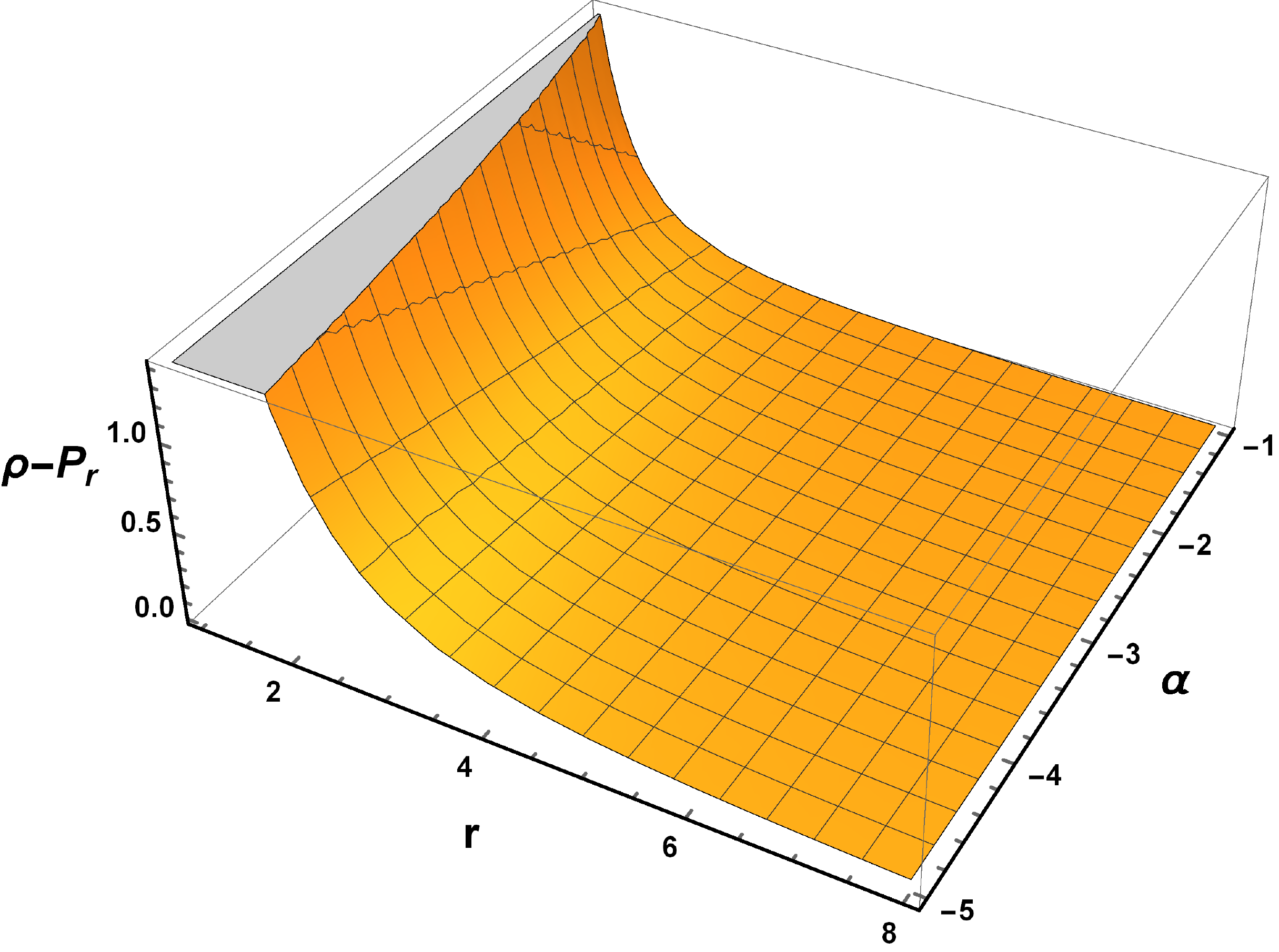}
}\hfill
\subfloat[$\rho-P_t$\label{sfig:testa}]{
  \includegraphics[scale =0.26]{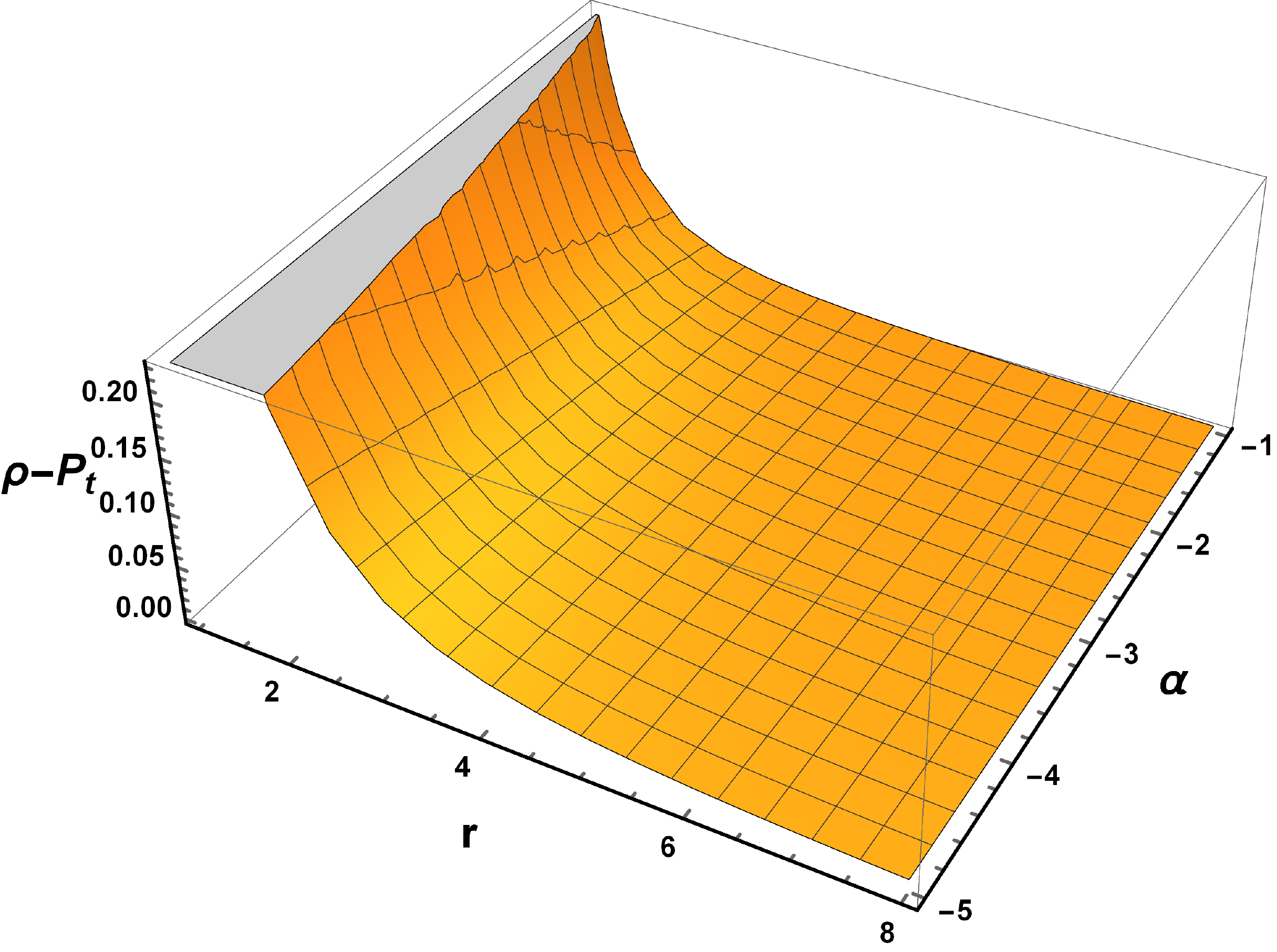}
}
\caption{Profiles of WEC, NEC and DEC w.r.t. $r$  and $\alpha$ with $m=-0.25$ for WH1.}
\label{f3}
\end{figure}

Figs. \ref{f2} and \ref{f3}, depicts the behavior of the energy conditions. It can be seen from Fig. \ref{f2} that the energy density, $\rho$ is positive throughout the spacetime. From Fig. \ref{f3}, one can observed that NEC for the radial pressure violates i.e. $\rho+P_r< 0$, whereas NEC for the lateral pressure i.e. $\rho+P_t\geq 0$ obeys. Also, DEC is satisfied i.e. $\rho-P_r\geq 0$ and $\rho-P_t\geq 0$. These profiles of energy conditions aligned with the properties of exotic matter which is  responsible for a traversable wormhole. From equations \eqref{22}-\eqref{24}, the strong energy condition (SEC) yields $\rho+P_r+2P_t=0$. This similar result obtained in \cite{Elizalde/2019}.

\subsubsection{\textbf{Wormhole (WH2) solution with $P_r=\omega\rho$}}

In this model, we assume a relation between $\rho$ and $P_r$ as a linear equation of state (EoS) such as \cite{kimet/2018,Lobo/2005,Francisco/2005,Mauricio/2013}.\\
\begin{equation}
\label{2a}
P_r=\omega\rho
\end{equation}
where $\omega$ is as equation of state (EoS).\\
 Therefore, using equations \eqref{17} and \eqref{18} in Eq. \eqref{2a}, one can obtain the following shape function
\begin{equation}
\label{29}
b(r)=cr^{-\frac{1}{\omega}}
\end{equation}
where $c$ is an integrating constant.\\
Note that, to satisfy the asymptotically flatness condition, i.e. $\frac{b(r)}{r}\rightarrow 0$ as $r\rightarrow \infty$, \,$\omega$ should be less than -1 i.e. $\omega<-1$. 

\begin{figure}[H]
\centering
	\includegraphics[scale=0.4]{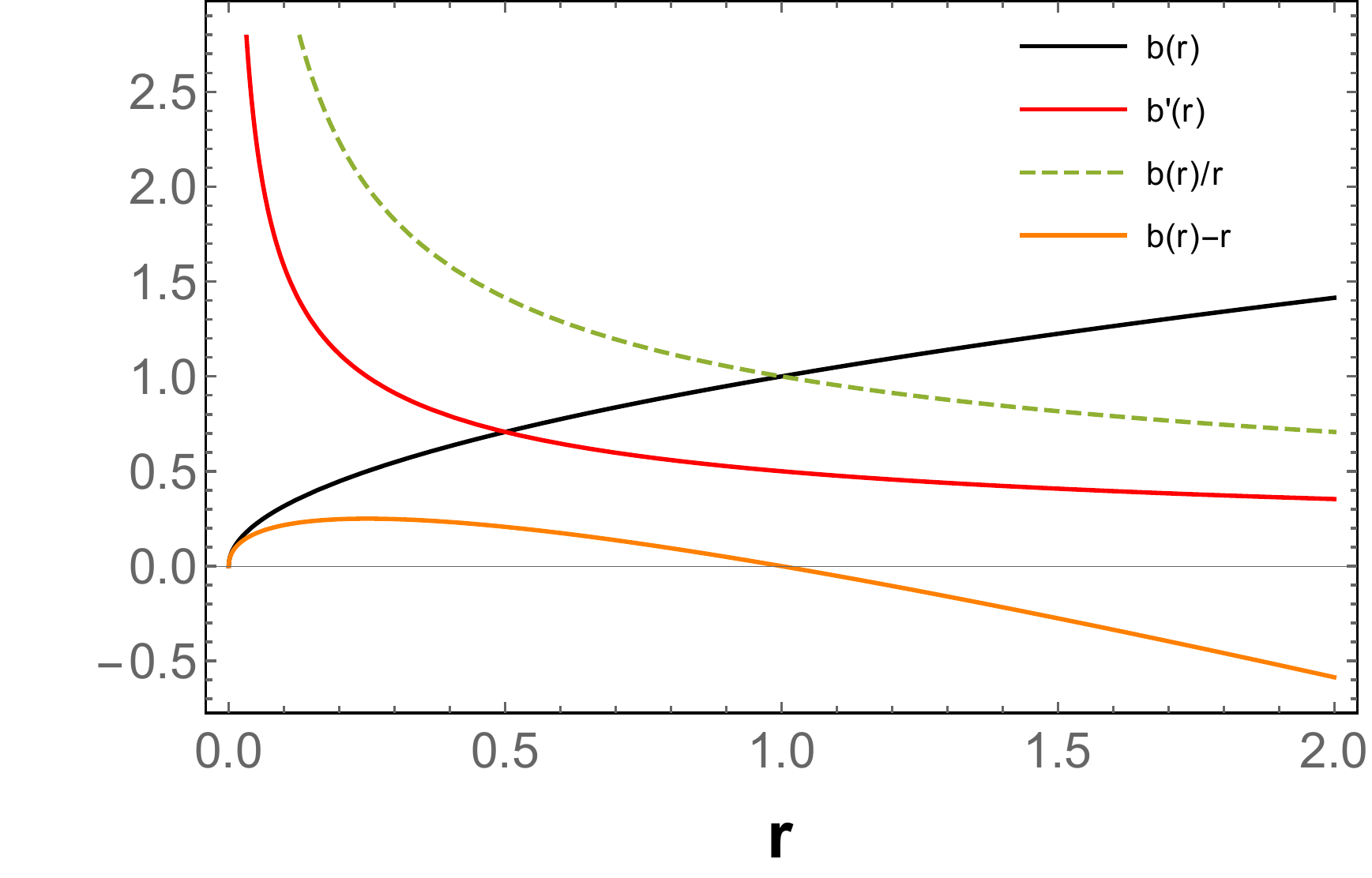}
	\caption{The behavior of shape function $b(r)$,  flaring out condition $b'(r)<1$, throat condition $b(r)-r<0$, and asymptotically flatness condition $\frac{b(r)}{r}\rightarrow0$ as $r\rightarrow\infty$ for $c=1, \omega=-2$ for WH2. }
	\label{f4}
\end{figure}

The profiles of the necessary conditions for a shape function i.e.  throat condition, flare out condition and asymptotically flatness conditions are depicted in Fig. \ref{f4}. From Fig. \ref{f4}, one can see that the shape function $b(r)$ is in the increasing direction as $r$ increases. For $r>r_0$, $b(r)-r<0$, which represents the consistent of throat condition for wormholes and from the same figure \ref{f4}, it is clear that $b(r)-r$ cuts the $r$-axis at $r_0=1$. Besides, the flaring out condition satisfied at $r_0$ i.e. $b^{'}(r_0)=b^{'}(1)\approx0.667<1$. Asymptotically flatness condition is also satisfied, i.e. $\frac{b(r)}{r}\rightarrow 0$ as $r\rightarrow \infty$ satisfied. Therefore, from Fig. \ref{f4}, one can notice that the shape function satisfies all the required conditions for a traversable WH.

Now, we discuss the traversable wormhole spacetime supported by the phantom energy with the presence of exotic matter. For phantom energy EoS, $P_r=\omega\rho$ with $\omega<-1$ and using the shape function \eqref{29} in field equations \eqref{17}-\eqref{19}, we get the stress-energy tensor components are given by\\
\begin{equation}
\label{30}
\rho=\frac{c\,\alpha}{\omega}\,r^{-\left(\frac{1}{\omega}+3\right)},
\end{equation}
\begin{equation}
\label{31}
\rho+P_r=c\,\alpha\left(\frac{1}{\omega}+1\right)\,r^{-\left(\frac{1}{\omega}+3\right)},
\end{equation}
\begin{equation}
\label{32}
\rho+P_t=\frac{1}{2}\,c\,\alpha\left(\frac{1}{\omega}-1\right)\,r^{-\left(\frac{1}{\omega}+3\right)},
\end{equation}
Now, the null energy condition (NEC) at the throat is given by
\begin{equation}
\label{33}
\rho+P_r\mid_{r_0}=c\,\alpha\left(\frac{1}{\omega}+1\right)\,r_{0}^{-\left(\frac{1}{\omega}+3\right)}.
\end{equation}
In this case, it is clear that $\omega$ should not be equal to $-1$ i.e. $\omega\neq{-1}$, so we consider $\omega<-1$ to imply the violation of NEC at the throat, i.e. the throat of the WH needs to open with phantom energy. 

The DEC for this model is
\begin{equation}
\label{34}
\rho-P_r=c\,\alpha\left(\frac{1}{\omega}-1\right)\,r^{-\left(\frac{1}{\omega}+3\right)},
\end{equation}
\begin{equation}
\label{35}
\rho-P_t=\frac{1}{2}c\,\alpha\left(\frac{3}{\omega}+1\right)\,r^{-\left(\frac{1}{\omega}+3\right)}.
\end{equation}

\begin{figure}[H]
   \centering
   \includegraphics[scale=0.3]{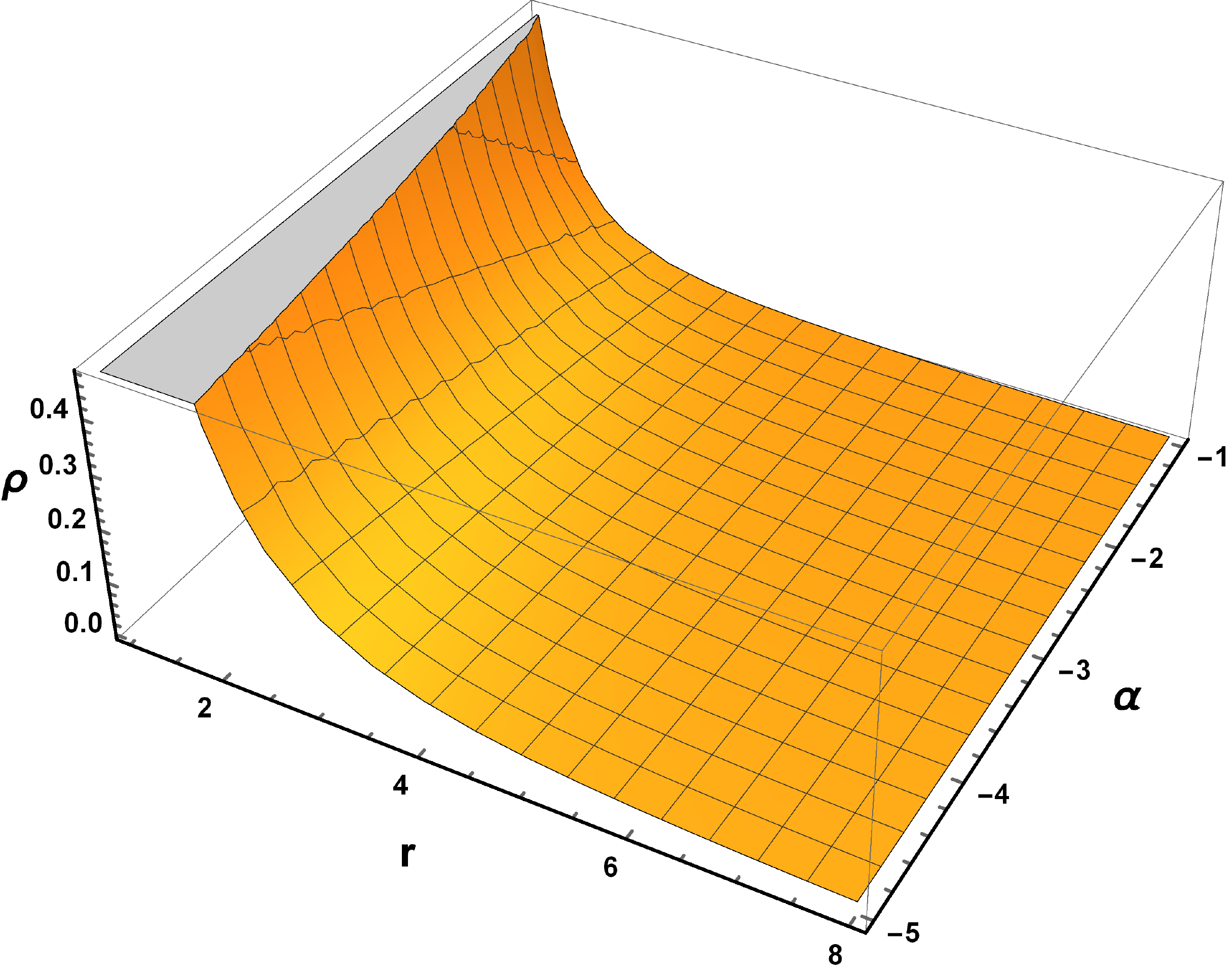}
   \caption{The profile of energy density, $\rho(r)$ w.r.t. $r$ and $\alpha$ with $\omega=-2,\,c=1$ for WH2.}
   \label{f5}
\end{figure}

\begin{figure}[]
\subfloat[$\rho+P_r$\label{sfig:testa}]{
  \includegraphics[scale =0.26]{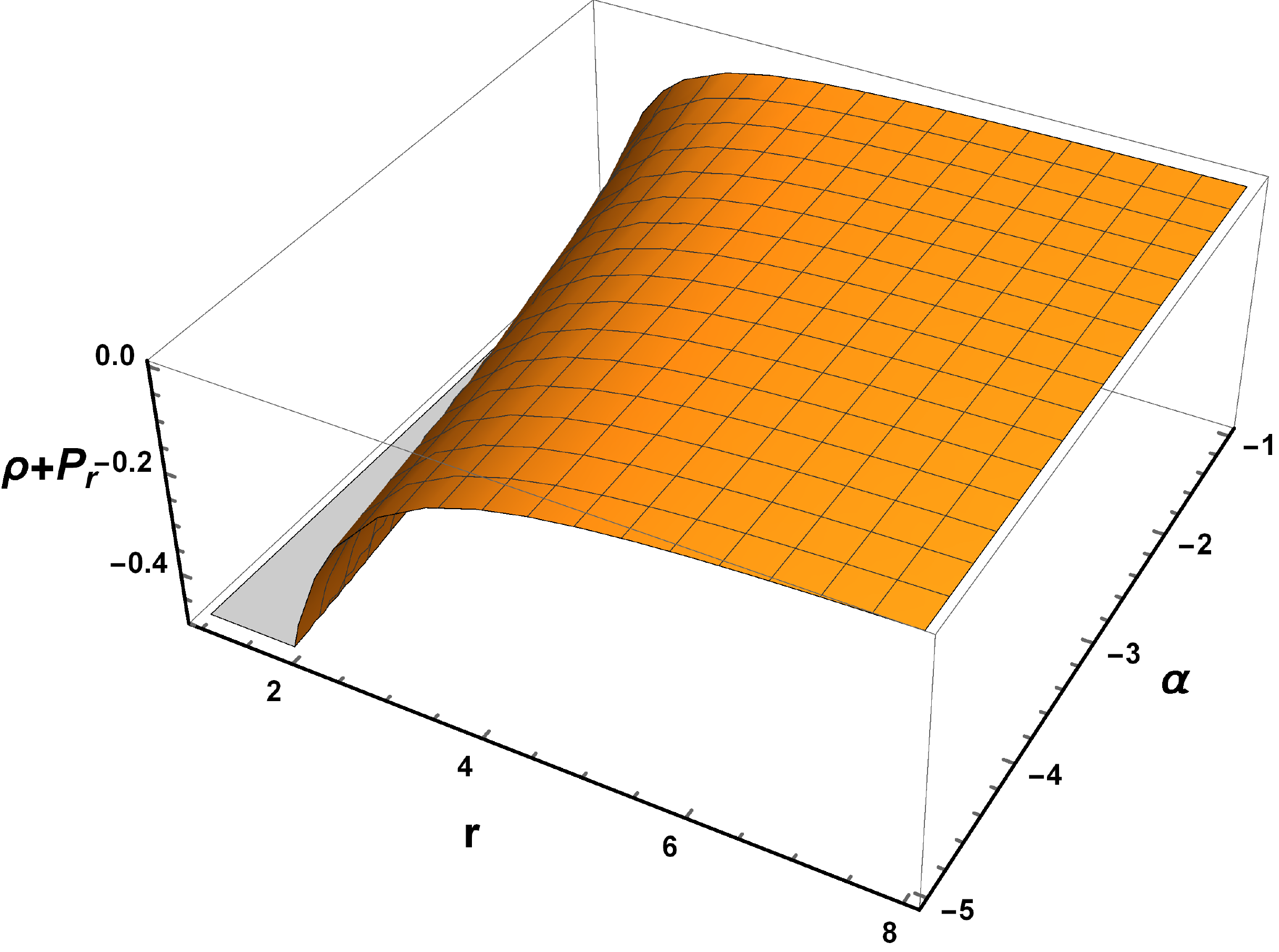}
}\hfill
\subfloat[$\rho+P_t$\label{sfig:testa}]{
  \includegraphics[scale =0.25]{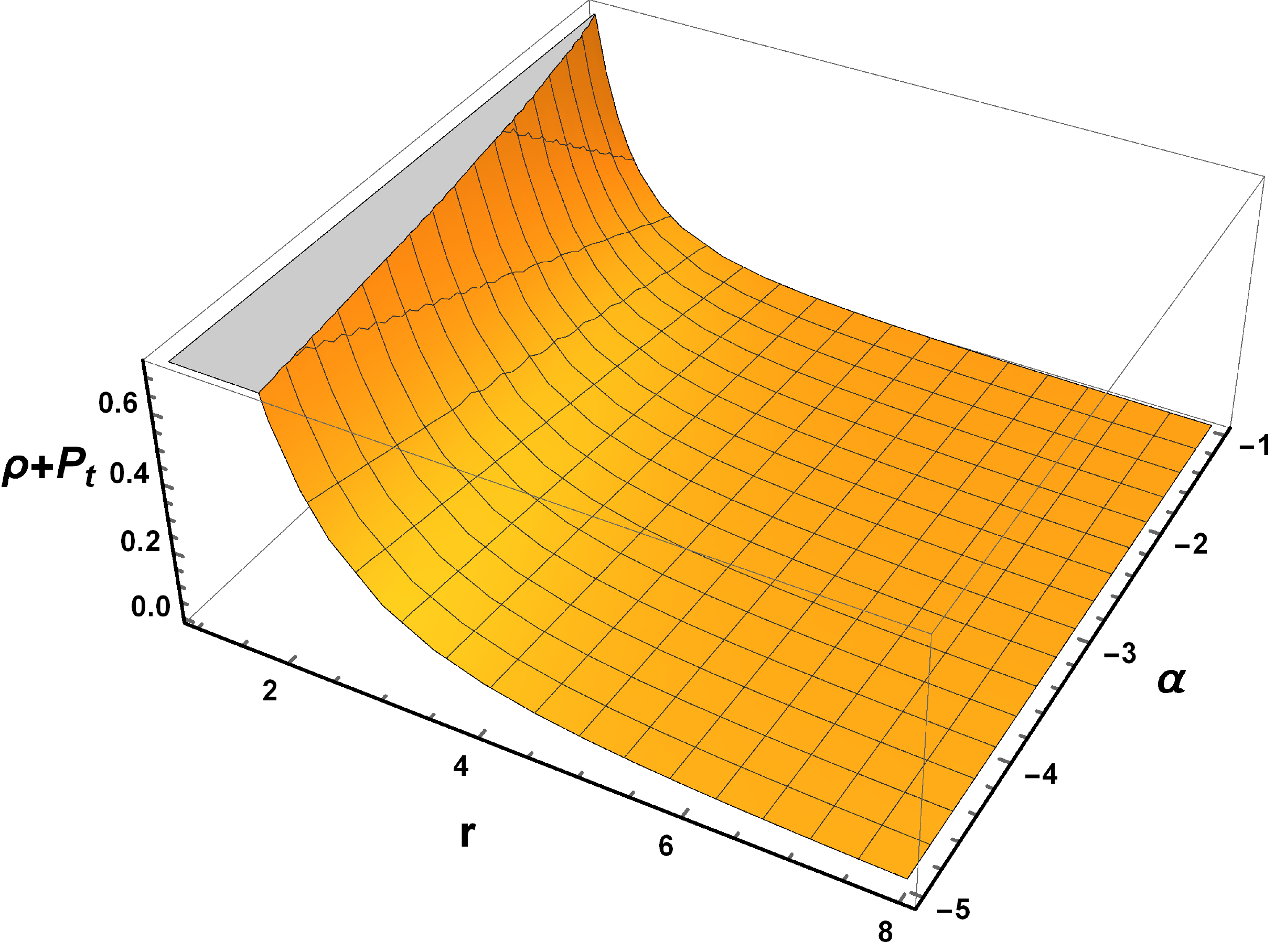}
}\hfill
\subfloat[$\rho-P_r$\label{sfig:testa}]{
  \includegraphics[scale =0.26]{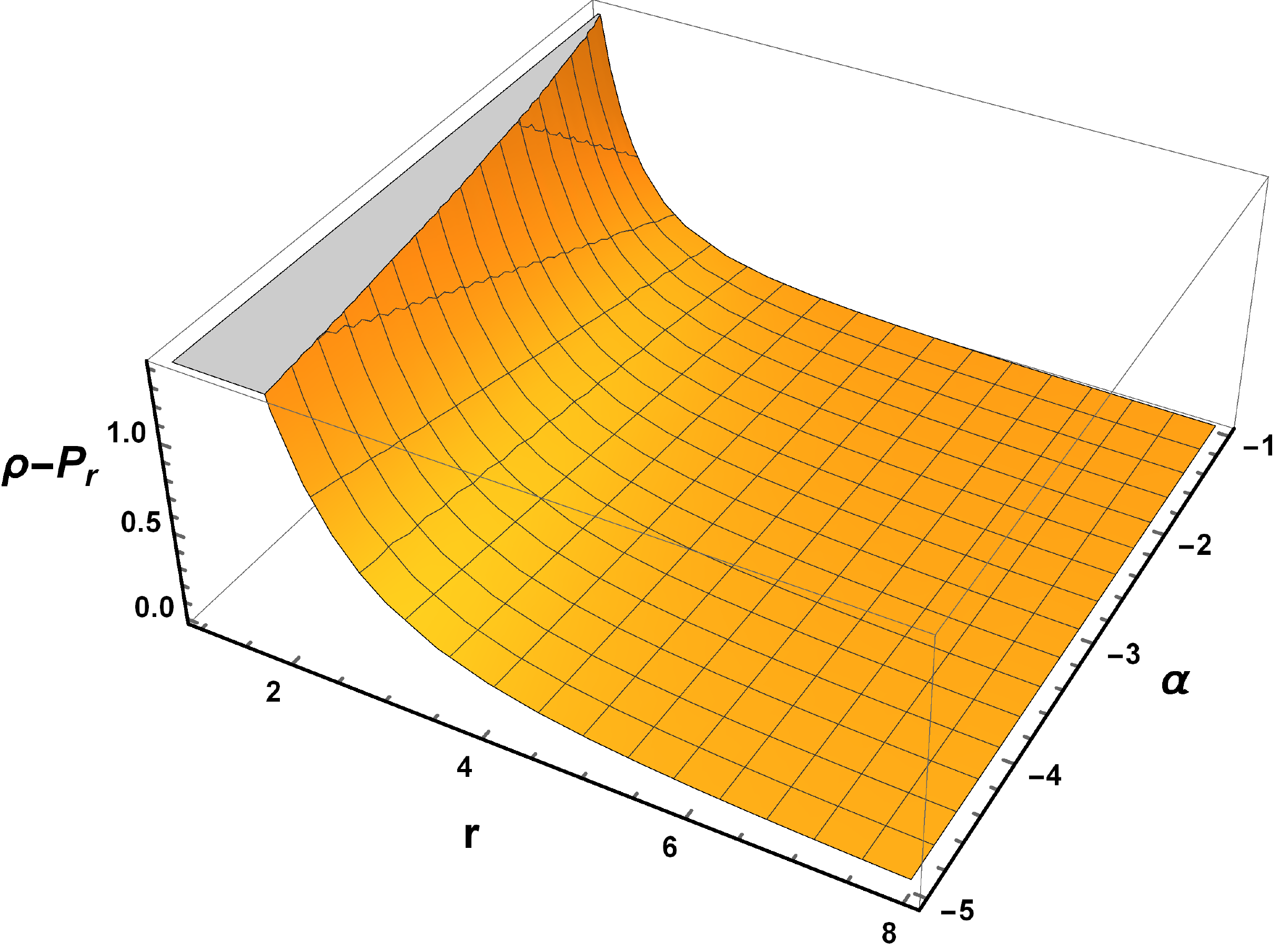}
}\hfill
\subfloat[$\rho-P_t$\label{sfig:testa}]{
  \includegraphics[scale =0.26]{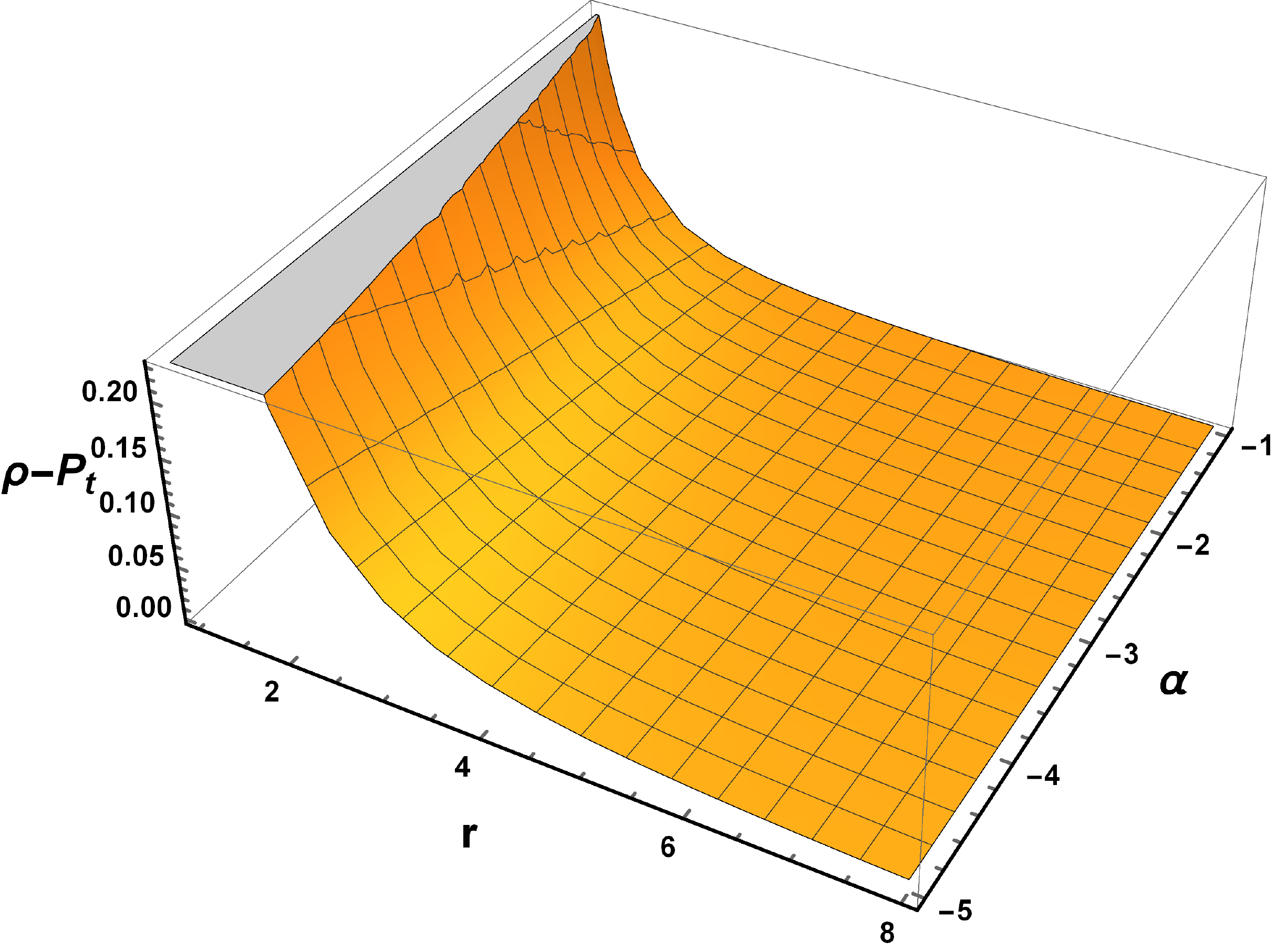}
}
\caption{Profiles of WEC, NEC and DEC w.r.t. $r$  and $\alpha$ with $\omega=-2, c=1$ for WH2.}
\label{f6}
\end{figure}

We draw the graphical behavior of energy conditions using equations \eqref{30}-\eqref{35} are shown in Figs. \ref{f5} and \ref{f6}. For this model, we have considered a  matter content that is related to phantom equation of state, i.e., $\omega <-1$. In Fig. \ref{f5}, we have shown the behavior of energy density and it takes positive values for all over the range. It can be seen from the Fig. \ref{f6} that DEC is satisfied i.e. $\rho-P_r\geq0$ and $\rho-P_t\geq0$ but we observe that NEC is violated due to $\rho+P_r<0$. The violation of NEC is proof of the presence of the exotic matter, which might be needed for the wormhole geometry.

\subsubsection{\textbf{Wormhole (WH3) solution with $b(r)=r_0\left(\frac{r}{r_0}\right)^n$}}

In this subsection, we have considered the specific shape function, $b(r)=r_0\left(\frac{r}{r_0}\right)^n$ and for this choice the corresponding stress energy components from Eqns. \eqref{17}-\eqref{19} are obtained as follows

\begin{equation}
\label{36}
\rho=-n\,\alpha\,\frac{r^{n-3}}{r_0^{n-1}}
\end{equation}
\begin{equation}
\label{37}
\rho+P_r=\left(1-n\right)\,\alpha\,\frac{r^{n-3}}{r_0^{n-1}}
\end{equation}
\begin{equation}
\label{38}
\rho+P_t=-\frac{1}{2}\left(n+1\right)\alpha\,\frac{r^{n-3}}{r_0^{n-1}}
\end{equation}
Now, at the throat i.e. at $r=r_0$, Eq. \eqref{37} reduce to
\begin{equation}
\label{39}
\rho+P_r\mid_{r_0}=\frac{\left(1-n\right)\alpha}{r_0^2}
\end{equation}
The dominant energy conditions (DEC) for this model are
\begin{equation}
\label{40}
\rho-P_r=-\left(n+1\right)\,\alpha\,\frac{r^{(n-3)}}{r_0^{(n-1)}}
\end{equation}
\begin{equation}
\label{41}
\rho-P_t=\left(\frac{1}{2}-\frac{3n}{2}\right)\,\alpha\,\frac{r^{(n-3)}}{r_0^{(n-1)}}
\end{equation}

\begin{figure}[H]
\centering
	\includegraphics[scale=0.3]{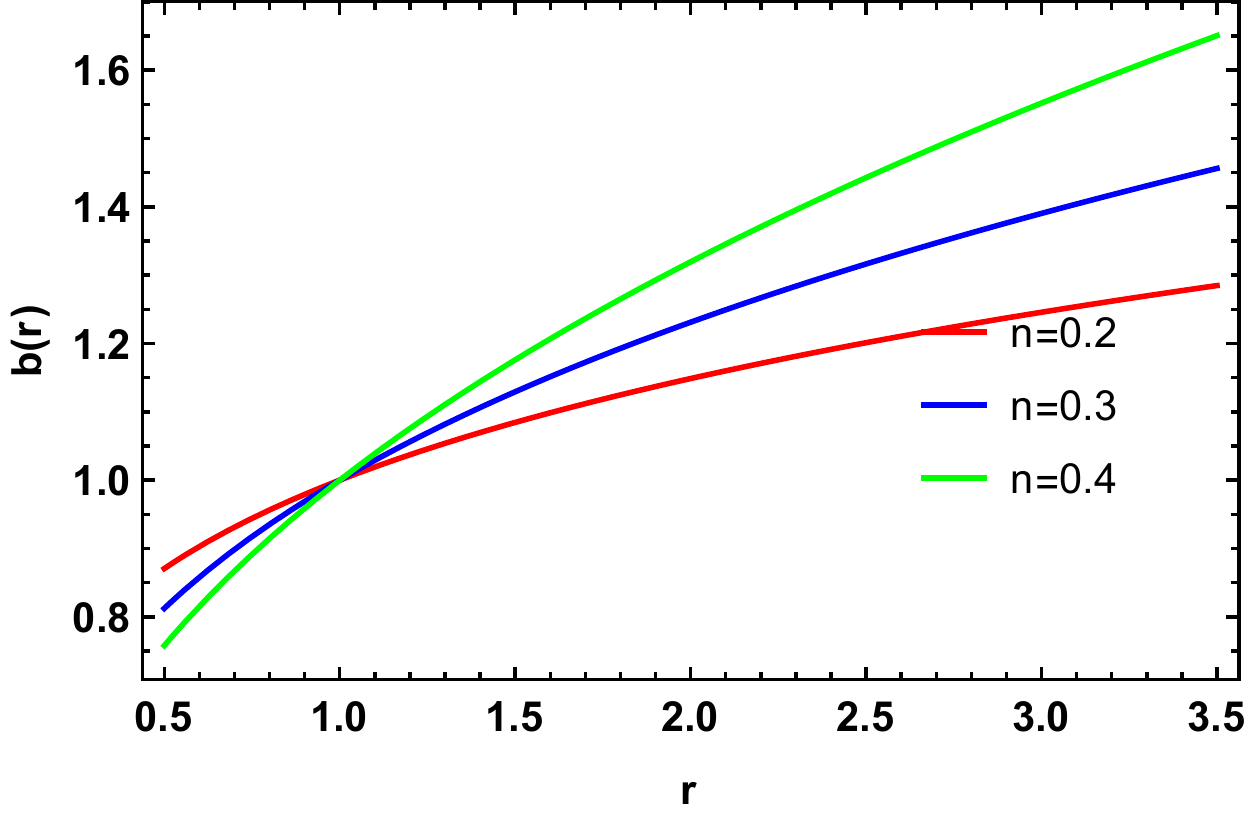}
	\includegraphics[scale=0.3]{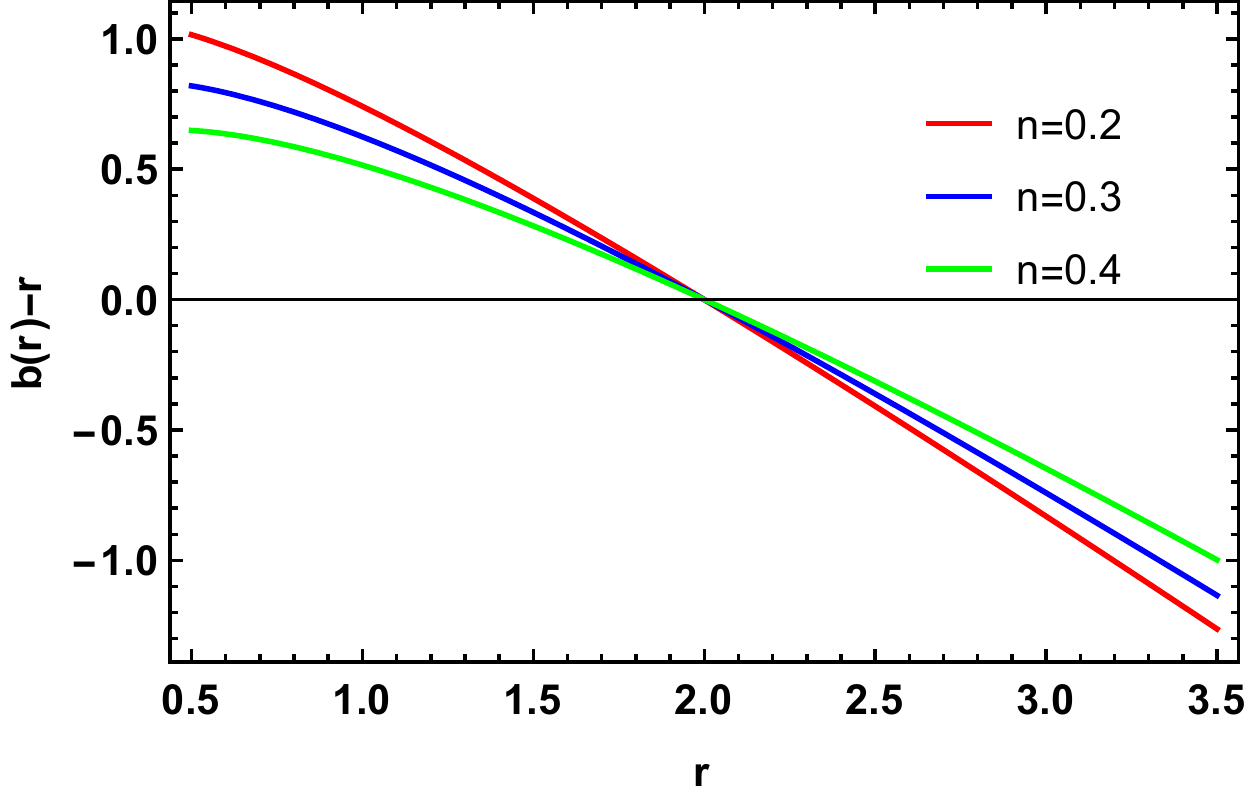}
	\includegraphics[scale=0.3]{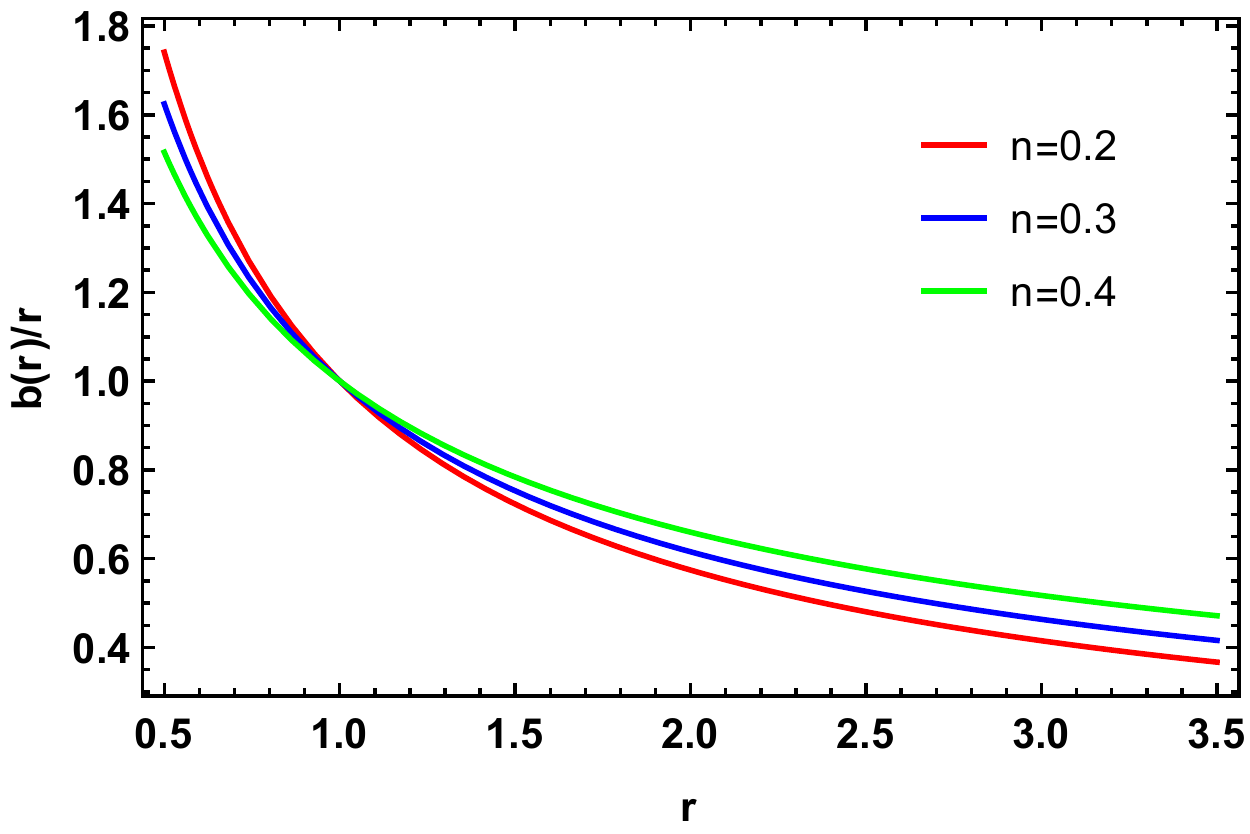}
	\includegraphics[scale=0.3]{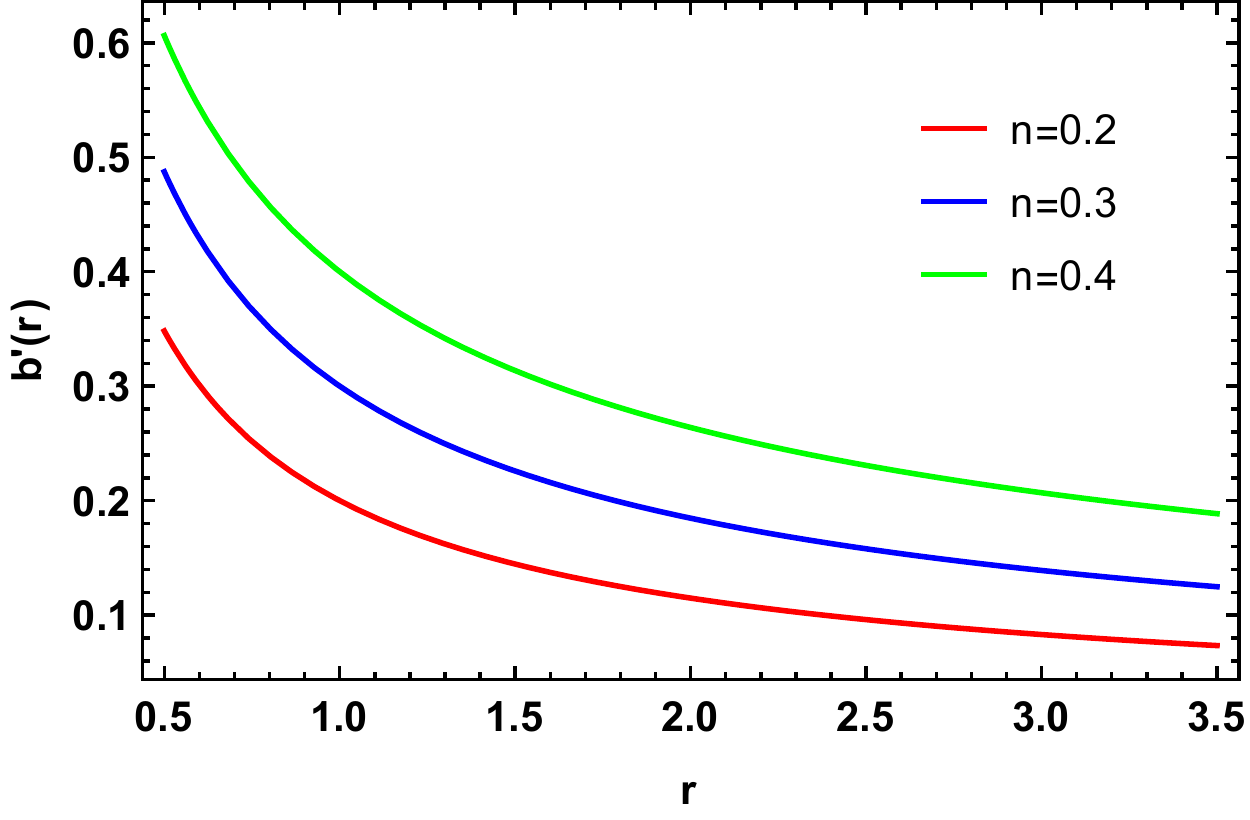}
	\caption{Characteristic of the shape functions for WH3.}
	\label{f7}
\end{figure}

\begin{figure}[H]
\centering
    \includegraphics[scale=0.5]{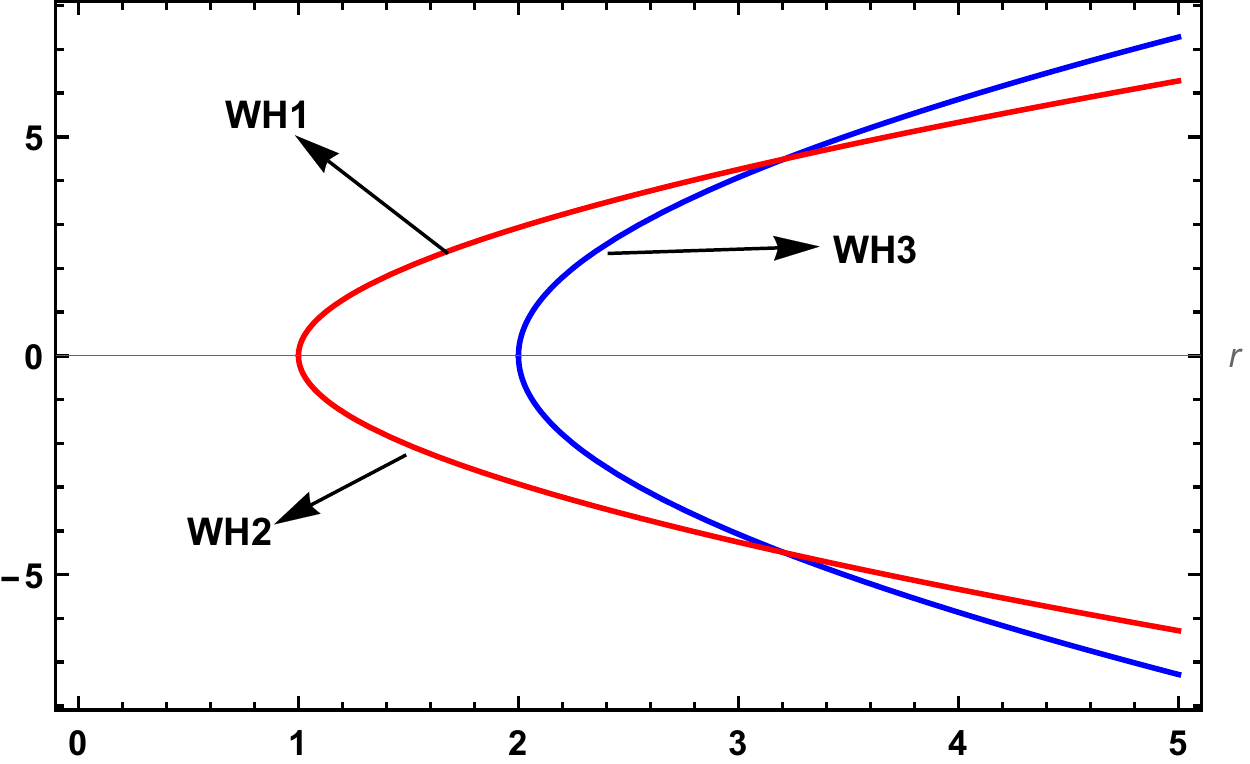}
    \caption{The profile shows two-dimensional embedding diagrams for WH1, WH2, and WH3 (here, WH1 and WH2 overlap each other).}
    \label{7a}
\end{figure}

Taking into account that the condition $\frac{b(r)}{r}\rightarrow0$  as  $r\rightarrow\infty$ will satisfy when $n<1$. In Fig. \ref{f7}, we have shown the behavior of $b(r),\,b(r)-r,\,\frac{b(r)}{r}$ and $b'(r)$ respectively and in this case $b(r)-r$ cuts the $r$-axis at $r_0$= 2, which is the throat radius of wormhole (see Fig. \ref{7a}). From Fig. \ref{f7}, one can see that $b(r)$ satisfies all the necessary conditions for a traversable wormhole.

\begin{figure}[H]
\centering
     \includegraphics[scale=0.3]{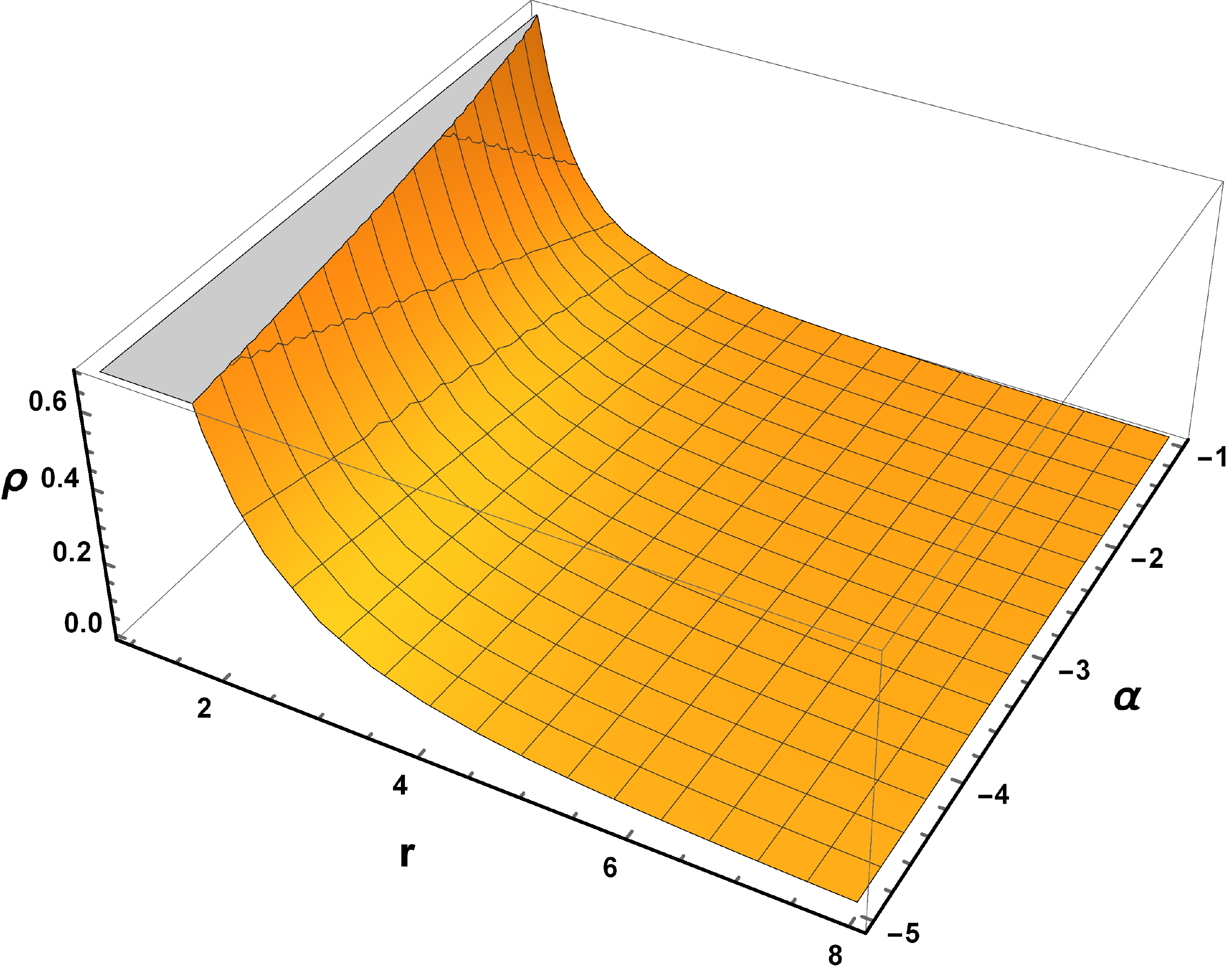}
     \caption{The energy density, $\rho(r)\geq0$ with $n=-0.5$, $r_0=2$ for WH3.}
     \label{f8}
\end{figure}

\begin{widetext}

\begin{figure}[]
\centering
\subfloat[$\rho+P_r$\label{sfig:testa}]{
  \includegraphics[scale =0.27]{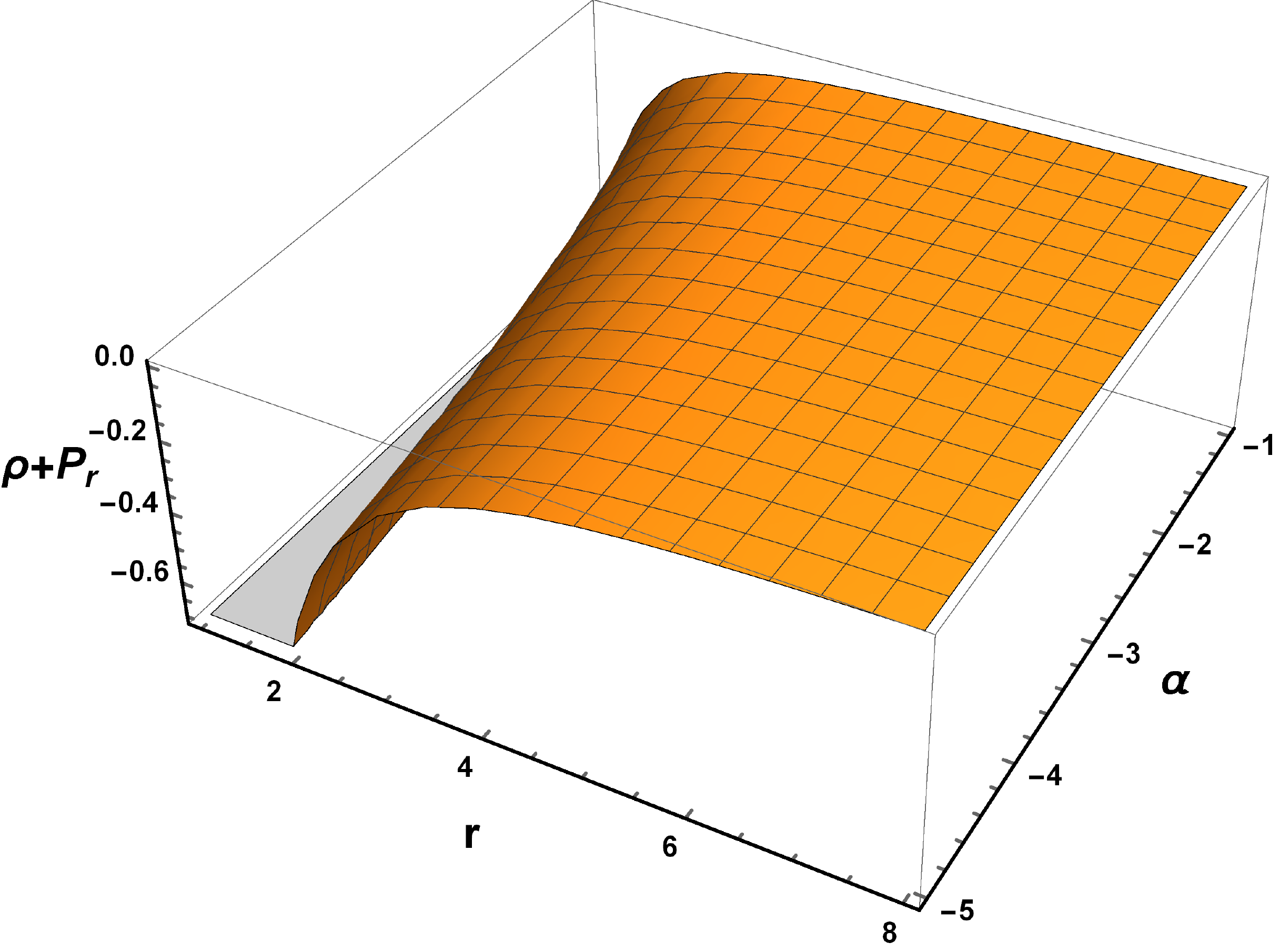}
}
\subfloat[$\rho+P_t$\label{sfig:testa}]{
  \includegraphics[scale =0.27]{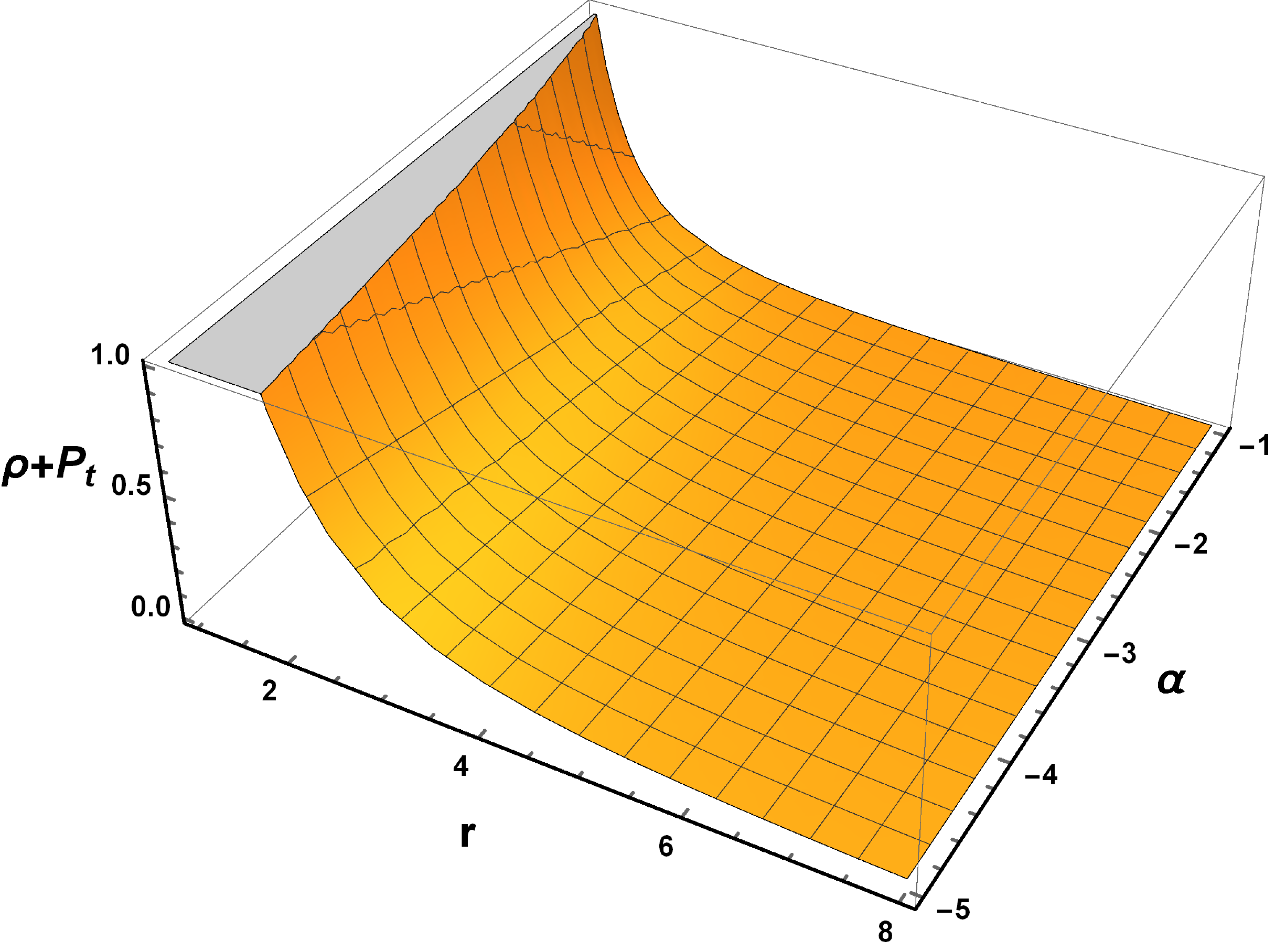}
}\hfill
\subfloat[$\rho-P_r$\label{sfig:testa}]{
  \includegraphics[scale =0.27]{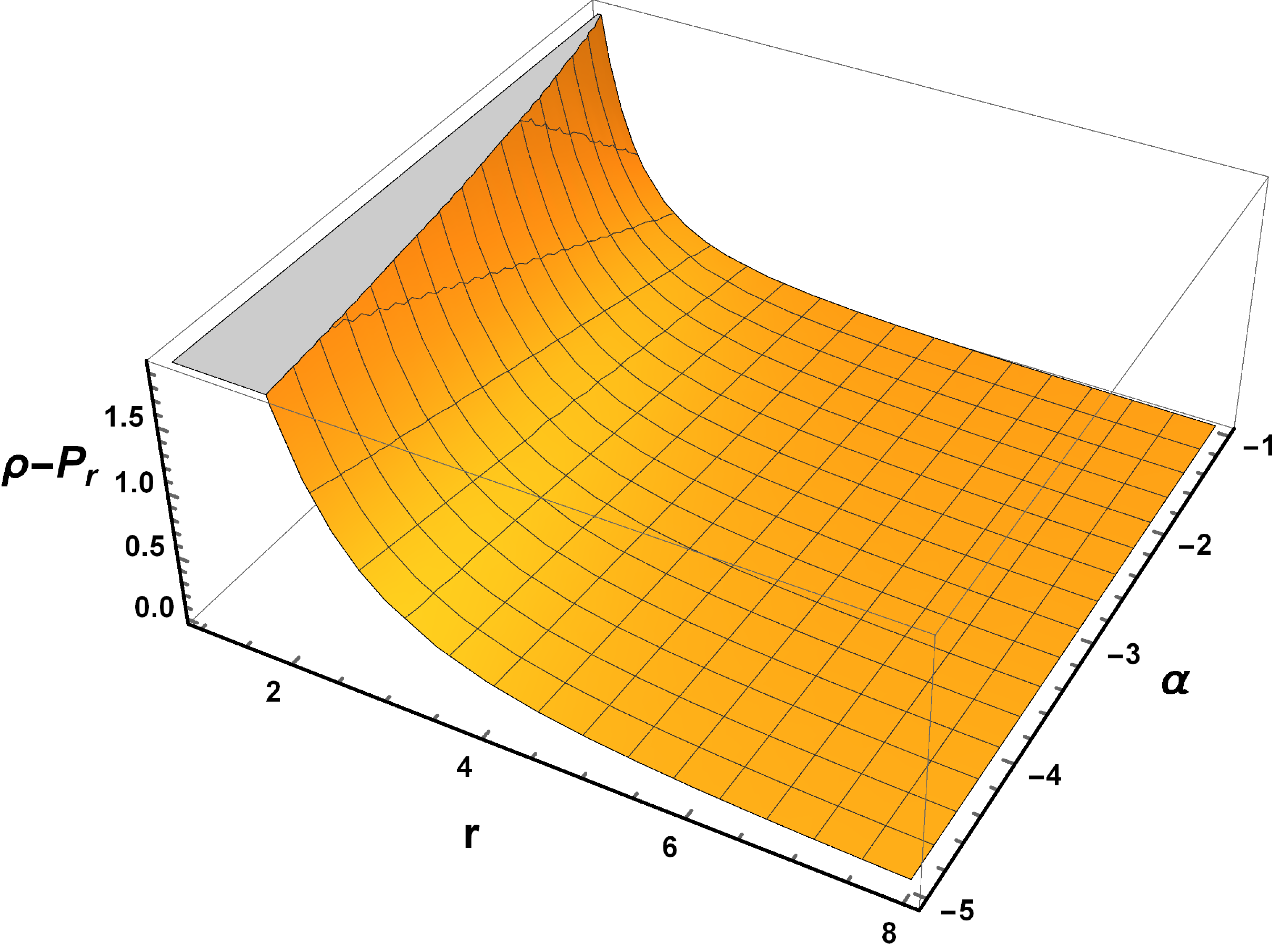}
}
\subfloat[$\rho-P_t$\label{sfig:testa}]{
  \includegraphics[scale =0.27]{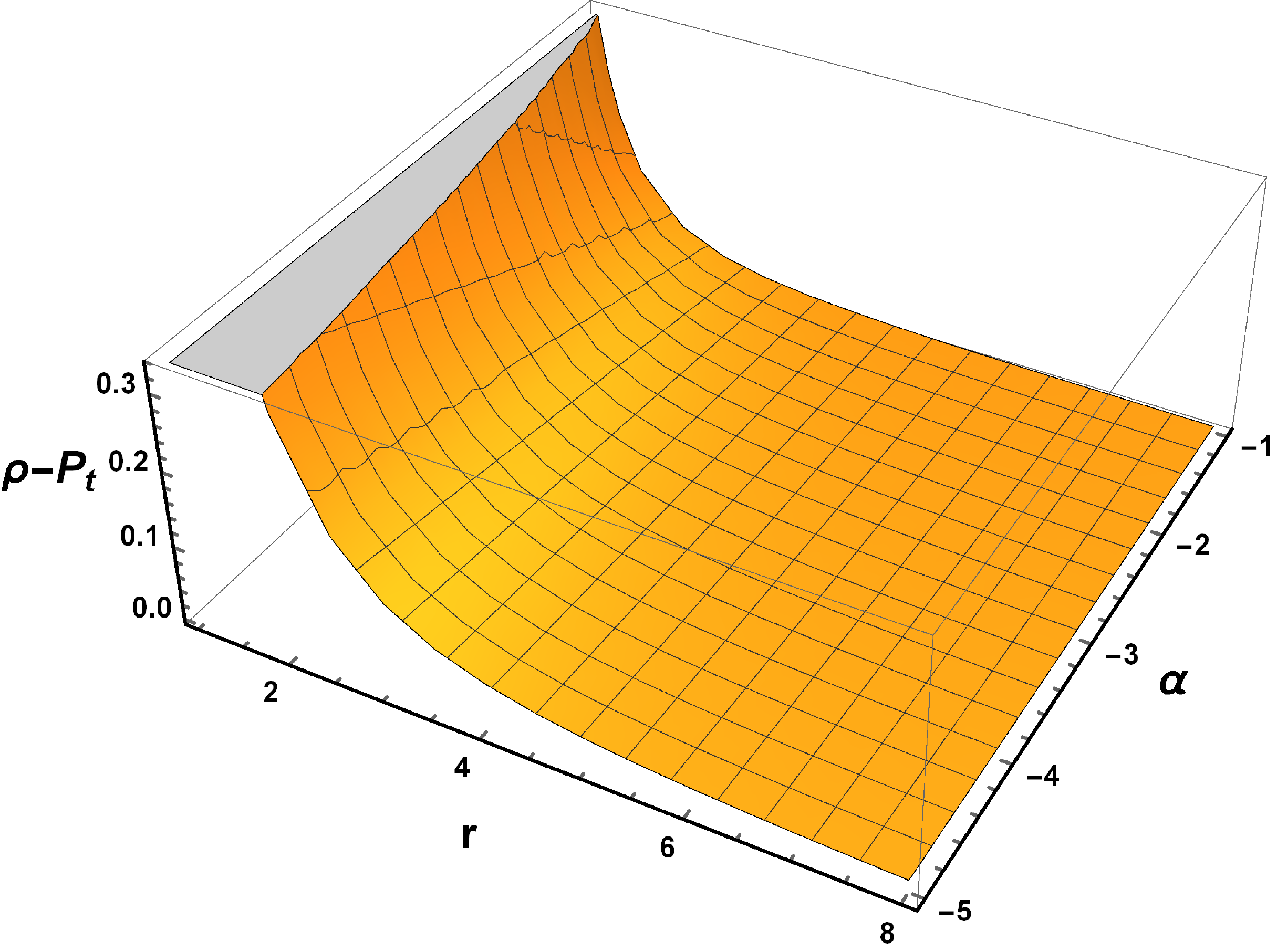}
}
\caption{Profiles of WEC, NEC and DEC w.r.t. $r$  and $\alpha$ with $n=-0.5$, $r_0=2$ for WH3.}
\label{f9}
\end{figure}

\end{widetext}

Moreover, from Fig. \ref{f8}, we see that energy density, $\rho$ is always positive throughout the spacetime. In Fig. \ref{f9} (b), (c) and (d), we observe that the energy condition tends to zero as the radial component $r$ goes on for the negative range of $\alpha$. But, NEC violates as $\rho+P_r< 0$.
 

\section{Wormhole Models With Non-linear $f(Q)$}\label{sec7}

In this section, by presuming a polynomial form of $f(Q)=a\,Q^2+B$, we continue trying to discuss the fundamental equations for the anisotropic fluid. For the choice of $P_t=mP_r$ and $P_r=w \rho$, the differential equations in \eqref{b1}-\eqref{b3} becomes highly non-linear and very complicated to solve and analyze these types of systems (see Ref. \cite{Newton/2020}) . Therefore, we consider two different specific shape functions in finding wormhole solutions.

\subsubsection{\textbf{Wormhole (WH4) solution with $b(r)=r_0\left(\frac{r}{r_0}\right)^n$}}

\begin{figure}[H]
\centering
	\includegraphics[scale=0.3]{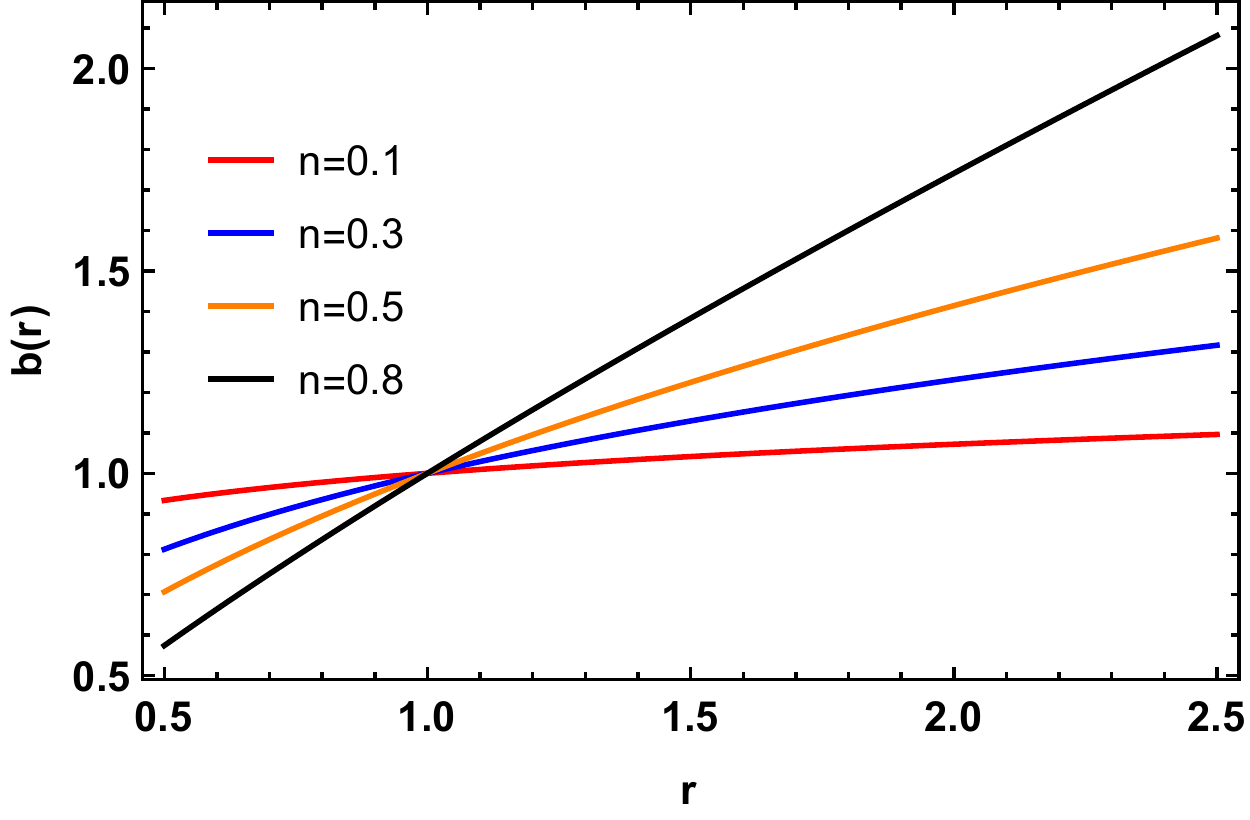}
	\includegraphics[scale=0.3]{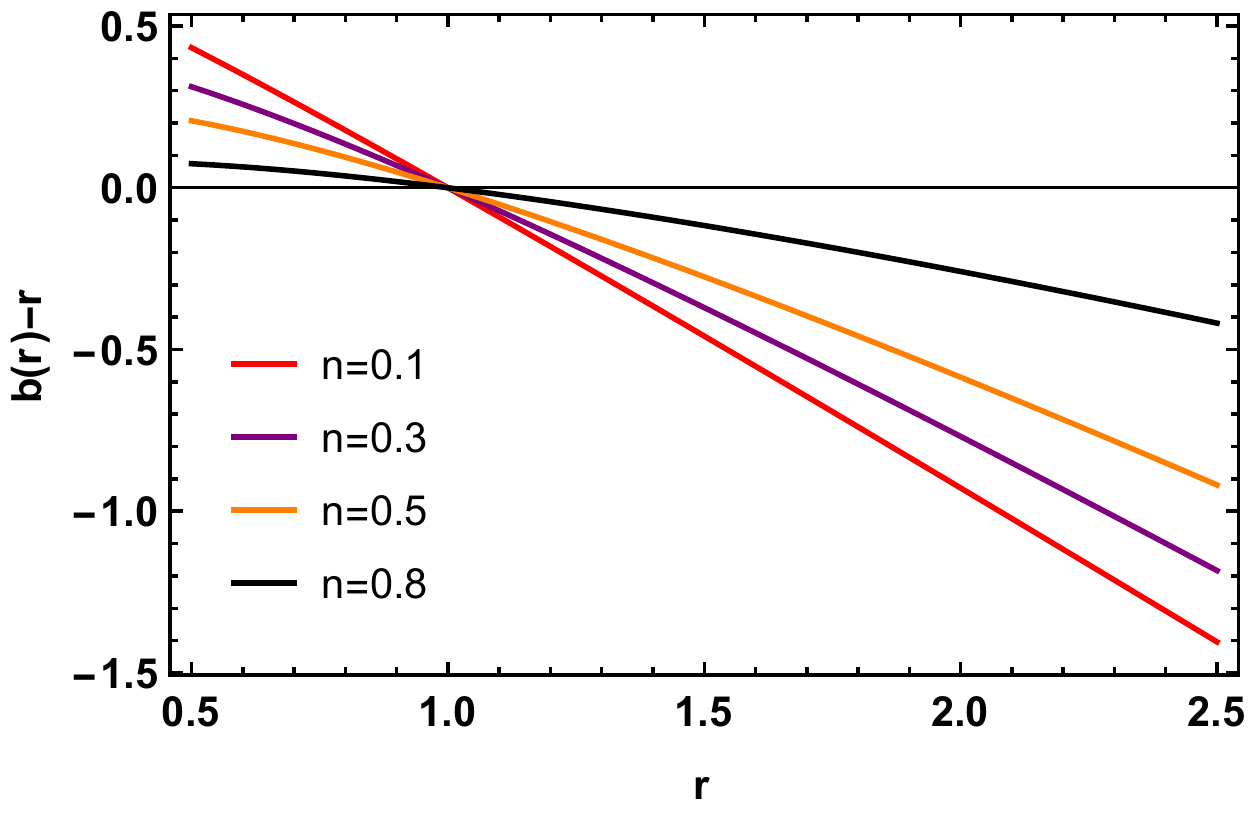}
	\includegraphics[scale=0.3]{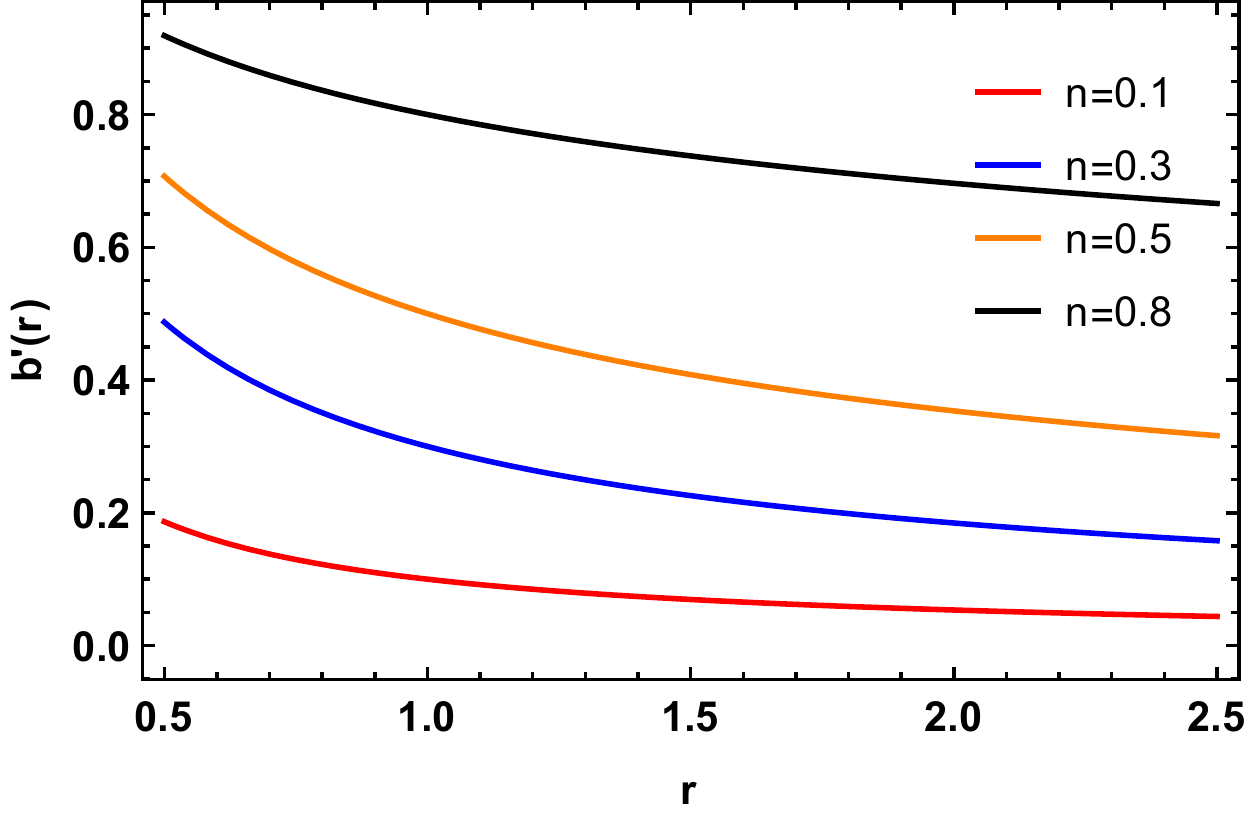}
	\includegraphics[scale=0.3]{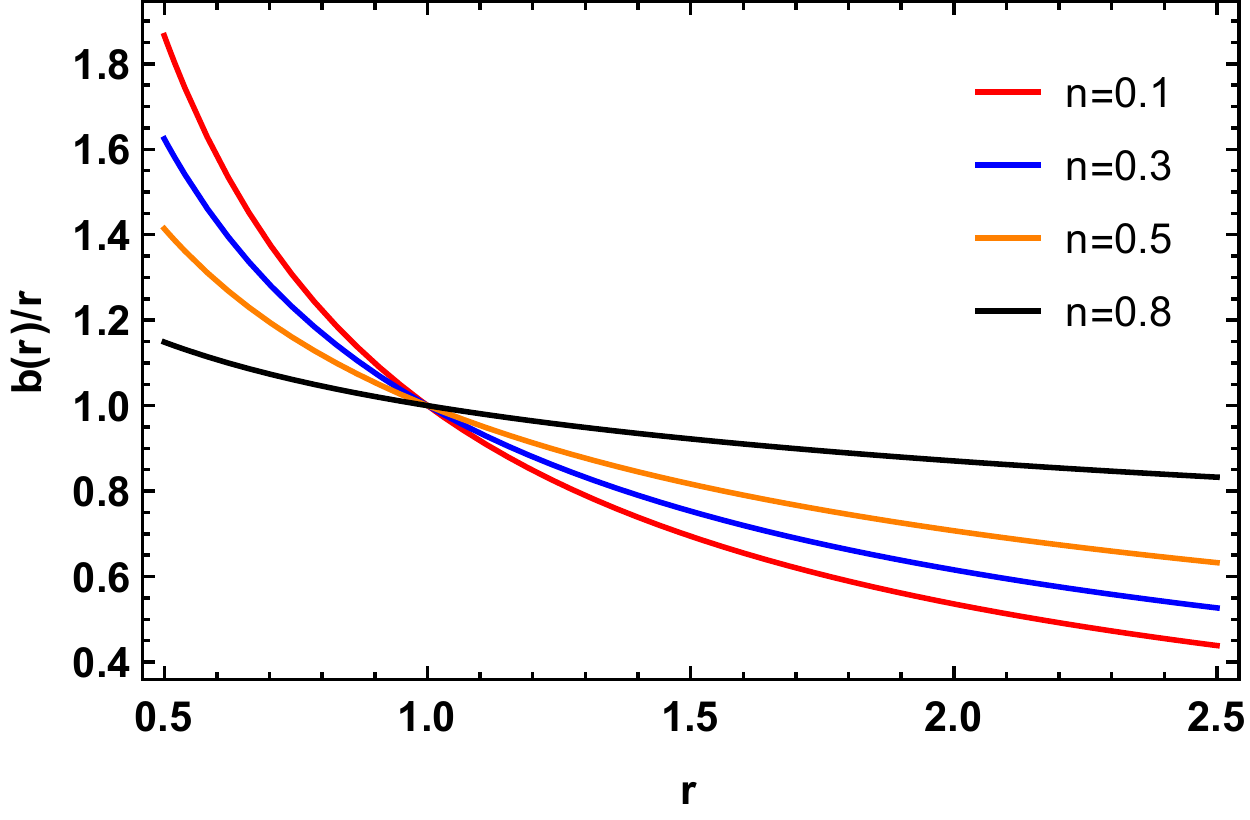}
	\caption{Characteristic of the shape functions for WH4}
	\label{f22}
\end{figure}

\begin{figure}[H]
\centering
     \includegraphics[scale=0.5]{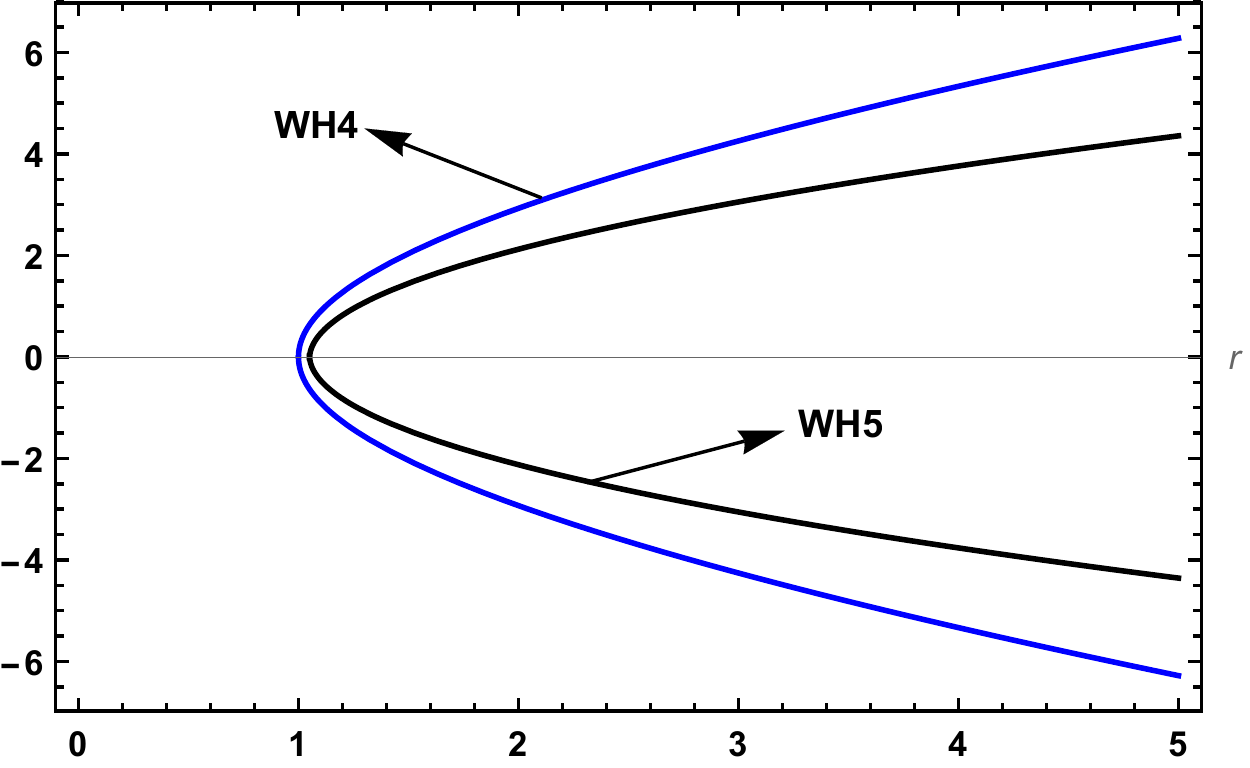}
     \caption{The profile shows two dimensional embedding diagram for WH4 and WH5}
     \label{f30}
\end{figure}

Here, in this subsection, we have assumed the specific shape function given by $b(r)=r_0\left(\frac{r}{r_0}\right)^n$ and for this choice the corresponding stress energy components from Eqns. \eqref{b1}-\eqref{b3} are obtained as follows\\
\begin{widetext}
\begin{equation}
\rho=\frac{2 a \left((6 n-11) r_0 \left(\frac{r}{r_0}\right){}^n+7 r\right) \left(r-r_0 \left(\frac{r}{r_0}\right){}^n\right)}{r^6}+\frac{B}{2}
\end{equation}
\begin{equation}
\rho+P_r= \frac{4 a \left(r-r_0 \left(\frac{r}{r_0}\right){}^n\right) \left((3 n-7) r_0 \left(\frac{r}{r_0}\right){}^n+4 r\right)}{r^6}
\end{equation}
\begin{equation}
\rho+P_t= \frac{2 a \left(r-r_0 \left(\frac{r}{r_0}\right){}^n\right) \left((3 n-5) r_0 \left(\frac{r}{r_0}\right){}^n+4 r\right)}{r^6}
\end{equation}
\begin{equation}
\rho-P_r= \frac{4 a r_0 \left((3 n-7) r-(3 n-4) r_0 \left(\frac{r}{r_0}\right){}^n\right) \left(\frac{r}{r_0}\right){}^n+12 a r^2+B r^6}{r^6}
\end{equation}
\begin{equation}
\rho-P_t= \frac{2 a r_0 \left(9 (n-3) r-(9 n-17) r_0 \left(\frac{r}{r_0}\right){}^n\right) \left(\frac{r}{r_0}\right){}^n+20 a r^2+B r^6}{r^6}
\end{equation}
\end{widetext}
The behavior of shape function is depicted in Fig. \ref{f22}. It can be seen from the Fig. \ref{f22} that shape function, $b(r)$ shows the increasing behavior for different values of $n$. Furthermore, to check the asymptotically flatness condition we plot $\frac{b(r)}{r}$ verses $r$ which gives $\frac{b(r)}{r}\rightarrow 0$ as $r\rightarrow \infty$. Here $b(r)-r$ cuts the $r$-axis at $r_0=1$, which is the throat radius (see Fig. \ref{f30}) for this WH4 when $n<1$. All of the above results shows the traversablity of WH4.
\begin{figure}[H]
\centering
     \includegraphics[scale=0.5]{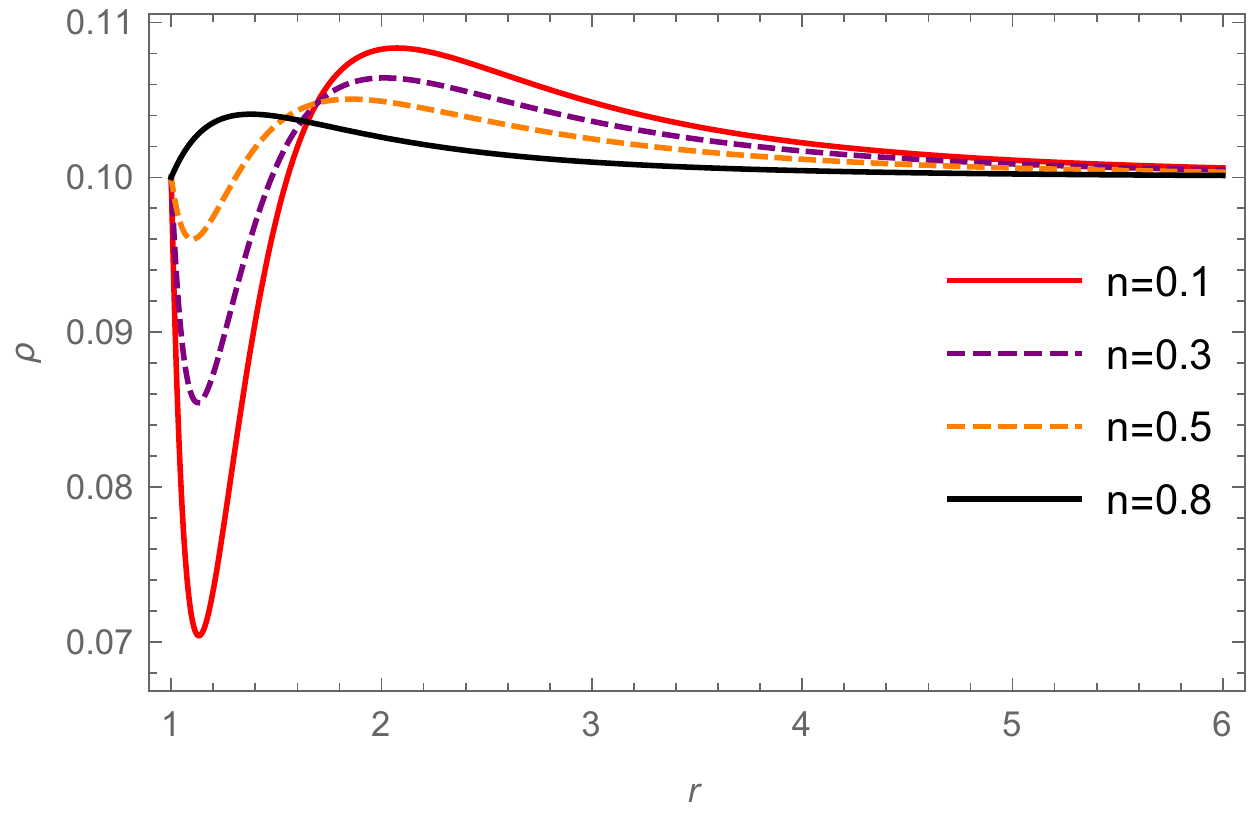}
     \caption{The plot of $\rho$ versus $r$ with $n=0.1 \ (Red)$, $n=0.3 \ (Purple)$, $n=0.5 \ (Orange)$, $n=0.8 \ (Black)$, $a=0.1$, $B=0.2$ and $r_0=1$ for WH4.}
     \label{f23}
\end{figure}

\begin{figure}[]
\subfloat[$\rho+P_r$\label{sfig:testa}]{
  \includegraphics[scale =0.5]{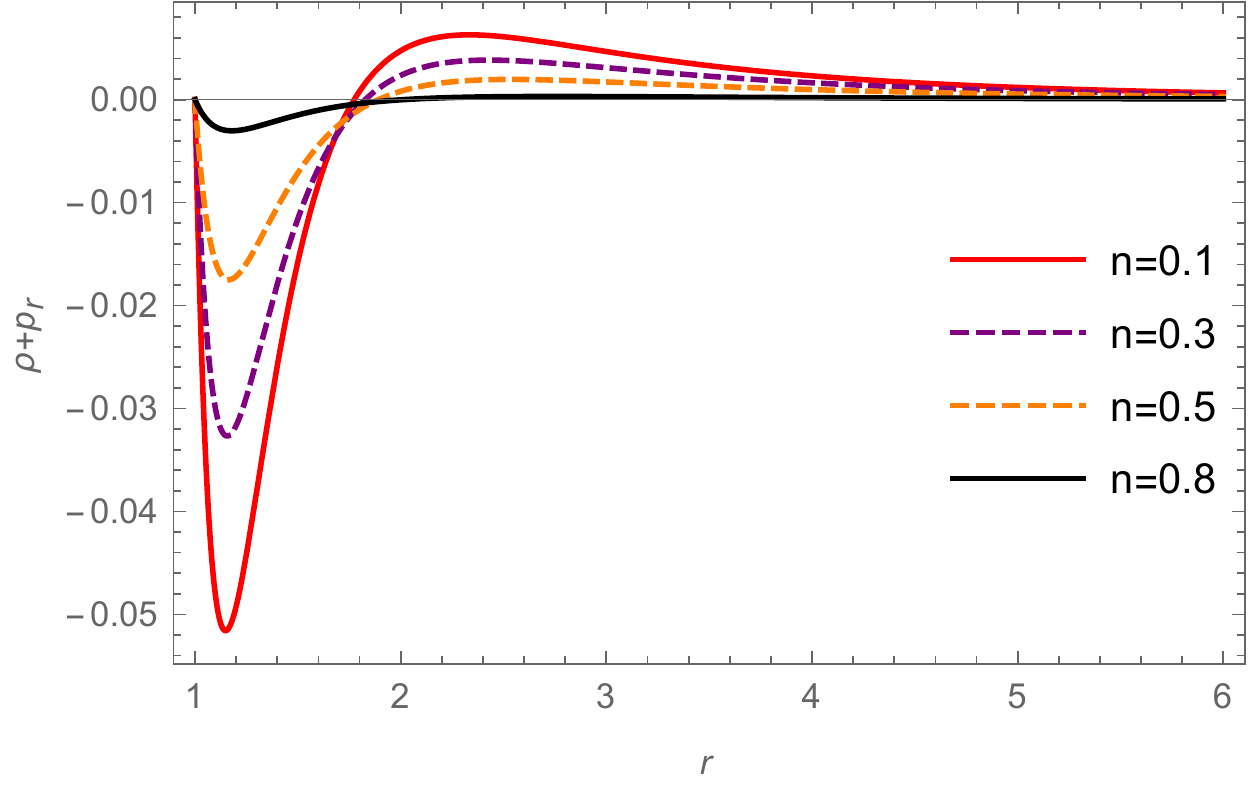}
}\hfill
\subfloat[$\rho+P_t$\label{sfig:testa}]{
  \includegraphics[scale =0.5]{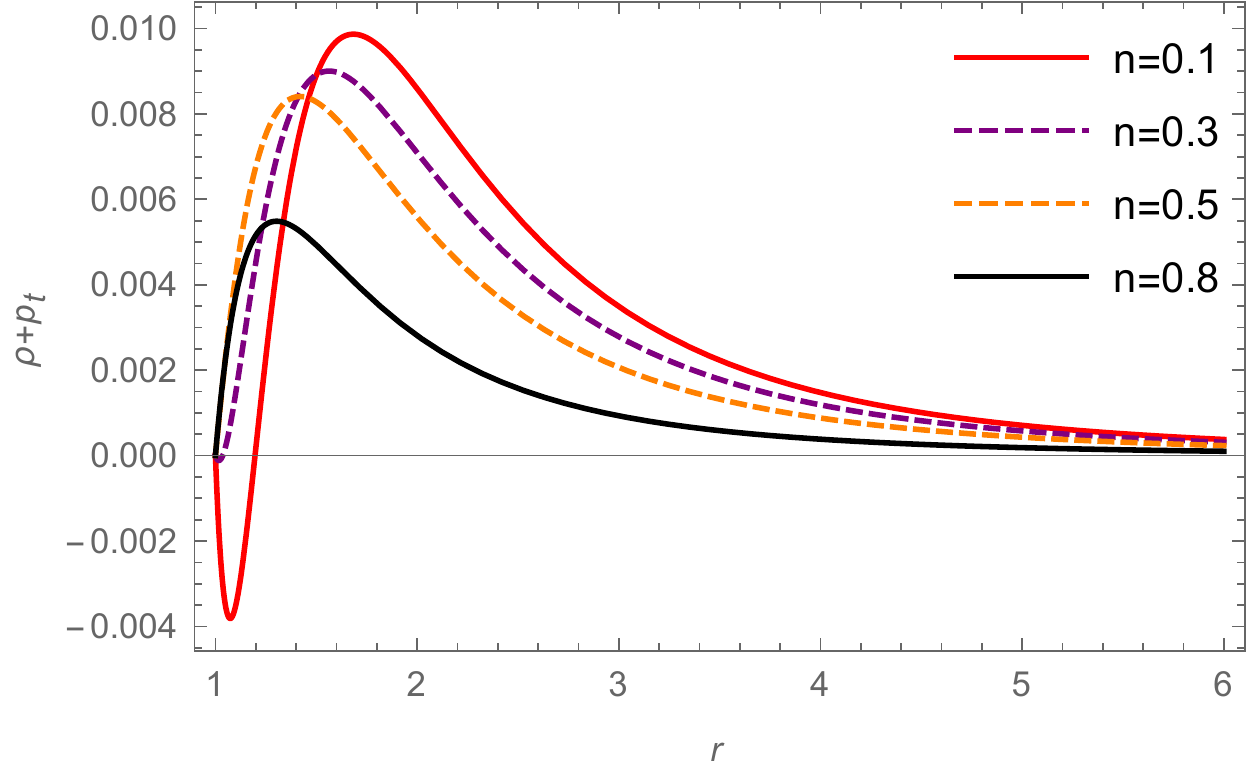}
}
\caption{The plots of $\rho+P_r$ and $\rho+P_t$ versus $r$ with $n=0.1\  (Red)$, $n=0.3 \ (Purple)$, $n=0.5 \ (Orange)$, $n=0.8 \ (Black)$, $a=0.1$, $B=0.2$ and $r_0=1$ for WH4.}
\label{f24}
\end{figure}
\begin{figure}[]
\subfloat[$\rho-P_r$\label{sfig:testa}]{
  \includegraphics[scale =0.5]{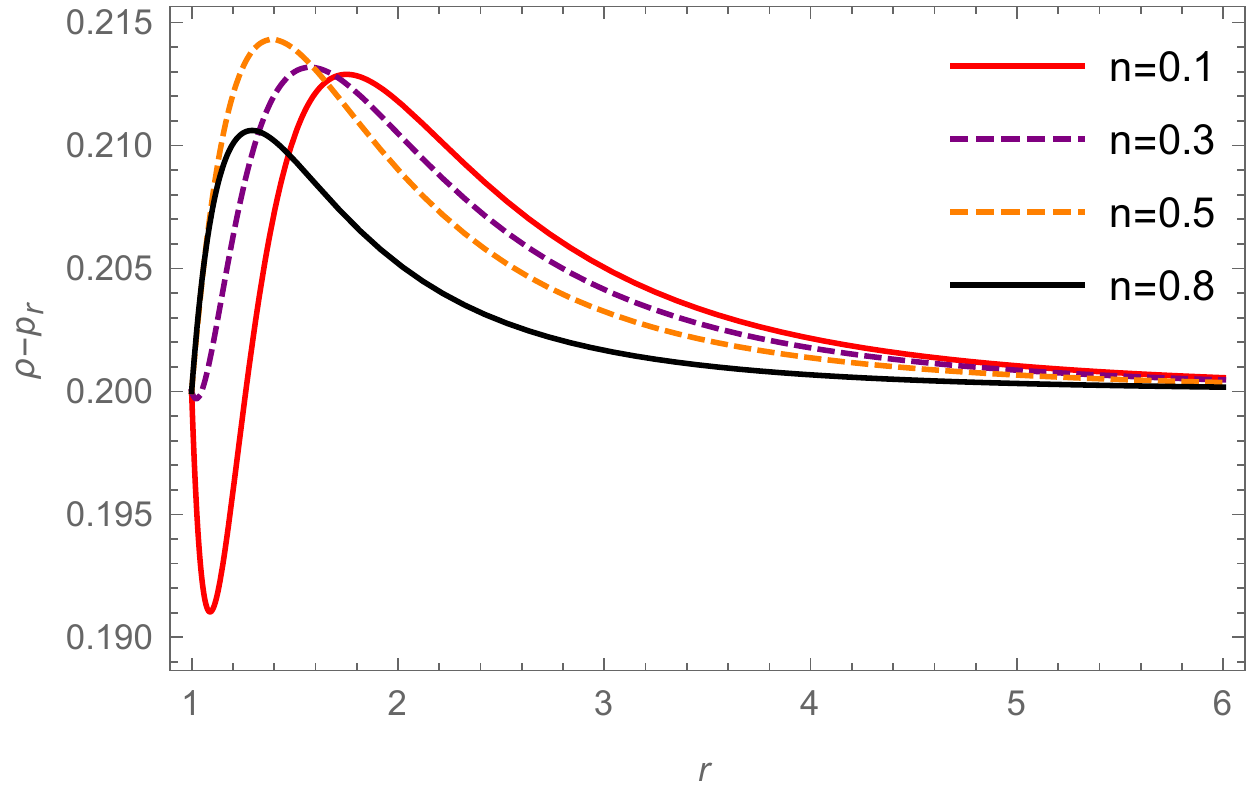}
}\hfill
\subfloat[$\rho-P_t$\label{sfig:testa}]{
  \includegraphics[scale =0.5]{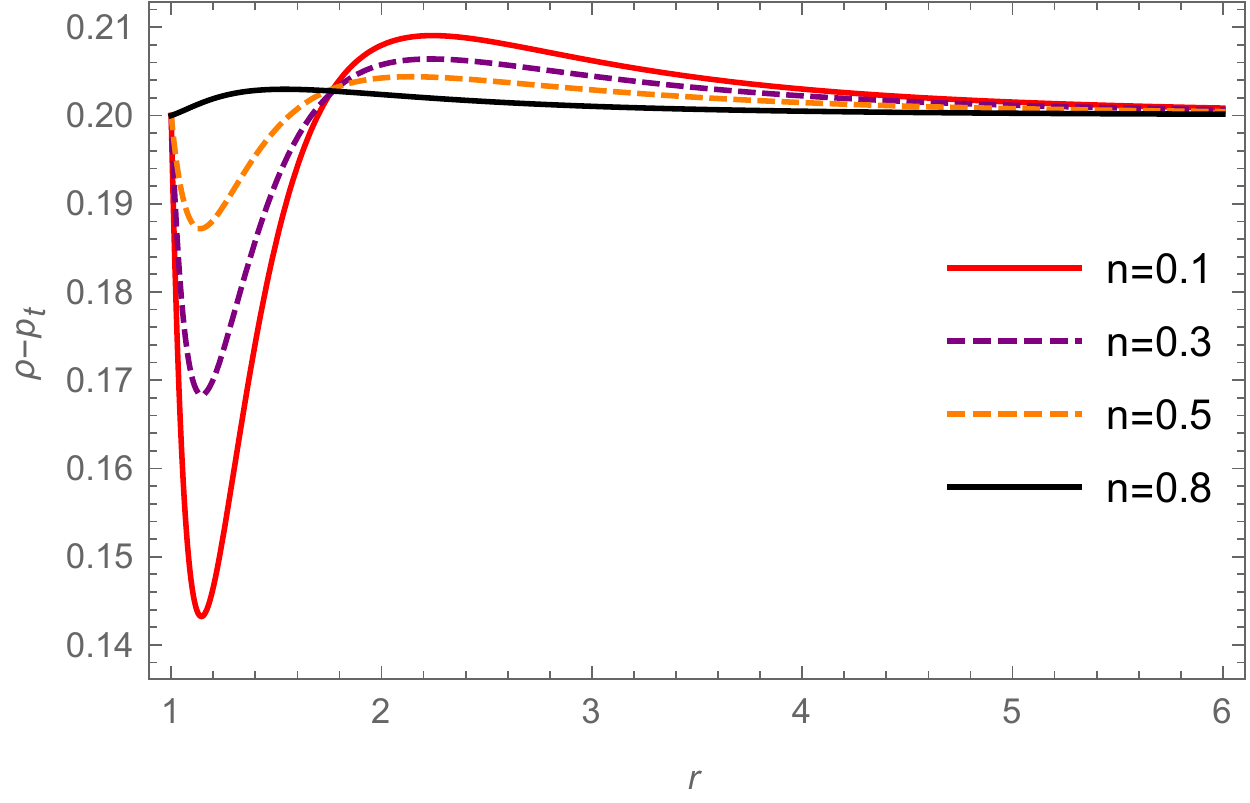}
}
\caption{The plots of $\rho-P_r$ and $\rho-P_t$ versus $r$ with $n=0.1\ (Red)$, $n=0.3 \ (Purple)$, $n=0.5\  (Orange)$, $n=0.8 \ (Black)$, $a=0.1$, $B=0.2$ and $r_0=1$ for WH4.}
\label{f25}
\end{figure}
The behavior of energy conditions (ECs) has shown in Figs. \ref{f23}-\ref{f25}. To verify the profiles of ECs, we have fixed the parameters as $a=0.1$, $B=0.2$ and $r_0=1$. As we know, energy density has to be positive everywhere, and its behaviors for WH4 are presented in Fig. \ref{f23} for different choices of $n$. Also, from Fig. \ref{f24}, we observed that $\rho+P_r<0$ and $\rho+P_t>0$, implying the violation of NEC at wormhole throat. Besides, we have shown the graph of DEC in Fig. \ref{f25}. The violation of NEC for different values of the parameter $n$ is the proof for the presence of exotic matter at the wormhole throat, which may be required for the wormhole geometry.

\subsubsection{\textbf{Wormhole (WH5) solution with $b(r)=\gamma\,r_0\,\left(1-\frac{r_0}{r}\right)+r_0$}}

\begin{figure}[H]
\centering
	\includegraphics[scale=0.3]{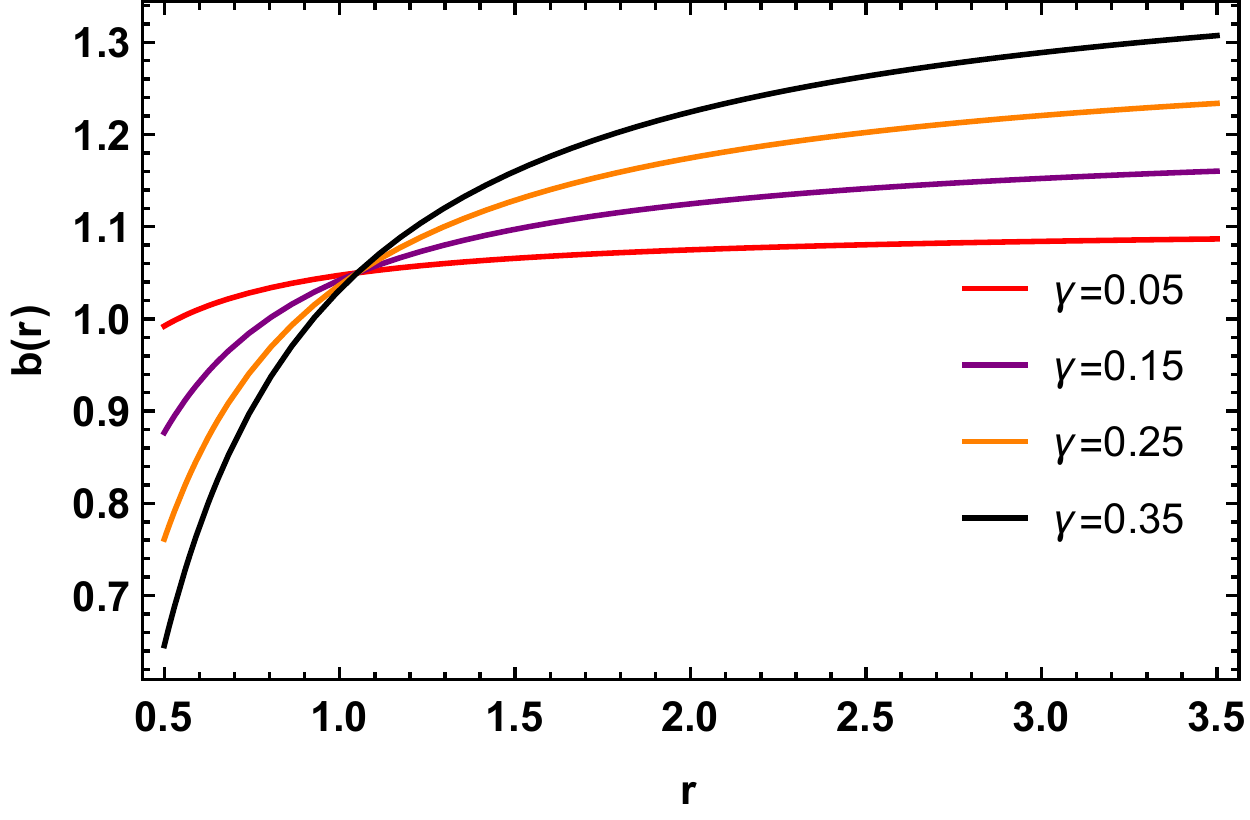}
	\includegraphics[scale=0.3]{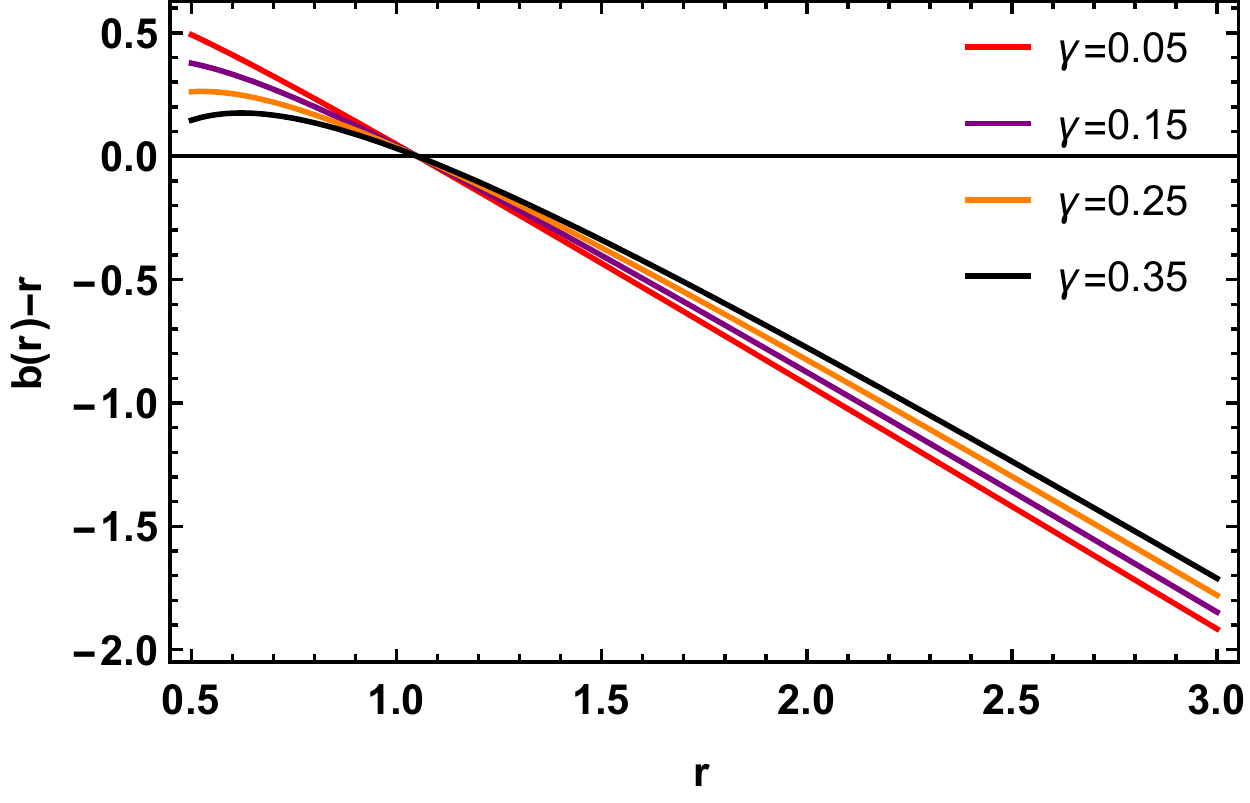}
	\includegraphics[scale=0.3]{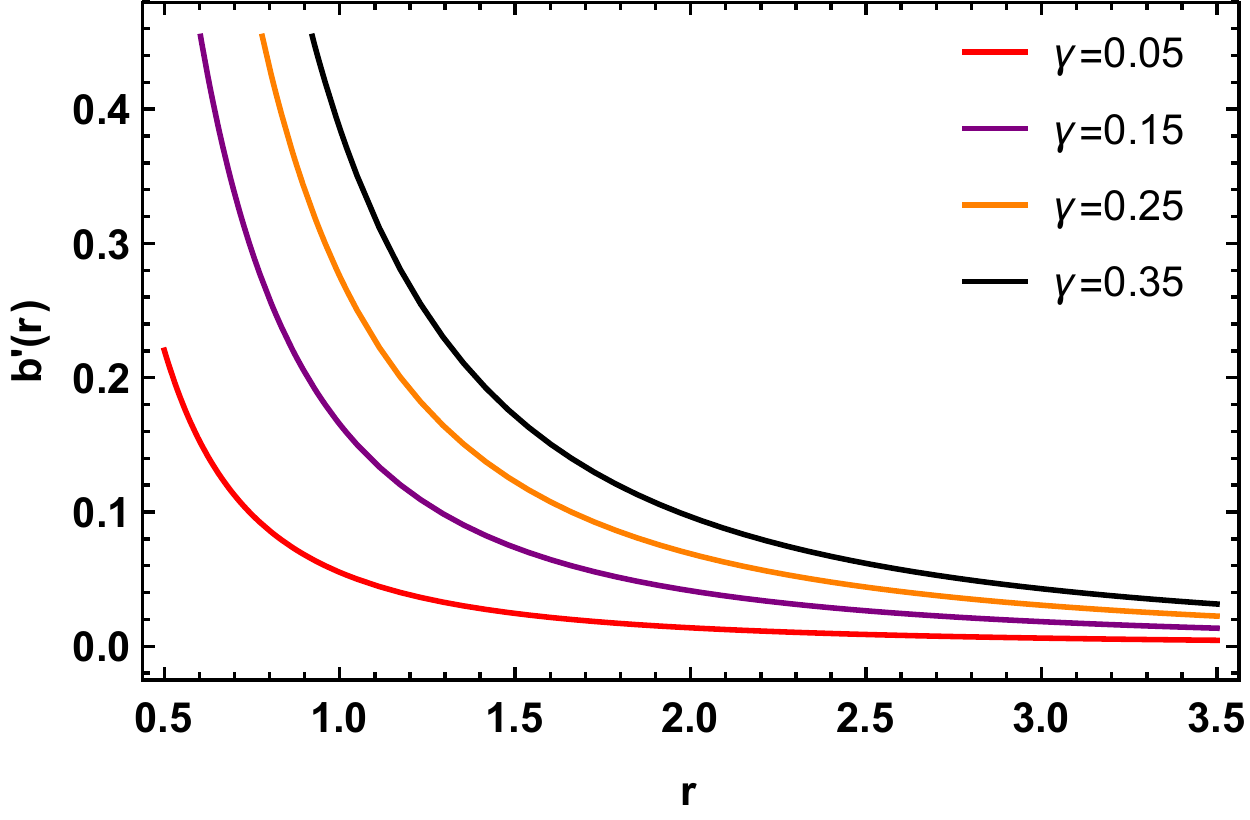}
	\includegraphics[scale=0.3]{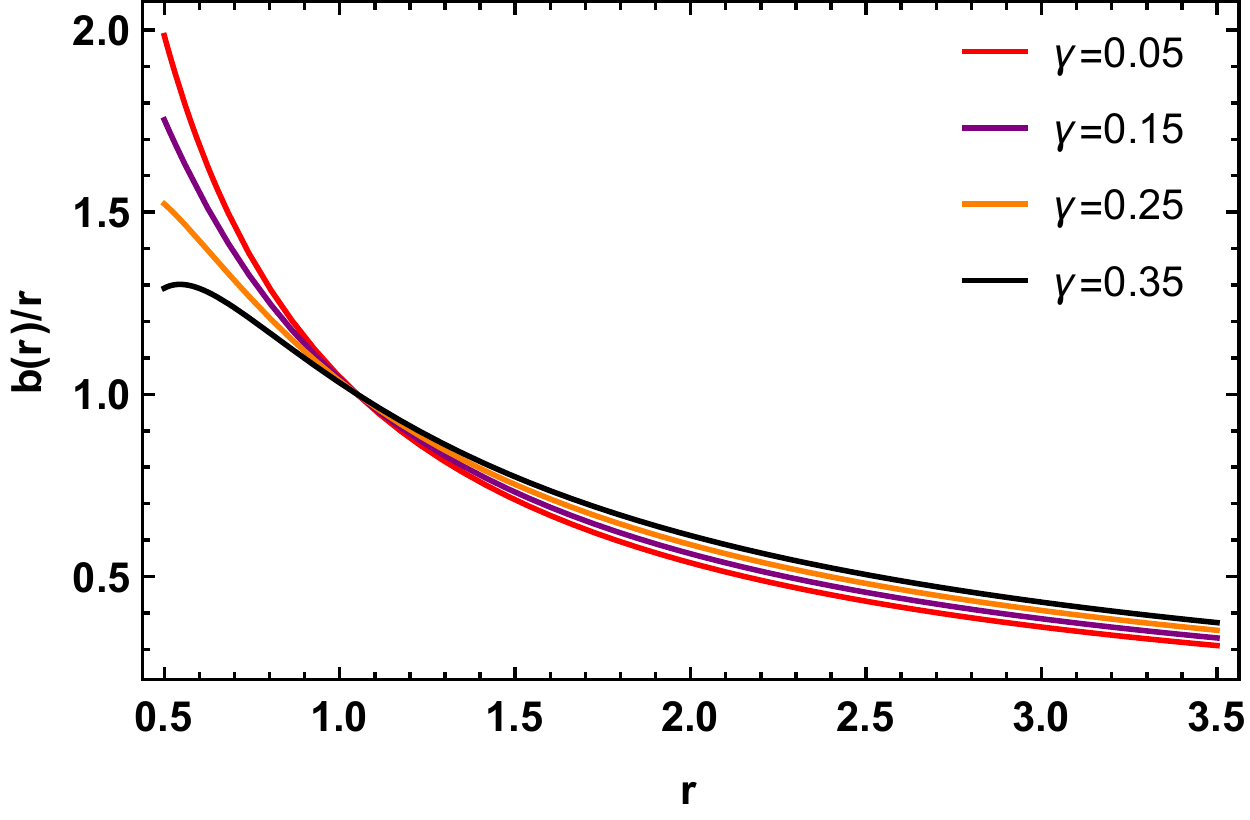}
	\caption{Characteristic of the shape functions for WH5}
	\label{f26}
\end{figure}

Here, we have considered a specific shape function given by $b(r)=\gamma\,r_0\,\left(1-\frac{r_0}{r}\right)+r_0$ and for this choice the corresponding stress energy components from Eqns. \eqref{b1}-\eqref{b3} are obtained as follows\\
\begin{widetext}
\begin{equation}
\rho=\frac{2 a \left(r-r_0\right) \left(r-\gamma  r_0\right) \left(7 r^2-11 (\gamma +1) r_0 r+17 \gamma  r_0^2\right)}{r^8}+\frac{B}{2}
\end{equation}
\begin{equation}
\rho+P_r= \frac{4 a \left(r-r_0\right) \left(r-\gamma  r_0\right) \left(4 r^2-7 (\gamma +1) r_0 r+10 \gamma  r_0^2\right)}{r^8}
\end{equation}
\begin{equation}
\rho+P_t= \frac{2 a \left(r-r_0\right) \left(r-\gamma  r_0\right) \left(4 r^2-5 (\gamma +1) r_0 r+8 \gamma  r_0^2\right)}{r^8}
\end{equation}
\begin{equation}
\rho-P_r= \frac{12 a r^4+4 a r_0 \left(r_0 \left(2 (\gamma  (2 \gamma +9)+2) r^2+\gamma  r_0 \left(7 \gamma  r_0-11 (\gamma +1) r\right)\right)-7 (\gamma +1) r^3\right)+B r^8}{r^8}
\end{equation}
\begin{equation}
\rho-P_t= \frac{20 a r^4+2 a r_0 \left(r_0 \left((\gamma  (17 \gamma +70)+17) r^2+\gamma  r_0 \left(26 \gamma  r_0-43 (\gamma +1) r\right)\right)-27 (\gamma +1) r^3\right)+B r^8}{r^8}
\end{equation}
\end{widetext}
The behavior of shape function is depicted in Fig. \ref{f26}. It can be seen from the Fig. \ref{f26} that shape function, $b(r)$ shows the increasing behavior for different values of $n$. We plot $\frac{b(r)}{r}$ with respect to $r$ to check the asymptotically flatness condition which gives $\frac{b(r)}{r}\rightarrow 0$ as $r\rightarrow \infty$. Here, in this case $b(r)-r$ cuts the $r$-axis at $r_0=1.05$, which is the throat radius (see Fig. \ref{f26}) for this WH5. Flare out condition is also satisfied as $b^{'}(r)<1$ for $r>r_0$. All the above profiles for the shape function of WH5 satisfies the necessary properties for a traversable wormhole.
\begin{figure}[H]
\centering
     \includegraphics[scale=0.5]{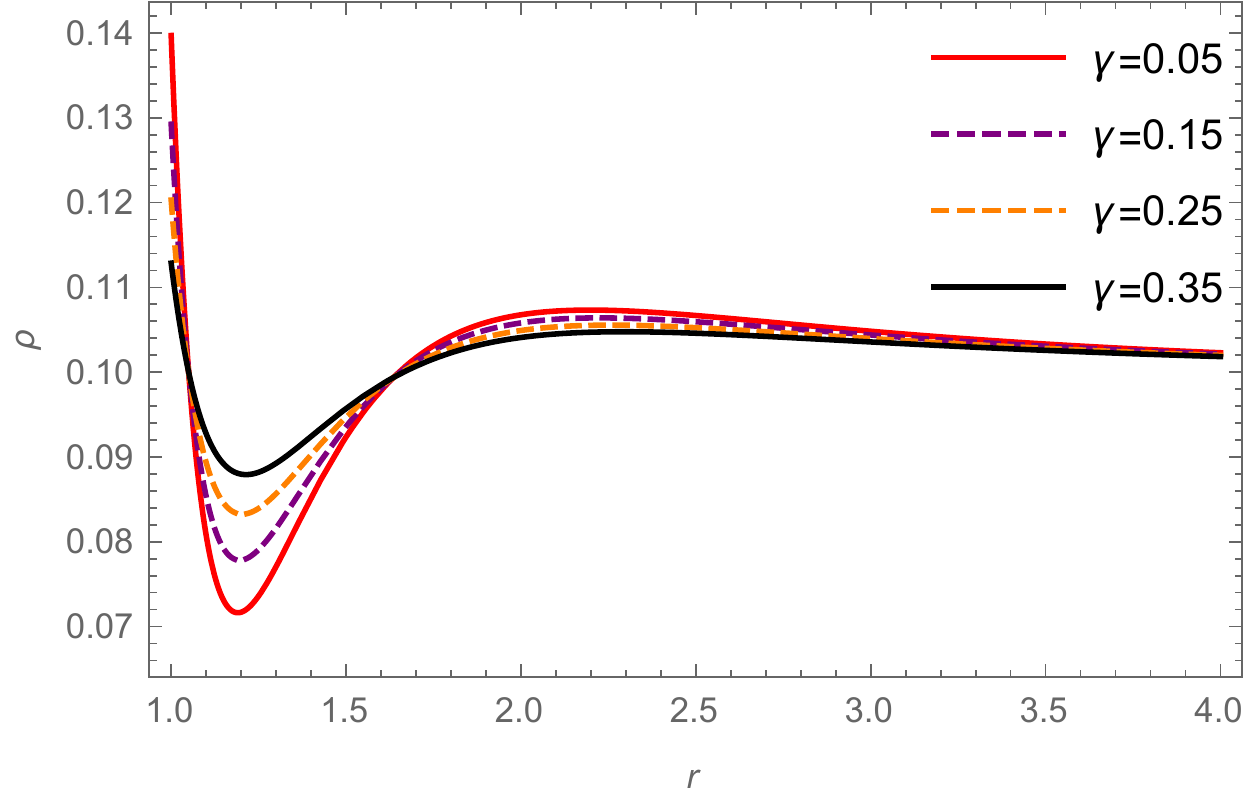}
     \caption{The plot of $\rho$ versus $r$ with $\gamma=0.05\ (Red)$, $\gamma=0.15 \ (Purple)$, $\gamma=0.25 \ (Orange)$, $\gamma=0.35 \ (Black)$, $a=0.1$, $B=0.2$ and $r_0=1.05$ for WH5.}
     \label{f27}
\end{figure}

\begin{figure}[]
\subfloat[$\rho+P_r$\label{sfig:testa}]{
  \includegraphics[scale =0.5]{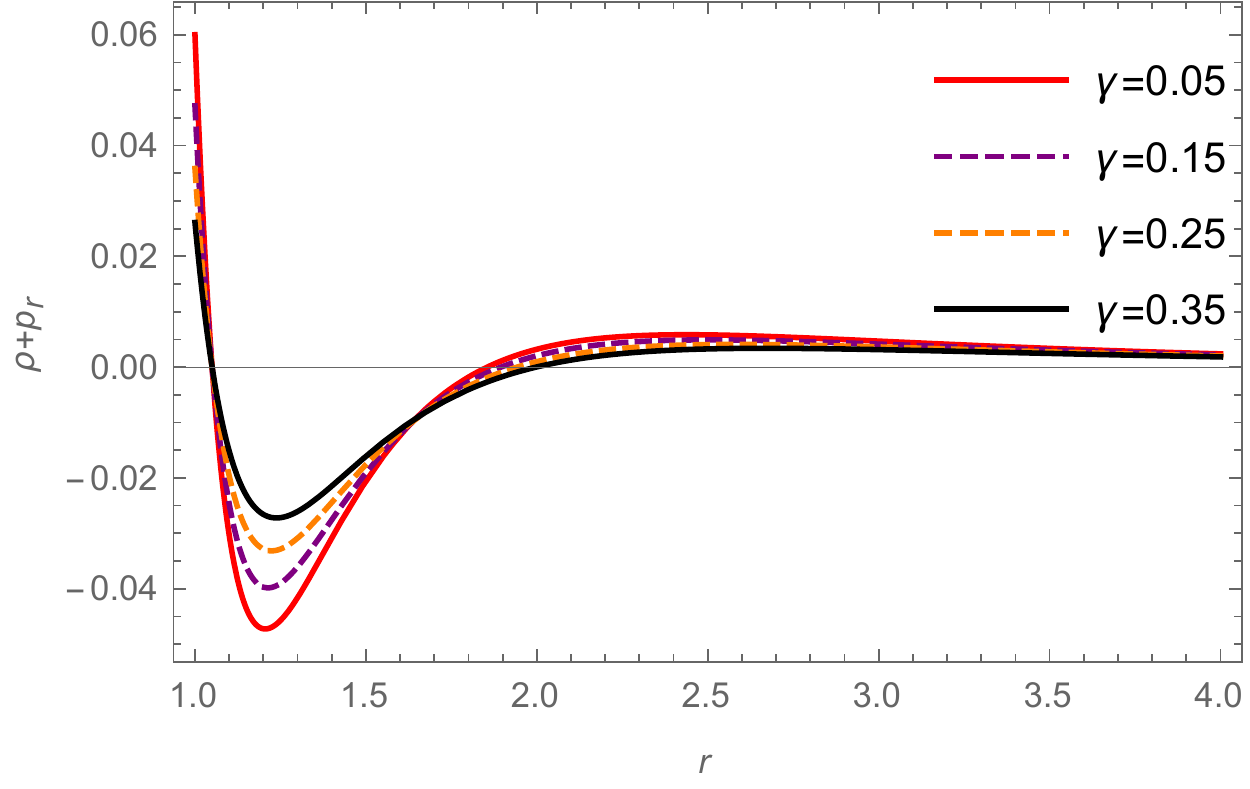}
}\hfill
\subfloat[$\rho+P_t$\label{sfig:testa}]{
  \includegraphics[scale =0.5]{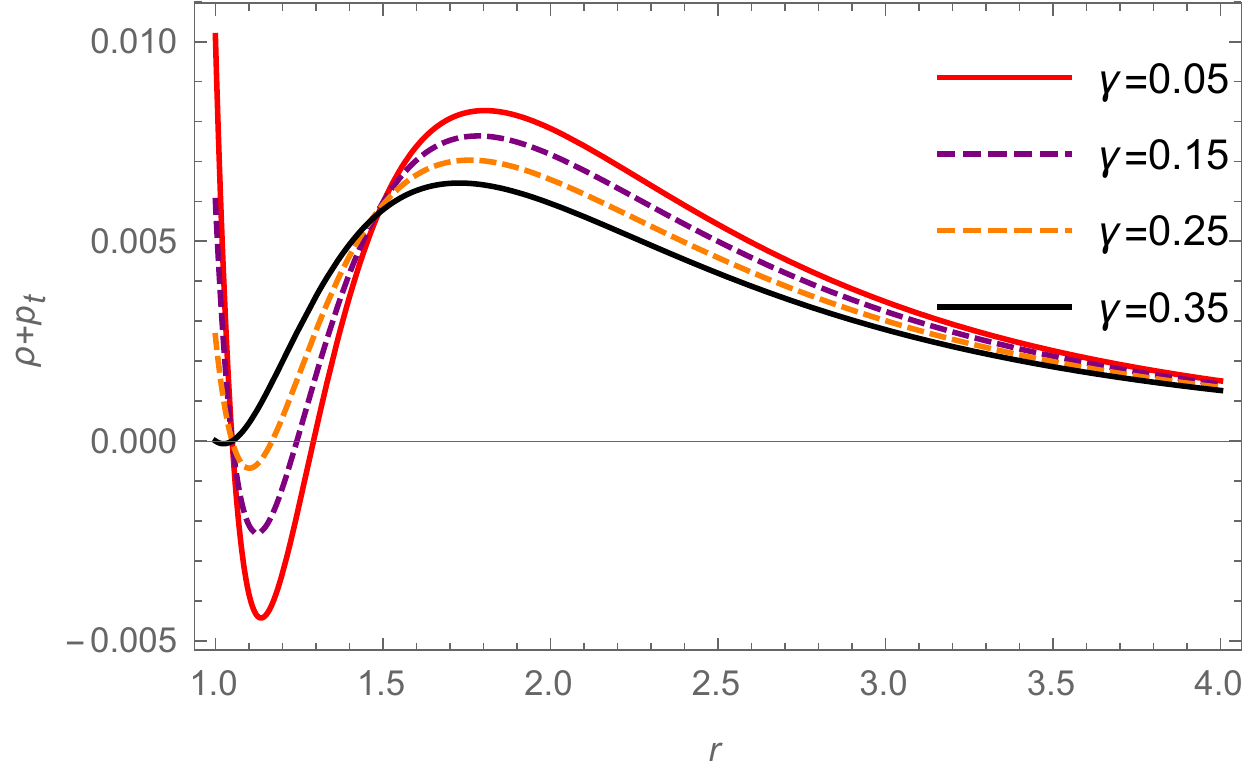}
}
\caption{The plots of $\rho+P_r$ and $\rho+P_t$ versus $r$ with $\gamma=0.05\ (Red)$, $\gamma=0.15 \ (Purple)$, $\gamma=0.25 \ (Orange)$, $\gamma=0.35 \ (Black)$, $a=0.1$, $B=0.2$ and $r_0=1.05$ for WH5.}
\label{f28}
\end{figure}
\begin{figure}[]
\subfloat[$\rho-P_r$\label{sfig:testa}]{
  \includegraphics[scale =0.5]{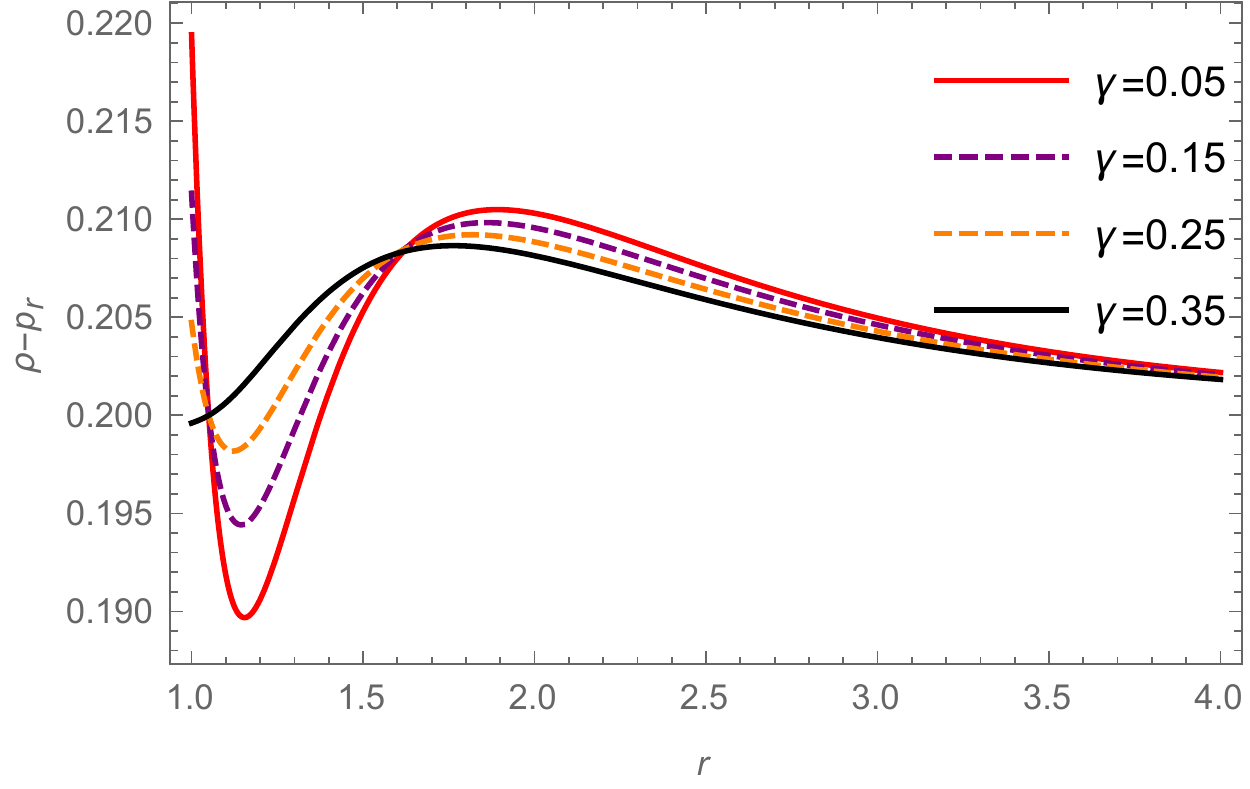}
}\hfill
\subfloat[$\rho-P_t$\label{sfig:testa}]{
  \includegraphics[scale =0.5]{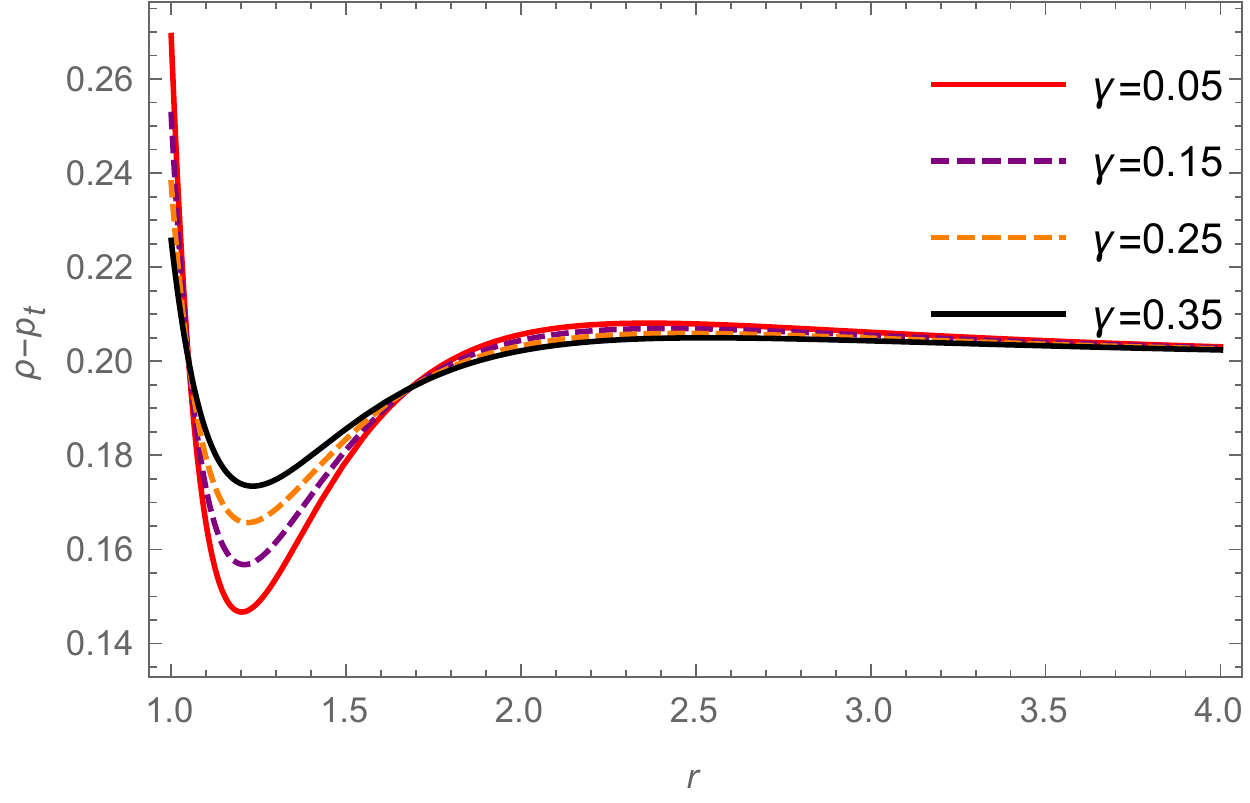}
}
\caption{The plots of $\rho+P_r$ and $\rho+P_t$ versus $r$ with $\gamma=0.05\ (Red)$, $\gamma=0.15 \ (Purple)$, $\gamma=0.25 \ (Orange)$, $\gamma=0.35\  (Black)$, $a=0.1$, $B=0.2$ and $r_0=1.05$ for WH5.}
\label{f29}
\end{figure}

Figs. \ref{f27}-\ref{f29} shows the behavior of energy conditions for WH5. Here, we have chosen the parameters as $a=0.1$, $B=0.2$ and $r_0=1.05$. In Fig. \ref{f27}, we have shown the graph of energy density, $\rho$ versus $r$, which shows a positive behavior throughout the space-time for different values of the parameter $\gamma$. Fig. \ref{f28} shows the behavior of NEC, and one can be seen from the figure that NEC is violated as $\rho+P_r<0$, which implying the violation of WEC. The violation of NEC is the primary requirement for a traversable wormhole. Moreover, we have shown the graph of DEC in Fig. \ref{f29}. These results tend to be determined that the wormhole solutions that have been obtained are acceptable in this symmetric teleparallel gravity.

\section{Embedding Diagram}\label{sec8}

\justifying
In this section, we are going to discuss the embedding diagrams that would help us to understand the wormhole spacetime in \eqref{9}. On the explicit interest in geometry, we composed some constraints on the coordinate system. We considered the equatorial slice $\theta=\pi/2$ for a fixed time i.e. $t=$constant. For this, Eq. \eqref{9} reduces to
\begin{equation}
\label{47}
ds^2=\left(1-\frac{b(r)}{r}\right)^{-1}+r^2 d\phi^2.
\end{equation}

This reformed metric can be embedded into three-dimensional Euclidean space with cylindrical coordinate $r,\,\ \phi$ and $z$ as
\begin{equation}
\label{48}
ds^2=dz^2+dr^2+r^2 d\phi^2.
\end{equation}

Now, comparing Eq. \eqref{47} and \eqref{48}, we can find the embedding surface $z(r)$, and we obtained a slope as

\begin{equation}
\label{49}
\frac{dz}{dr}=\pm \sqrt{\frac{r}{r-b(r)}-1}
\end{equation}

\begin{figure}[H]
\subfloat[WH1\label{sfig:testa}]{
  \includegraphics[scale =0.26]{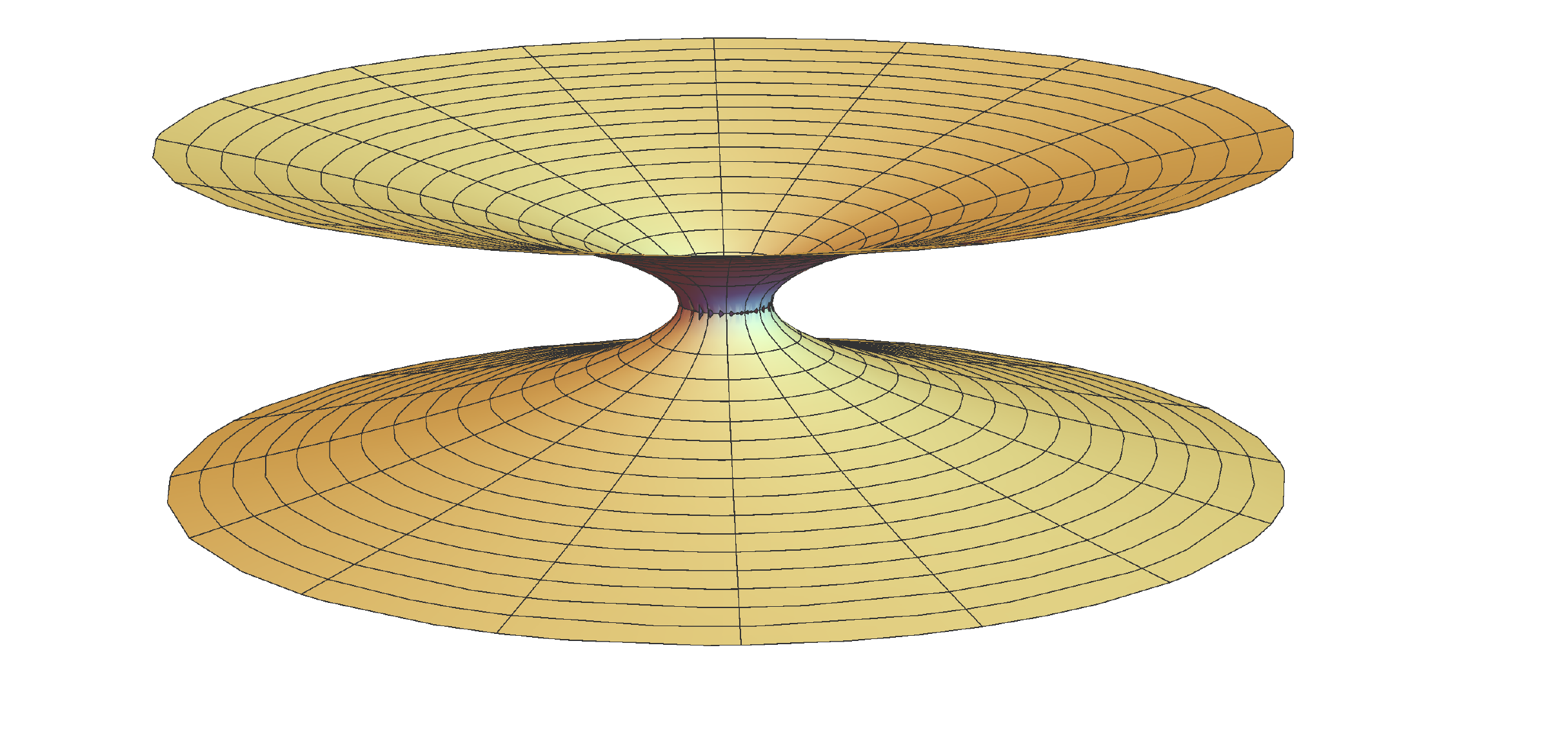}
}\hfill
\subfloat[WH2\label{sfig:testa}]{
  \includegraphics[scale =0.26]{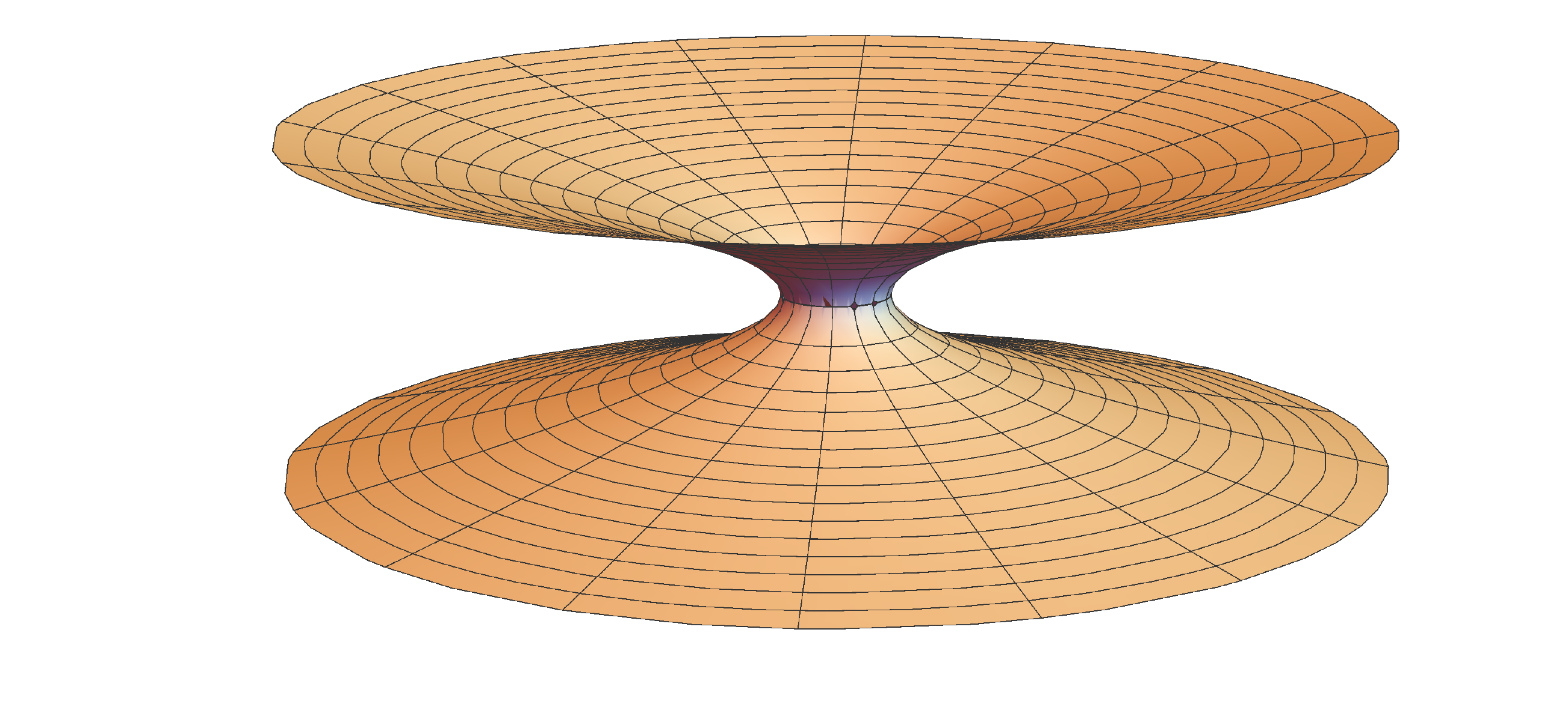}
}\hfill
\subfloat[WH3\label{sfig:testa}]{
  \includegraphics[scale =0.26]{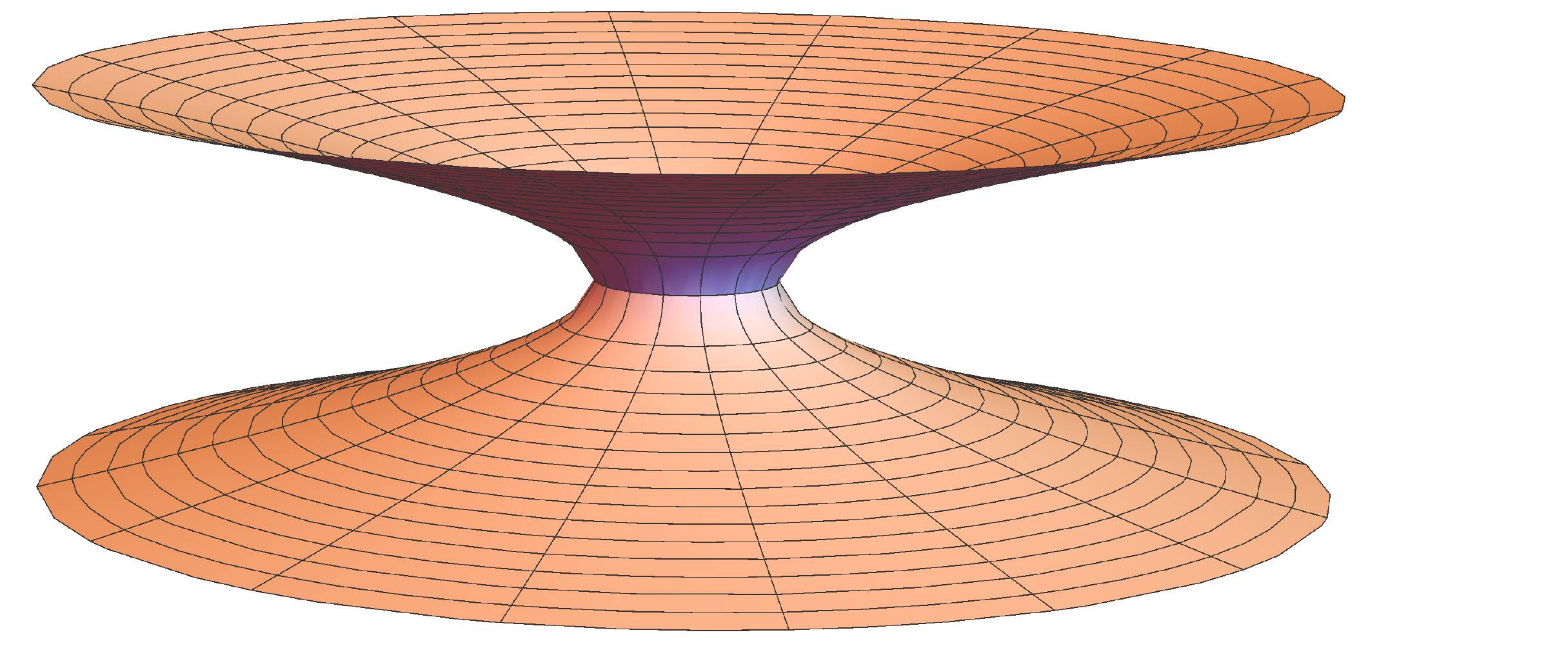}
}
\hfill
\subfloat[WH4\label{sfig:testa}]{
  \includegraphics[scale =0.26]{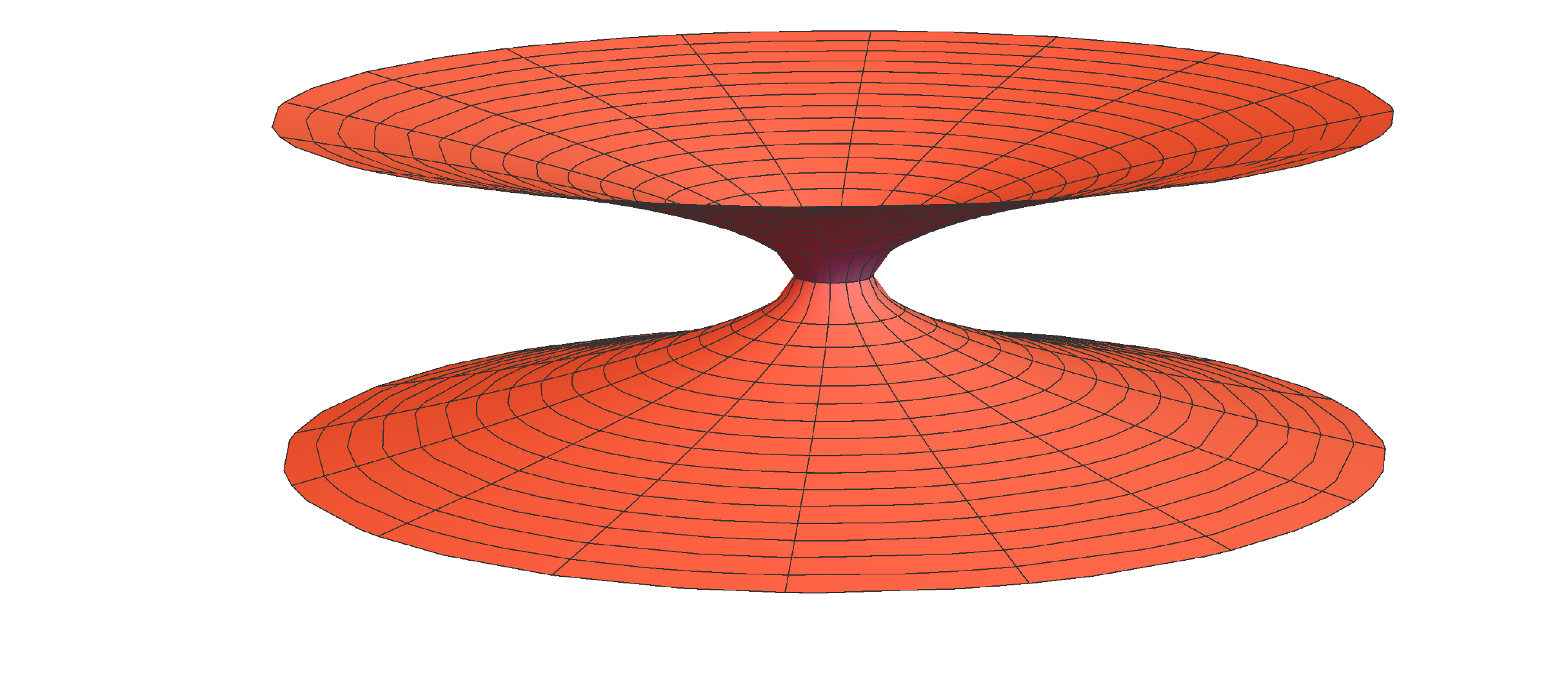}
}
\hfill
\subfloat[WH5\label{sfig:testa}]{
  \includegraphics[scale =0.26]{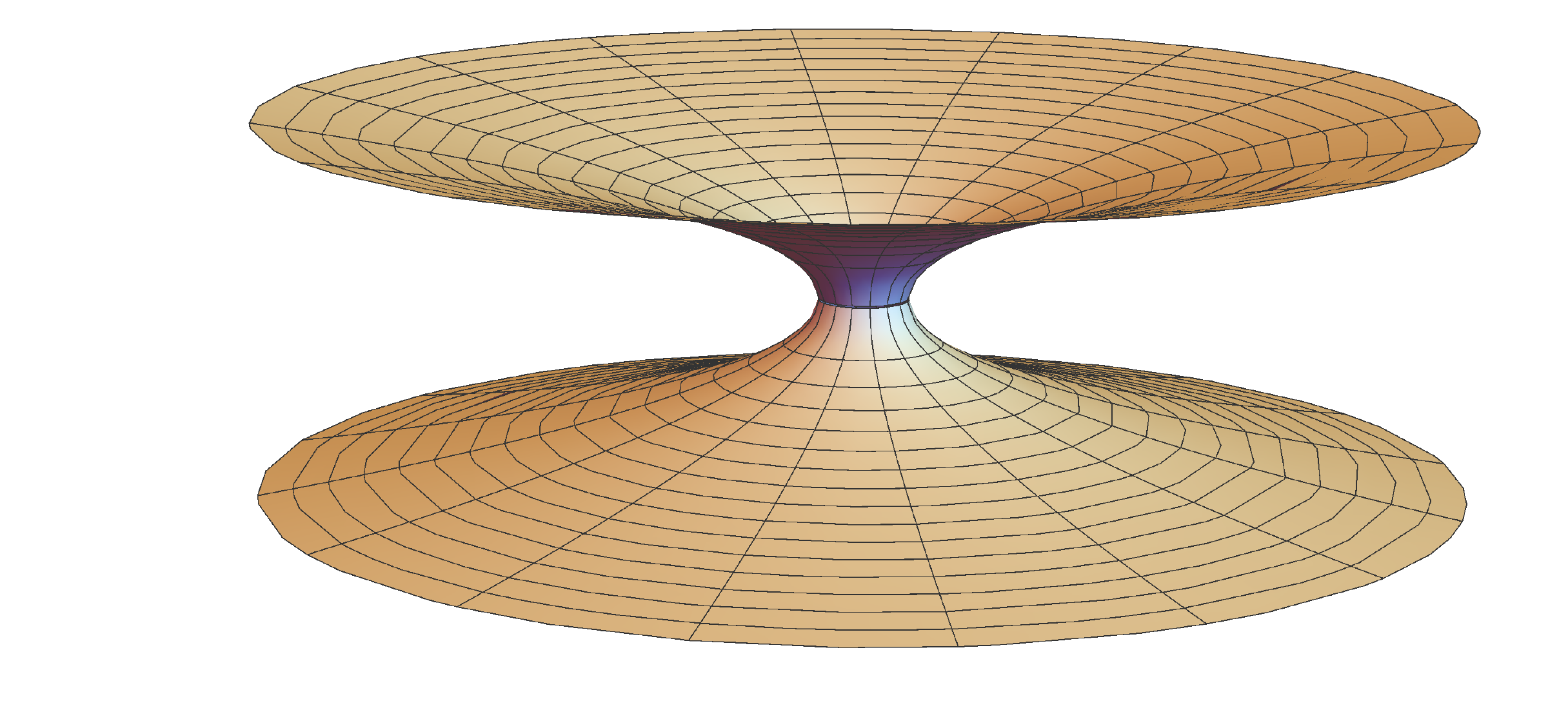}
}
\caption{Embedding diagrams.}
\label{f10}
\end{figure}

In Eq. \eqref{49}, we used the solutions for $b(r)$, which are evaluated for three WH model to draw the embedding surfaces. In Fig. \ref{f10}, we have shown the embedding surfaces for WH1, WH2, WH3, WH4 and WH5 concerning five models. The values of free parameters are the same as those used to discuss the energy conditions for respective models.

\section{Volume Integral Quantifier}  \label{sec9}

\begin{figure}[H]
\centering
	\includegraphics[scale=0.27]{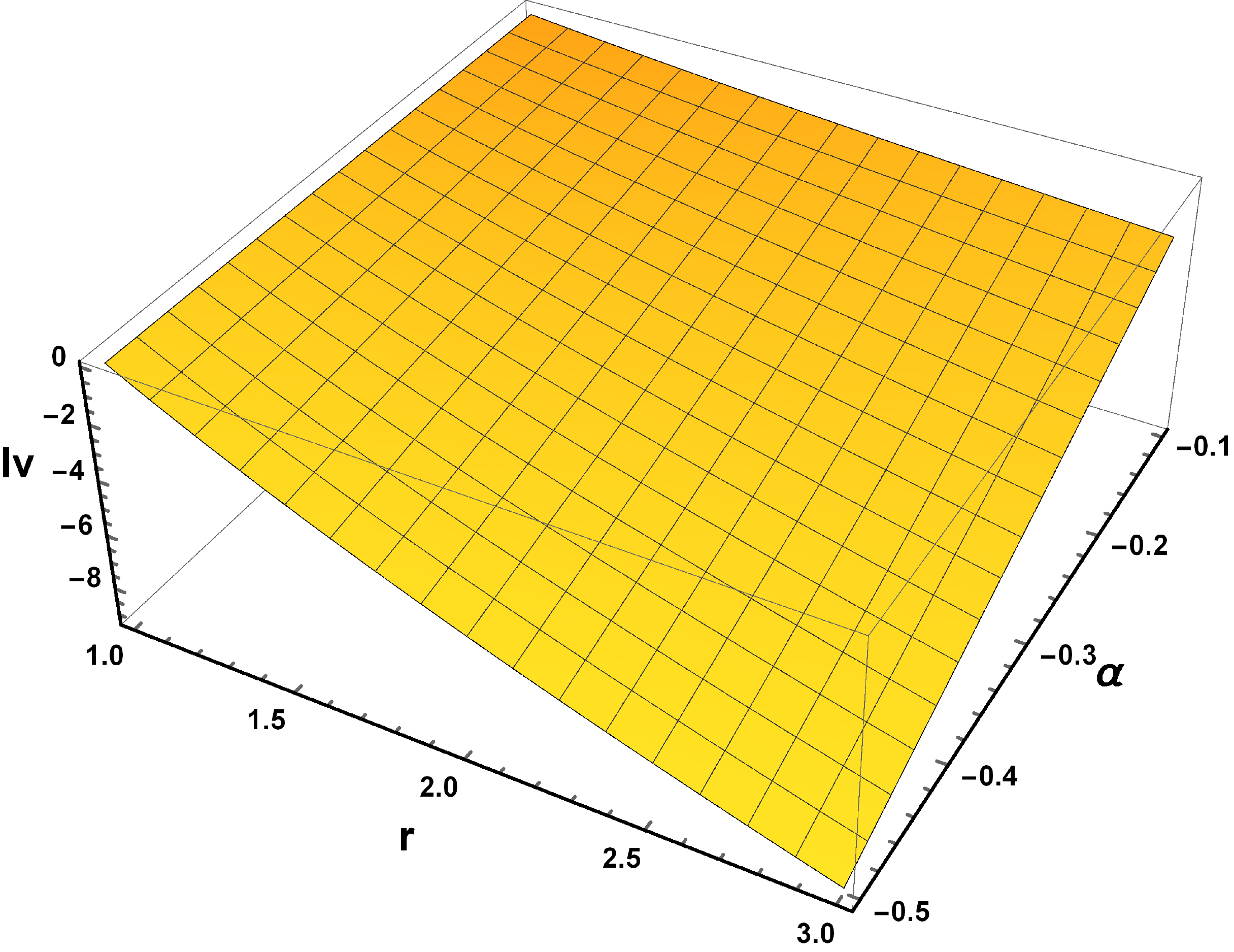}
	\caption{3D Plot for volume integral quantifier associated with model 1 and here it is clear that when $a\rightarrow r_0^+$ then $Iv\rightarrow0$ i.e. minimize the violation of NEC would be possible. For this model we consider $n=-0.25$, $r_0=1$ and $a=3$.}
	\label{f19}
	\end{figure}

\begin{figure}[H]
\centering
	\includegraphics[scale=0.27]{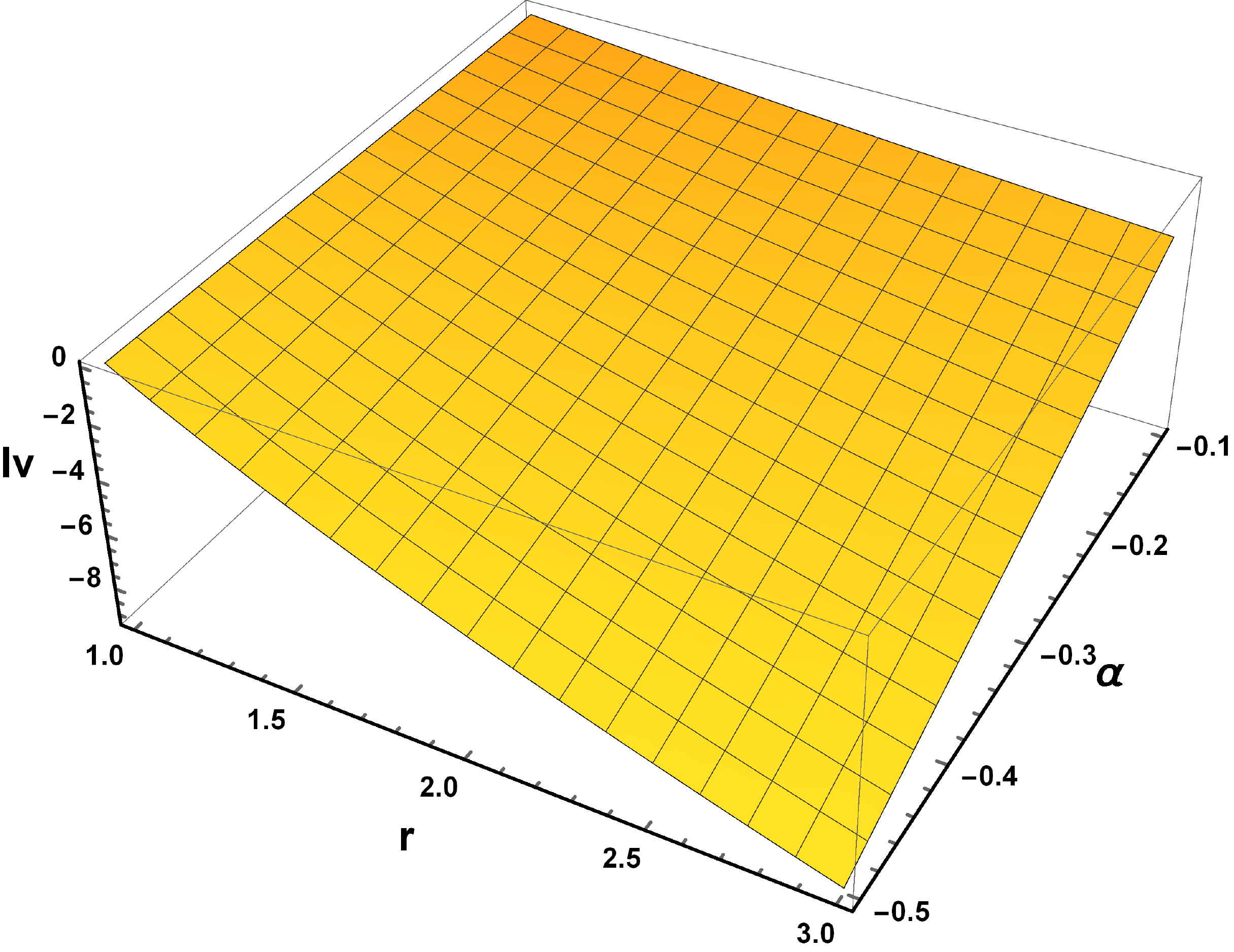}
	\caption{3D Plot for volume integral quantifier associated with model 2 and here it is clear that when $a\rightarrow r_0^+$ then $Iv\rightarrow0$. For this model we consider $\omega=-2$, $r_0=1$ and $a=3$.}
	\label{f20}
	\end{figure}

Volume integral quantifier provides the information about the "total amount of exotic matter" required for wormhole maintenance. To do this, one may compute the definite integral $\int T_{\mu\nu}U^\mu\,U^\nu$ and $\int T_{\mu\nu}k^\mu\,k^\nu$, where $U^\mu$ is the four-velocity \cite{Visser/2003}. For spherically symmetry and average null energy condition (ANEC) violating matter related to the radial component is defined as
\begin{equation}
Iv=\oint[\rho+P_r]dV
\end{equation}
where $dV=r^2\sin\theta\,dr\,d\theta\,d\phi$\\
it can also be written as
\begin{equation}
Iv=8\pi\int_{r_0}^{\infty}\,(\rho+P_r)r^2\,dr
\end{equation}
Now suppose that the wormhole enlarge from the throat, $r_0$, with a cutoff of the stress energy tensor at certain radius $a$, then it reduces to
\begin{equation}
Iv=8\pi\int_{r_0}^a\,(\rho+P_r)r^2\,dr
\end{equation}
where $r_0$ is the throat of the wormhole, which is the minimum value of $r$. The main point of this discussion is when limiting $a\rightarrow r_0^+$, one can verify that $Iv\rightarrow 0$. From Figs. \ref{f19},\ref{f20} and \ref{f21} we found that for each wormhole solution one may construct wormhole solutions with small quantities of exotic matter which need to open the wormhole throat.
\begin{figure}[H]
\centering
	\includegraphics[scale=0.27]{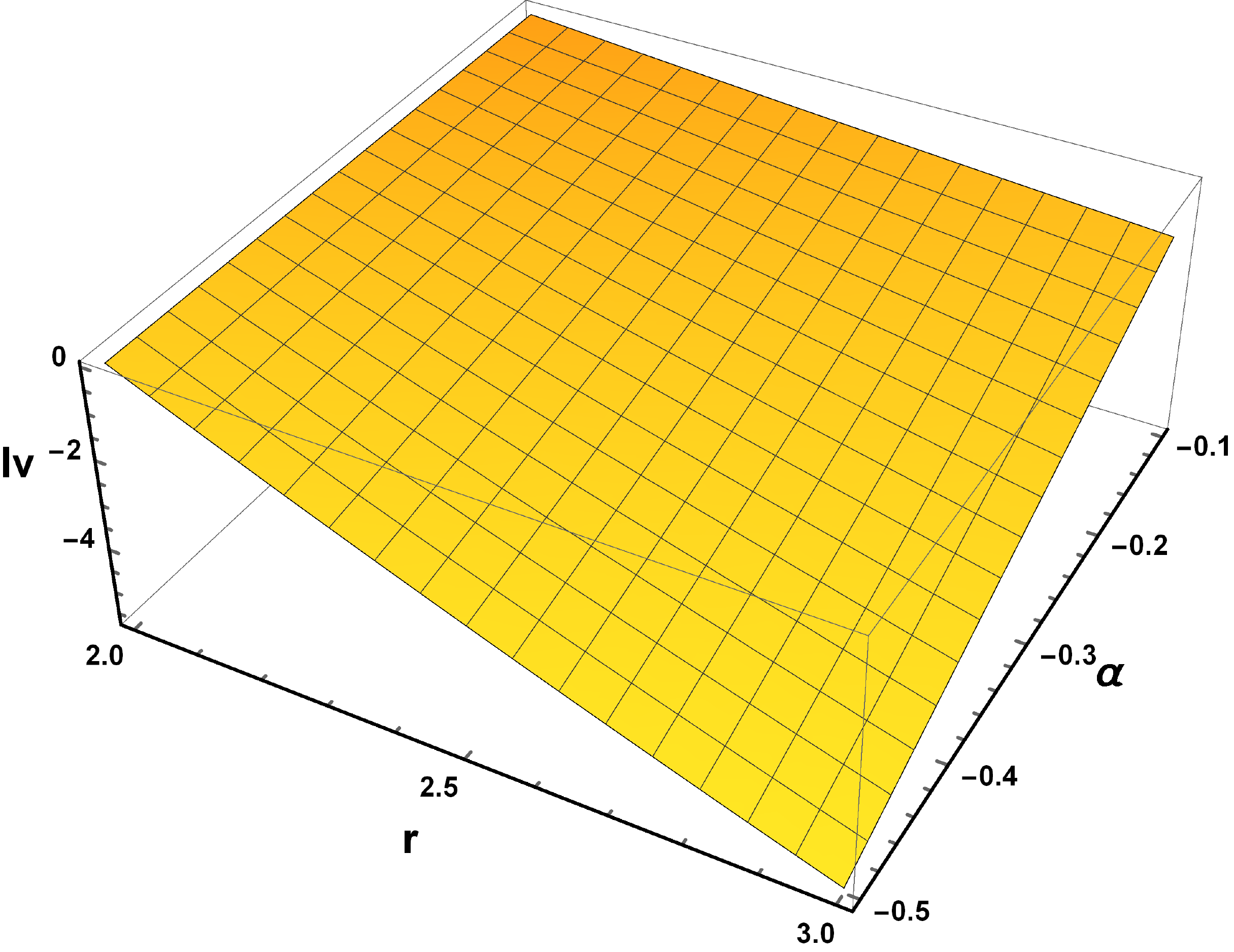}
	\caption{3D Plot for volume integral quantifier associated with model 3 and here it is clear that when $a\rightarrow r_0^+$ then $Iv\rightarrow0$ and we consider $n=-0.5$, $r_0=2$ and $a=3$.}
	\label{f21}
	\end{figure}
	\begin{figure}[H]
\centering
	\includegraphics[scale=0.27]{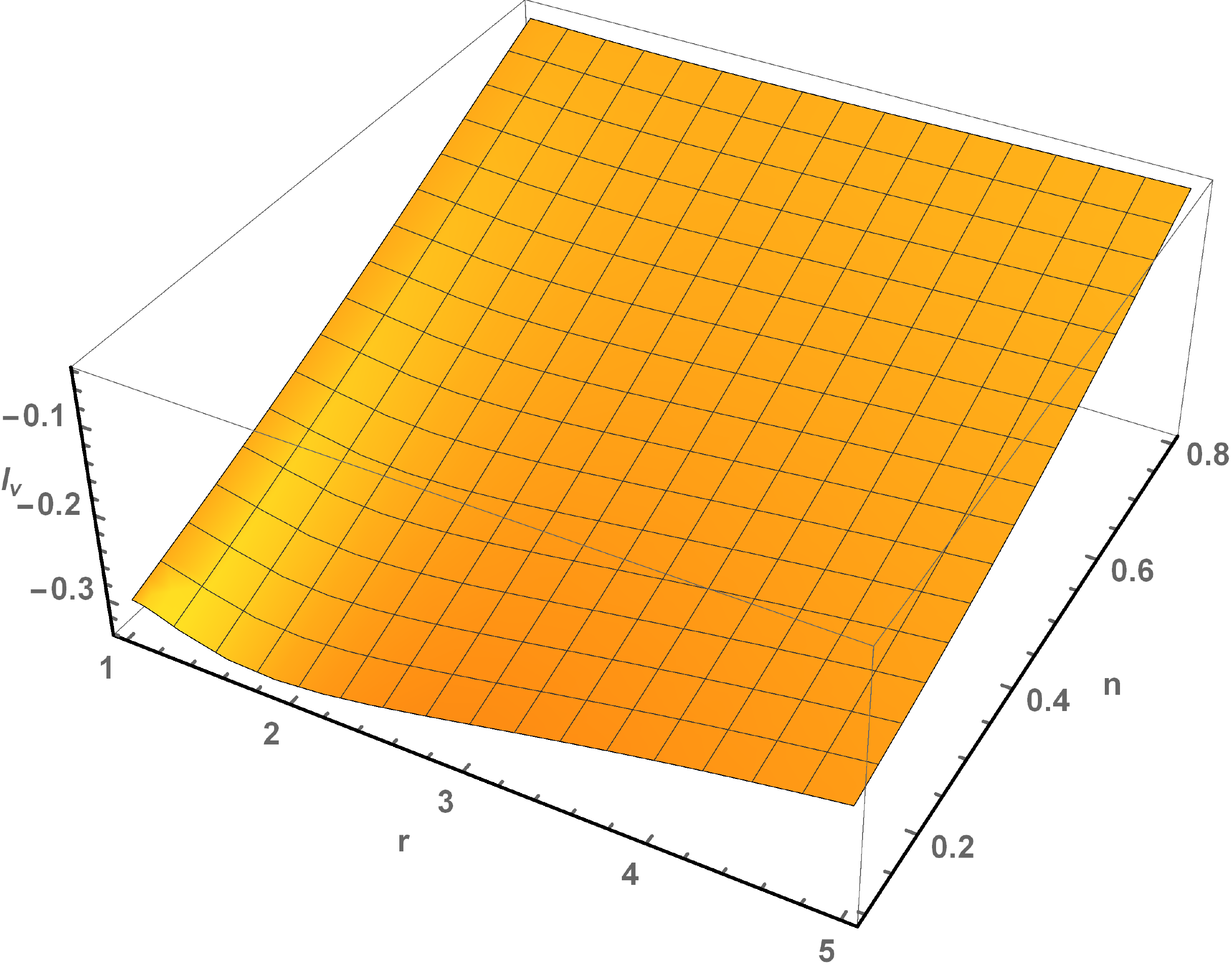}
	\caption{3D Plot for volume integral quantifier associated with model 4 and here it is clear that when $a\rightarrow r_0^+$ then $Iv\rightarrow0$ and we consider $a=0.1$ and $B=0.2$.}
	\label{f21}
	\end{figure}
	\begin{figure}[H]
\centering
	\includegraphics[scale=0.27]{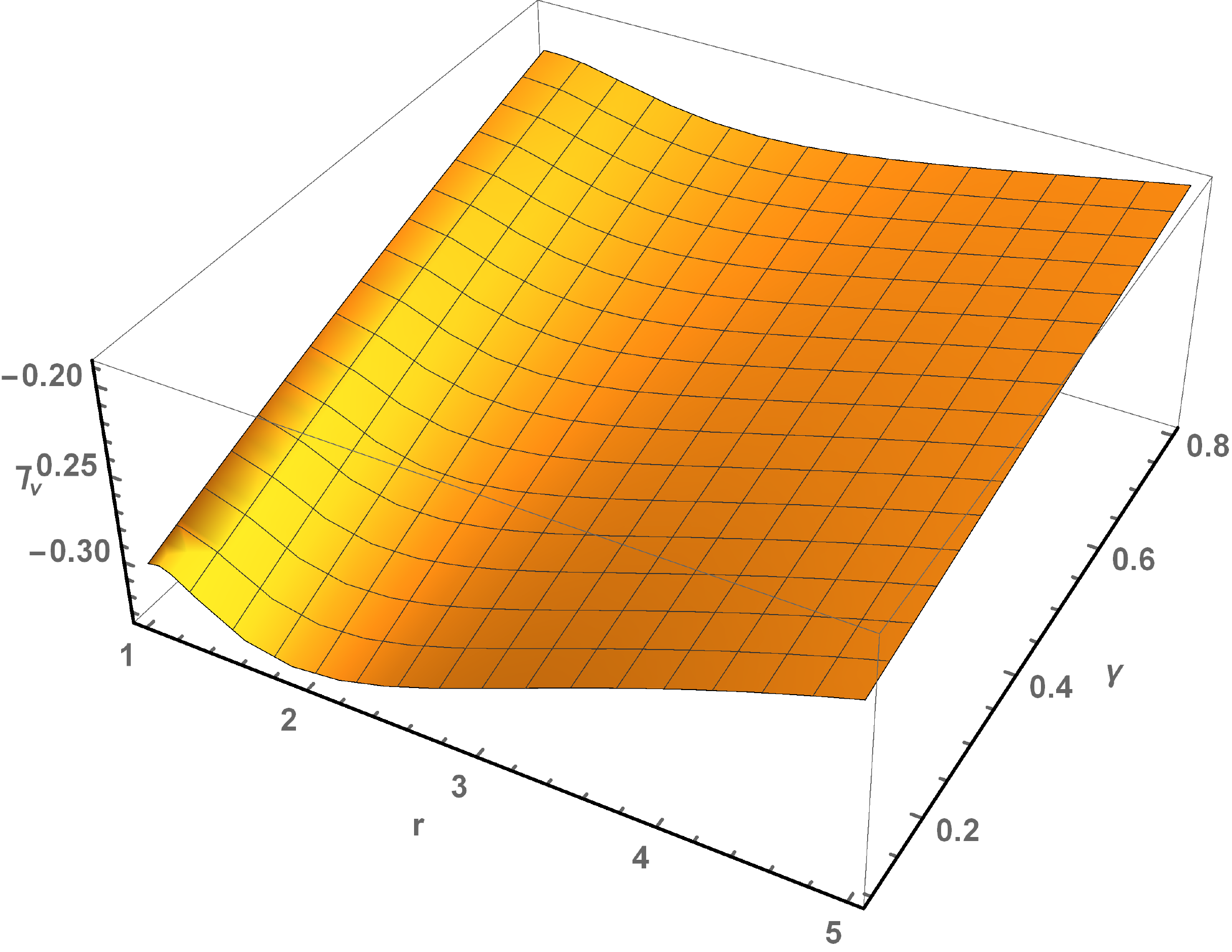}
	\caption{3D Plot for volume integral quantifier associated with model 5 and here it is clear that when $a\rightarrow r_0^+$ then $Iv\rightarrow0$ and we consider $a=0.1$ and $B=0.2$.}
	\label{f21}
	\end{figure}

\section{Concluding Remarks}\label{sec10}

In this manuscript, we have discussed Morris-Throne wormholes, i.e., static and spherically symmetric traversable wormholes in the framework of symmetric teleparallel gravity (i.e., $f(Q)$ gravity), where the gravitational interaction is described by the non-metricity term $Q$. $f(Q)$ gravity is a recently developed modified theory, so many investigations are going on to explore the current interests of the cosmological scenarios. And, the study of WH solutions in $f(Q)$ is a novel approach. Besides this, to have a traversable WH, the NEC has to be violated. That is possible in the presence of exotic matter in WH-throat, which is physically unrealistic. To explore a realistic model, it is better to minimize the usage of exotic matter. In this view, we explored three traversable WH solutions by considering a linear functional form of $Q$ (i.e., $f(Q)\propto Q$) and two WH solutions for non-linear form of $f(Q)$. Also, we considered a constant redshift function, i.e., $\Phi'(r)=0$ throughout our calculation, which simplifies our calculations and provides some exciting WH solutions.

For the first case, we have considered a relation between the radial and lateral pressure. We calculated the solution for shape function $b(r)$ in the power-law form. The parameter $'n'$ has to be negative to satisfy the asymptotically flatness condition. Keeping this in mind, we have tested all the necessary requirements of a shape function and the energy conditions. We found that the null energy condition is violated. For the second WH solution, we have considered the phantom energy equation of state (EoS), which also violates the NEC. We have explored the WH solution by taking the range of $\omega$ as $ \omega< -1$. It is interesting to note here that the energy density $\rho >0$ throughout the spacetime and the phantom energy may support traversable wormholes. In the third model, we have considered a specific shape function for $b(r)$. For this, we have discussed its' stability through the energy conditions and in this case, also NEC is violated. Thus, the violation of NEC for each model defines the possibilities of the presence of exotic matter at the wormhole's throat.

Moreover, we have discussed two WH geometries by considering a quadratic Lagrangian $f(Q)$ with two specific shape functions. To test the traversability and stability for these two models, we have tested the flaring out condition and energy conditions. Both the models successfully passed all the tests. The advantage of these types of models is that it minimizes the usage of the unknown form of matter called "exotic matter" to have a traversable wormhole in comparison to the models for linear $f(Q)$ case (as it mimics the fundamental interaction of gravity, i.e., GR). Furthermore, in the case of the non-linear form of $f(Q)$, the profiles of NEC are violated near the throat of the WHs and then satisfied (see Fig. \ref{f24} and \ref{f28}). Whereas, for linear $f(Q)$ case, it violated throughout the evolution of $r$ (see Fig. \ref{f3}, \ref{f6}, and \ref{f9}). It is worthy of exploring the wormhole geometries in the non-linear framework of $f(Q)$ gravity.

Finally, we have discussed the volume integral quantifier to measure the exotic matter required for a traversable WH. We estimate that a small amount of exotic matter is required to have a traversable WH for our three solutions. This results are aligned with \cite{Jusufi/2020}.

Therefore, it is safe to conclude that in $f(Q)$ gravity, we found suitable geometries for traversable WH that violates the NEC at its' throat. It will be interesting to explore the wormhole geometries in a more generalized $f(Q)$ form in the future, which may help to construct the traversable wormhole without presence of exotic matter.

\section*{Acknowledgements}

S.M. acknowledges Department of Science \& Technology (DST), Govt. of India, New Delhi, for awarding INSPIRE Fellowship (File No. DST/INSPIRE Fellowship/2018/IF180676). PKS acknowledges CSIR, New  Delhi, India for financial support to carry out the Research project [No.03(1454)/19/EMR-II, Dt. 02/08/2019]. We are very much grateful to the honorable referee and the editor for the illuminating suggestions that have significantly improved our work in terms of research quality and presentation.

\end{document}